%% file: NMAssiInfGen.tex
\documentclass[a4paper,11pt]{article}
\pdfoutput=1 % if your are submitting a pdflatex (i.e. if you have
\usepackage{jcappub} % for details on the use of the package, please
\usepackage[T1]{fontenc} % if needed

\title{\boldmath Nonminimally Assisted Inflation:\\ A General Analysis}

\author[a]{Sang Chul Hyun,}
\author[b,*]{Jinsu Kim,}
\author[c]{Tatsuki Kodama,} 
\author[a,d,*]{Seong Chan Park,}
\author[e]{and Tomo Takahashi}
\note[*]{Corresponding author.}

\affiliation[a]{
	Department of Physics \& IPAP \& Lab for Dark Universe, 
	Yonsei University,\\
	Seoul 03722, Korea
}
\affiliation[b]{
	School of Physics Science and Engineering, 
	Tongji University,\\
	Shanghai 200092, China
}
\affiliation[c]{
	Graduate School of Science and Engineering, 
	Saga University\\
	Saga 840-8502, Japan
}
\affiliation[d]{
	Korea Institute for Advanced Study,\\
	Seoul 02455, Korea
}
\affiliation[e]{
	Department of Physics,
	Saga University\\
	Saga 840-8502, Japan
}

\emailAdd{bsg04103@yonsei.ac.kr}
\emailAdd{kimjinsu@tongji.edu.cn}
\emailAdd{namer-namer@outlook.jp}
\emailAdd{sc.park@yonsei.ac.kr}
\emailAdd{tomot@cc.saga-u.ac.jp}

\abstract{
The effects of a scalar field, known as the ``assistant field,'' which nonminimally couples to gravity, on single-field inflationary models are studied. The analysis provides analytical expressions for inflationary observables such as the spectral index ($n_s$), the tensor-to-scalar ratio ($r$), and the local-type nonlinearity parameter ($f_{\rm NL}^{(\rm local)}$). The presence of the assistant field leads to a lowering of $n_s$ and $r$ in most of the parameter space, compared to the original predictions. In some cases, $n_s$ may increase due to the assistant field. This revives compatibility between ruled-out single-field models and the latest observations by Planck-BICEP/Keck. The results are demonstrated using three example models: loop inflation, power-law inflation, and hybrid inflation.
}

\begin{document}
\maketitle
\flushbottom

%%%%%%%%%%%%%%%%%%%%%%%%%%%%%%%%%%%%%%%%%%
\section{Introduction}
\label{sec:intro}
%%%%%%%%%%%%%%%%%%%%%%%%%%%%%%%%%%%%%%%%%%
The latest joint analysis of the Planck results \cite{Planck:2018jri} and BICEP/Keck (BK) data \cite{BICEP:2021xfz} has put a strong constraint on the spectral index, $0.958 \leq n_s \leq 0.975$ (95\% C.L.), and a stringent upper bound on the tensor-to-scalar ratio, $r \leq 0.036$ (95\% C.L.).
As a consequence, a plethora of single-field inflationary models have been ruled out. Notably, the chaotic inflation model \cite{Linde:1983gd} with a power-law potential $V(\phi) \sim \phi^p$, the power-law inflation model \cite{Lucchin:1984yf} with an exponential potential $V(\phi) \sim \exp(-\lambda\phi)$, the hybrid inflation model \cite{Cortes:2009ej} with an effectively single-field potential $V(\phi) \sim 1 + (\phi/\mu)^2$, and the loop inflation (also known as spontaneously broken SUSY) model \cite{Dvali:1994ms} with $V(\phi) \sim 1 + \lambda\log(\phi/M_{\rm P})$ are ruled out as either the tensor-to-scalar ratio $r$ or the spectral index $n_s$ is too large to be allowed by the latest bounds.

To save single-field inflationary models, many mechanisms have been proposed. For instance, an introduction of the so-called nonminimal coupling of the inflaton field to gravity of the form $\xi\phi^2 R$ \cite{Futamase:1987ua,Fakir:1990eg,Cervantes-Cota:1995ehs,Komatsu:1999mt,Bezrukov:2007ep,Park:2008hz}, where $R$ is the Ricci scalar, is known to reduce the tensor-to-scalar ratio\footnote{
See also, {\it e.g.}, Refs.~\cite{Lerner:2011ge,Linde:2011nh,Kim:2014kok,Boubekeur:2015xza,Kim:2016bem,Tenkanen:2017jih,Ferreira:2018nav,Antoniadis:2018yfq,Takahashi:2018brt,Shokri:2019rfi,Takahashi:2020car,Reyimuaji:2020goi,Cheong:2021kyc,Kodama:2021yrm} for the analysis of the effects of the nonminimal coupling to gravity in single-field inflation models for observables such as the spectral index and tensor-to-scalar ratio.
}. The nonminimally-coupled models predict similar tensor-to-scalar ratio and spectral index values as Starobinsky's $R^2$ model \cite{Starobinsky:1980te} as they are approximately equivalent to each other under the slow-roll assumption for a particular combination of the inflaton potential and the nonminimal coupling function\footnote{
See, {\it e.g.}, Refs.~\cite{Gialamas:2019nly,Gialamas:2020snr,Gialamas:2021enw,Gialamas:2022xtt} for the $R^2$-type models in the Palatini formulation. 
}. Alternatively, one may modify the inflaton kinetic term. The $\alpha$-attractor model \cite{Ferrara:2013rsa,Kallosh:2013yoa} is a well-known example where the kinetic term has a pole.
All of the aforementioned models, the nonminimally-coupled model, Starobinsky's $R^2$ model, and the $\alpha$-attractor model, have one thing in common: The potential in the Einstein frame which can be achieved via the Weyl rescaling features a plateau region in the large-field limit where the inflaton slowly rolls during inflation.

In Ref.~\cite{Hyun:2022uzc}, an alternative approach has been proposed. Instead of directly modifying the inflaton sector in such a way that the Einstein-frame potential becomes flat, an extra scalar field $s$, dubbed an assistant field, is introduced. The assistant field does not directly interact with the inflaton field $\phi$, while it talks to gravity through the nonminimal coupling $\sim s^m R$ with $m>0$. The assistant field is further assumed to be effectively massless. Therefore, the potential in the original Jordan frame is given only by the original $\phi$ field. Even though the additional assistant field $s$ has no direct interaction with the original inflaton field $\phi$, due to its nonminimal coupling to the Ricci scalar, the potential in the Einstein frame and the dynamics of inflation become nontrivial. It was shown in Ref.~\cite{Hyun:2022uzc} that the chaotic inflation model, when assisted by the assistant field, may be revived and become compatible with the latest observational constraints.

In this work, we generalise the analysis of Ref.~\cite{Hyun:2022uzc} to a general inflationary model for the original inflaton field $\phi$. We remain agnostic as to the form of the $\phi$-field potential. We follow the same nomenclature as Ref.~\cite{Hyun:2022uzc} and call the extra scalar field $s$ the assistant field. The characteristics of the assistant field are: (i) it does not directly couple to the original inflaton field $\phi$, (ii) it nonminimally couples to gravity through $s^m R$, and (iii) it is effectively massless. Including the assistant field $s$ and its explicit nonminimal coupling term, we perform a general analytical study about inflationary observables such as the spectral index and the tensor-to-scalar ratio. Our analytical formulae for the spectral index and the tensor-to-scalar ratio can readily be applied to a vast range of inflationary models, and one may easily see whether an otherwise ruled-out model can become revived with the help of the assistant field. We also discuss the non-Gaussianity and running of the spectral index that may become sizeable for multifield models, providing the corresponding analytical formula.
To demonstrate how the assistant field affects observables in concrete models, we consider as an example three models for the original $\phi$ field, namely the loop inflation model, the power-law inflation model, and the hybrid inflation model, and show how the presence of the assistant field may bring the models to the observationally-favoured region.

The rest of the paper is organised as follows. In Sec.~\ref{sec:model}, we start with a generic action for a multifield inflation model with a nonminimal coupling to gravity and perform the Weyl rescaling to the Einstein frame, setting the notations. Defining properties for the assistant field are introduced, and we set our model with a general potential for the original $\phi$ field. Section~\ref{sec:genanalysis} presents the general analytical study for the spectral index, the tensor-to-scalar ratio, and the non-Gaussianity. We show how the inflationary observables are modified in comparison with the original predictions due to the presence of the assistant field. As an application of our general analytical formulae, three examples, the loop inflation model, power-law inflation, and hybrid inflation, are discussed in Sec.~\ref{sec:examples}. We show how these models become compatible with the latest observational data. We conclude in Sec.~\ref{sec:conc}. Appendix~\ref{apdx:SIrunning} summarises the computation and general behaviour of the running of the spectral index.

%%%%%%%%%%%%%%%%%%%%%%%%%%%%%%%%%%%%%%%%%%
\section{Model}
\label{sec:model}
%%%%%%%%%%%%%%%%%%%%%%%%%%%%%%%%%%%%%%%%%%
Let us consider the following generic action for multifield inflation:
\begin{align}\label{eqn:genSJordan}
S = \int d^4x\,\sqrt{-g_{\rm J}}\left[
f(\varphi^i)g_{\rm J}^{\mu\nu}R_{{\rm J}\mu\nu}(\Gamma_{\rm J})
-\frac{1}{2}G_{{\rm J}ij}g_{\rm J}^{\mu\nu}
\partial_\mu \varphi^i \partial_\nu \varphi^j
-V_{\rm J}(\varphi^i)
\right]\,,
\end{align}
where the subscript J indicates that the action is written in the Jordan frame, $f(\varphi^i)$ is a function of fields that represents the nonminimal coupling, $G_{{\rm J}ij}$ denotes the kinetic mixing between fields, and $V_{\rm J}(\varphi^i)$ is the Jordan-frame potential\footnote{
Multifield inflation models with the nonminimal coupling to gravity have been discussed in a certain range of contexts. See, {\it e.g.}, Refs.~\cite{Kaiser:2010yu,White:2012ya,Kaiser:2012ak,Greenwood:2012aj,Kaiser:2013sna,White:2013ufa,Schutz:2013fua,Kawai:2014gqa,Kawai:2015ryj,Karamitsos:2017elm,Pi:2017gih,Cheong:2019vzl,Liu:2020zzv,Lee:2021rzy,Cheong:2022gfc,Kubota:2022pit,Geller:2022nkr,Kawai:2022emp} for such works.
}.
Here, the Ricci tensor $R_{{\rm J}\mu\nu}$ is given as a function of the connection $\Gamma_{\rm J}$ to capture not only the standard metric formulation, but also the Palatini formulation. In the Palatini formulation, the metric and the connection are {\it a priori} independent to each other, while in the metric formulation, the connection is related to the metric, becoming the Levi-Civita connection. The connection is assumed to be torsion free.

Through the Weyl rescaling, $g_{{\rm J}\mu\nu} \rightarrow g_{{\rm E}\mu\nu} = \Omega^2 g_{{\rm J}\mu\nu}$, the Jordan-frame action \eqref{eqn:genSJordan} can be brought to the Einstein frame, denoted by the subscript E. The conformal factor $\Omega^2$ is chosen to be $2f/M_{\rm P}^2$ with $M_{\rm P}$ being the reduced Planck mass.
After the Weyl rescaling, we obtain the Einstein-frame action as follows:
\begin{align}\label{eqn:genSEinstein}
S = \int d^4x \, \sqrt{-g_{\rm E}} \left[
\frac{M_{\rm P}^2}{2}g_{\rm E}^{\mu\nu}R_{{\rm E}\mu\nu}(\Gamma_{\rm E})
-\frac{1}{2}G_{{\rm E}ij}g_{\rm E}^{\mu\nu}
\partial_\mu \varphi^i \partial_\nu \varphi^j
-V_{\rm E}
\right]\,,
\end{align}
where $V_{\rm E}$ is the Einstein-frame potential given by
\begin{align}
V_{\rm E} = 
\frac{V_{\rm J}}{\Omega^4} = 
\frac{M_{\rm P}^4 V_{\rm J}}{4f^2}
\,,
\end{align}
and
\begin{align}
G_{{\rm E}ij} = \frac{M_{\rm P}^2}{2f}\left(
G_{{\rm J}ij} + \kappa \frac{3f_{,i}f_{,j}}{f}
\right)\,,
\end{align}
with $f_{,i}\equiv \partial f/\partial \varphi^i$, represents the field-space metric. The parameter $\kappa$ parametrises which framework we work in; $\kappa=0$ denotes the Palatini formulation while $\kappa=1$ corresponds to the metric formulation. Hereinafter, we omit the subscript E for brevity.

Our interests are the cases where the system is comprised of two fields, namely the original inflaton field $\phi$ and the assistant field $s$. The assistant field is effectively massless and does not directly couple to the $\phi$ field in the Jordan frame. These properties indicate that $G_{{\rm J}ij} = \delta_{ij}$, {\it i.e.}, no kinetic mixing between the two scalar fields in the original Jordan frame, and that the Jordan-frame potential $V_{\rm J}$ becomes a function of only the $\phi$ field, {\it i.e.}, $V_{\rm J} = V_{\rm J}(\phi)$. Furthermore, the assistant field nonminimally couples to gravity, {\it i.e.}, $f = f(s)$.
Upon imposing a $Z_2$ symmetry, a natural choice for the nonminimal coupling from the dimensional analysis viewpoint would be
\begin{align}\label{eqn:NMcouplingGen}
f(s) = \frac{M_{\rm P}^2}{2}\left[
1 + \xi_2\left(\frac{s}{M_{\rm P}}\right)^2
\right]\,,
\end{align}
where $\xi_2$ is a dimensionless coupling. Moreover, in order for the assistant field not to significantly alter, but to merely assist the inflationary dynamics, the nonminimal coupling term takes not too large values, {\it i.e.}, $\xi_2(s/M_{\rm P})^2 \ll 1$. We further note that the $\xi_2(s/M_{\rm P})^2 \rightarrow 0$ limit corresponds to the original $\phi$-field inflation. In the current work, we focus on the quadratic nonminimal coupling of the form \eqref{eqn:NMcouplingGen}. However, our results can easily be extended to $f(s)=(M_{\rm P}^2/2)[1+\xi_m(s/M_{\rm P})^m]$ with $m>2$ in a straightforward manner.

The action in the Einstein frame \eqref{eqn:genSEinstein} is therefore obtained as follows:
\begin{align}\label{eqn:EinsteinAction}
S = \int d^4x \, \sqrt{-g} \, \left[
\frac{M_{\rm P}^2}{2}g^{\mu\nu}R_{\mu\nu}(\Gamma)
-\frac{1}{2}\mathcal{K}_1g^{\mu\nu}\partial_\mu\phi\partial_\nu\phi
-\frac{1}{2}\mathcal{K}_2g^{\mu\nu}\partial_\mu s \partial_\nu s
-V
\right]\,,
\end{align} 
where
\begin{align}
\mathcal{K}_1 &=
\frac{M_{\rm P}^2}{2f}
\,,\\
\mathcal{K}_2 &=
\frac{M_{\rm P}^2}{2f}
+\kappa \frac{3M_{\rm P}^2}{2}\left(\frac{f_{,s}}{f}\right)^2
\,,
\end{align}
and the Einstein-frame potential is given by
\begin{align}\label{eqn:EFpot}
V = \frac{M_{\rm P}^4}{4f^2(s)}V_{\rm J}(\phi)\,.
\end{align}
We note that the potential is product-separable, $V=K(s)V_{\rm J}(\phi)$, where $K(s)=M_{\rm P}^4/(4f^2(s))$.
Upon canonically normalising the $s$ field through
\begin{align}\label{eqn:sigma-and-s}
\left(\frac{\partial\sigma}{\partial s}\right)^2 = \mathcal{K}_2\,,
\end{align}
we may rewrite the action as
\begin{align}\label{eqn:twofieldS}
S = \int d^4x \, \sqrt{-g} \, \left[
\frac{M_{\rm P}^2}{2}g^{\mu\nu}R_{\mu\nu}(\Gamma)
-\frac{1}{2}e^{2b}g^{\mu\nu}\partial_\mu\phi\partial_\mu\phi
-\frac{1}{2}g^{\mu\nu}\partial_\mu \sigma \partial_\nu \sigma
-V
\right]\,,
\end{align}
where $b=b(\sigma(s))$ is defined via
\begin{align}
e^{2b} \equiv \mathcal{K}_1 = \frac{M_{\rm P}^2}{2f}\,.
\end{align}
We note that the action \eqref{eqn:twofieldS} has been explored in detail in Refs.~\cite{Choi:2007su,Kim:2013ehu}.
For brevity, in the following, we set $M_{\rm P} = 1$.

%%%%%%%%%%%%%%%%%%%%%%%%%%%%%%%%%%%%%%%%%%
\section{General Analysis}
\label{sec:genanalysis}
%%%%%%%%%%%%%%%%%%%%%%%%%%%%%%%%%%%%%%%%%%
Inflationary observables such as the scalar power spectrum $\mathcal{P}_\zeta$, the scalar spectral index $n_s$, the tensor-to-scalar ratio $r$, and the local-type shape-independent nonlinearity parameter $f_{\rm NL}^{\rm (local)}$ for the action \eqref{eqn:twofieldS} have been obtained in, {\it e.g.}, Refs.~\cite{Kim:2013ehu,Hyun:2022uzc} by using the $\delta N$ formalism~\cite{Starobinsky:1985ibc,Salopek:1990jq,Sasaki:1995aw,Sasaki:1998ug,Lyth:2004gb} under the slow-roll approximation. 
We do not repeat the derivation here and simply list the resultant expressions.
The spectral index and the tensor-to-scalar ratio are given by
\begin{align}
n_s &= 1 - 2\epsilon^\sigma_* - 2\epsilon^\phi_*
-\frac{4e^{-2X}}{u^2\alpha^2/\epsilon_*^\sigma+v^2/\epsilon_*^\phi}
-\frac{1}{12}\frac{\eta_*^b+2\epsilon_*^b}{u^2\alpha^2/\epsilon_*^\sigma+v^2/\epsilon_*^\phi}\left(
u\alpha\sqrt{\frac{\epsilon_*^\phi}{\epsilon_*^\sigma}}
-v\sqrt{\frac{\epsilon_*^\sigma}{\epsilon_*^\phi}}
\right)^2
\nonumber\\
&\quad
+\frac{2}{u^2\alpha^2/\epsilon_*^\sigma+v^2/\epsilon_*^\phi}\left[
u^2\alpha^2\frac{\eta_*^{\sigma\sigma}}{\epsilon_*^\sigma}
+v^2\frac{\eta_*^{\phi\phi}}{\epsilon_*^\phi}
+4uv\alpha
+\frac{1}{2}s_*^bs_*^\sigma\sqrt{\epsilon_*^b\epsilon_*^\sigma}v
\left(
\frac{v}{\epsilon_*^\phi} - \frac{2u\alpha}{\epsilon_*^\sigma}
\right)
\right]
\,,\label{eqn:scalarSI}\\
r &=
\frac{16e^{-2X}}{u^2\alpha^2/\epsilon_*^\sigma + v^2/\epsilon_*^\phi}
\,,\label{eqn:TTSratio}
\end{align}
where
\begin{align}
u \equiv \frac{\epsilon_e^\sigma}{\epsilon^\sigma_e+\epsilon^\phi_e}
\,,\;\;
v \equiv \frac{\epsilon_e^\phi}{\epsilon^\sigma_e+\epsilon^\phi_e}
\,,\;\;
X \equiv
2b_e - 2b_*
\,,\;\;
\alpha \equiv
e^{2b_* - 2b_e}\left[
1 + \frac{\epsilon_e^\phi}{\epsilon_e^\sigma}\left(
1-e^{2b_e - 2b_*}
\right)
\right]
\,.\label{eqn:misc}
\end{align}
Here, $s^\sigma \equiv {\rm sgn}(V_{,\sigma})$ and $s^b \equiv {\rm sgn}(b_{,\sigma})$ denote the signs of the potential and $b$ derivatives with respect to the field $\sigma$, respectively.
The subscript $*$ ($e$) indicates that the quantities are evaluated at the pivot scale (end of inflation). The slow-roll parameters are defined as
\begin{gather}
\epsilon^\sigma \equiv 
\frac{1}{2}\left(
\frac{V_{,\sigma}}{V}
\right)^2 = 
\frac{1}{2}\left(
\frac{K_{,\sigma}}{K}
\right)^2
\,,\qquad
\epsilon^\phi \equiv 
\frac{1}{2}\left(
\frac{V_{,\phi}}{V}e^{-b}
\right)^2 = 
\frac{1}{2}\left(
\frac{V_{{\rm J},\phi}}{V_{\rm J}}\right)^2e^{-2b}
\,,\nonumber\\
\eta^{\sigma\sigma} \equiv
\frac{V_{,\sigma\sigma}}{V}
= \frac{K_{,\sigma\sigma}}{K}
\,,\qquad
\eta^{\phi\phi} \equiv
\frac{V_{,\phi\phi}}{V}e^{-2b}
= \frac{V_{{\rm J},\phi\phi}}{V_{\rm J}}e^{-2b}
\,,\nonumber\\
\eta^{\phi\sigma} \equiv
\frac{V_{,\phi\sigma}}{V}e^{-b}
\,,\qquad
\epsilon^b \equiv 8b_{,\sigma}^2
\,,\label{eqn:SRparams}
\end{gather}
and $\eta^b \equiv 16b_{,\sigma\sigma}$.
The local-type, shape-independent, nonlinearity parameter is given by
\begin{align}
-\frac{6}{5}f_{\rm NL}^{\rm (local)} &=
\frac{2e^{-X}}{(u^2\alpha^2/\epsilon_*^\sigma + v^2/\epsilon_*^\phi)^2}\Bigg[
\left(1-\frac{\eta_*^{\sigma\sigma}}{2\epsilon_*^\sigma}\right)\frac{u^3\alpha^3}{\epsilon_*^\sigma}
+\left(1-\frac{\eta_*^{\phi\phi}}{2\epsilon_*^\phi}\right)\frac{v^3}{\epsilon_*^\phi}
\nonumber\\
&\qquad
+\frac{1}{2}s_*^bs_*^\sigma \frac{u^2v\alpha^2}{\epsilon_*^\sigma}\sqrt{\frac{\epsilon_*^b}{\epsilon_*^\sigma}}
+\left(\frac{u\alpha}{\epsilon_*^\sigma} - \frac{v}{\epsilon_*^\phi}\right)^2e^X \mathcal{C}
\Bigg]
\,,\label{eqn:fNLlocal}
\end{align}
where
\begin{align}
\mathcal{C} &\equiv
\frac{\epsilon_e^\sigma \epsilon_e^\phi}{\epsilon_e^2}\left(
\frac{\epsilon_e^\sigma \eta_e^{\phi\phi} + \epsilon_e^\phi\eta_e^{\sigma\sigma}}{\epsilon_e} 
- 4\frac{\epsilon_e^\phi \epsilon_e^\sigma}{\epsilon_e} 
- \frac{1}{2}s_e^\sigma s_e^b \sqrt{\frac{\epsilon_e^b}{\epsilon_e^\sigma}}\frac{(\epsilon_e^\phi)^2}{\epsilon_e}
\right)
\,.
\end{align}
We leave detailed discussion on the running of the spectral index $\alpha_s$ in Appendix \ref{apdx:SIrunning}.

The Einstein-frame potential for the assistant field is given by
\begin{align}
K(s) = \frac{1}{(1+\xi_2 s^2)^2}\,,
\end{align}
as can be seen from Eqs.~\eqref{eqn:NMcouplingGen} and \eqref{eqn:EFpot}. Thus, it is straightforward to obtain the slow-roll parameters for the $s$ field or, equivalently, its canonically-normalised version $\sigma$. They are given by
\begin{gather}
\epsilon^\sigma = \epsilon^b =
\frac{8\xi_2^2 s^2}{1+\xi_2 s^2 (1+6\kappa\xi_2)}
\,,\quad
\eta^{\sigma\sigma} =
\frac{4\xi_2[-1 + 3\xi_2 s^2 + 4 \xi_2^2 s^4 (1+6\kappa\xi_2)]
}{[1+\xi_2 s^2(1+6\kappa\xi_2)]^2}
\,,
\end{gather}
where we have used Eq.~\eqref{eqn:sigma-and-s}.
Note also that $\eta^b =
-16\xi_2(1+\xi_2 s^2)/[1+\xi_2 s^2(1+6\kappa\xi_2)]^2$.
One may regard the $s$-field value at the pivot scale, $s_*$, as an input parameter\footnote{
Quantum kick for the assistant field in the de Sitter phase can be estimated as $(\Delta \sigma)^2 = \mathcal{K}_2 (\Delta s)^2 \simeq N^2H_*^2/(2\pi)^2 \simeq N^2rA_s/8$, where $A_s\approx2\times10^{-9}$ is the magnitude of the scalar power spectrum at the pivot scale and $N$ is the number of $e$-folds. In the $\xi_2 s^2\ll 1$ limit, we find $\Delta s \lesssim 4.2\times10^{-4}$ ($1.8\times10^{-4}$) for $r<0.2$
(0.035) with $N=60$. As far as a relatively larger value of $s$ is considered, such fluctuations can safely be neglected.
}. The $s$-field value at the end of inflation is then given by requiring the number of $e$-folds, $N$, to be $N_e$; in this work, we choose $N_e=60$ from the end of inflation.
The number of $e$-folds can be solely given by the assistant field $s$,
\begin{align}\label{eqn:efoldings}
N = \int_e^* \frac{K}{K_{,\sigma}} \, d\sigma
= \frac{1}{4\xi_2}\left[
\ln\left(\frac{s_e}{s_*}\right)
+3\kappa\xi_2\ln\left(
\frac{1+\xi_2 s_e^2}{1+\xi_2 s_*^2}
\right)
\right]
\,.
\end{align}
In the Palatini formulation, $\kappa = 0$, and one may obtain an exact analytical solution for $s_e$ as
\begin{align}\label{eqn:send-spivot-P}
s_e = s_* e^{4\xi_2 N_e}\,.
\end{align}
In the metric formulation, on the other hand, such a closed analytical form for the solution does not exist. However, an approximated form may be found. 
In this paper, we assume that the nonminimal coupling of the assistant field $s$ is small. Thus, we can take the small nonminimal coupling limit where $\xi_2 s^2 \ll 1$. In this limit, we may approximate the number of $e$-folds as
\begin{align}
N \approx \frac{1}{4\xi_2}
\ln\left(\frac{s_e}{s_*}\right)
+\frac{3}{4}\xi_2\left(
s_e^2 - s_*^2
\right)
\,.
\end{align}
Requiring $N = N_e$ then gives
\begin{align}\label{eqn:send-spivot-M}
s_e \approx s_* e^{4\xi_2 N_e}
\left[
1 - 3\xi_2^2 s_*^2 \left(
e^{8\xi_2 N_e} - 1
\right)
\right]
\,.
\end{align}
\begin{figure}[t!]
\centering
\includegraphics[width=0.9\linewidth]{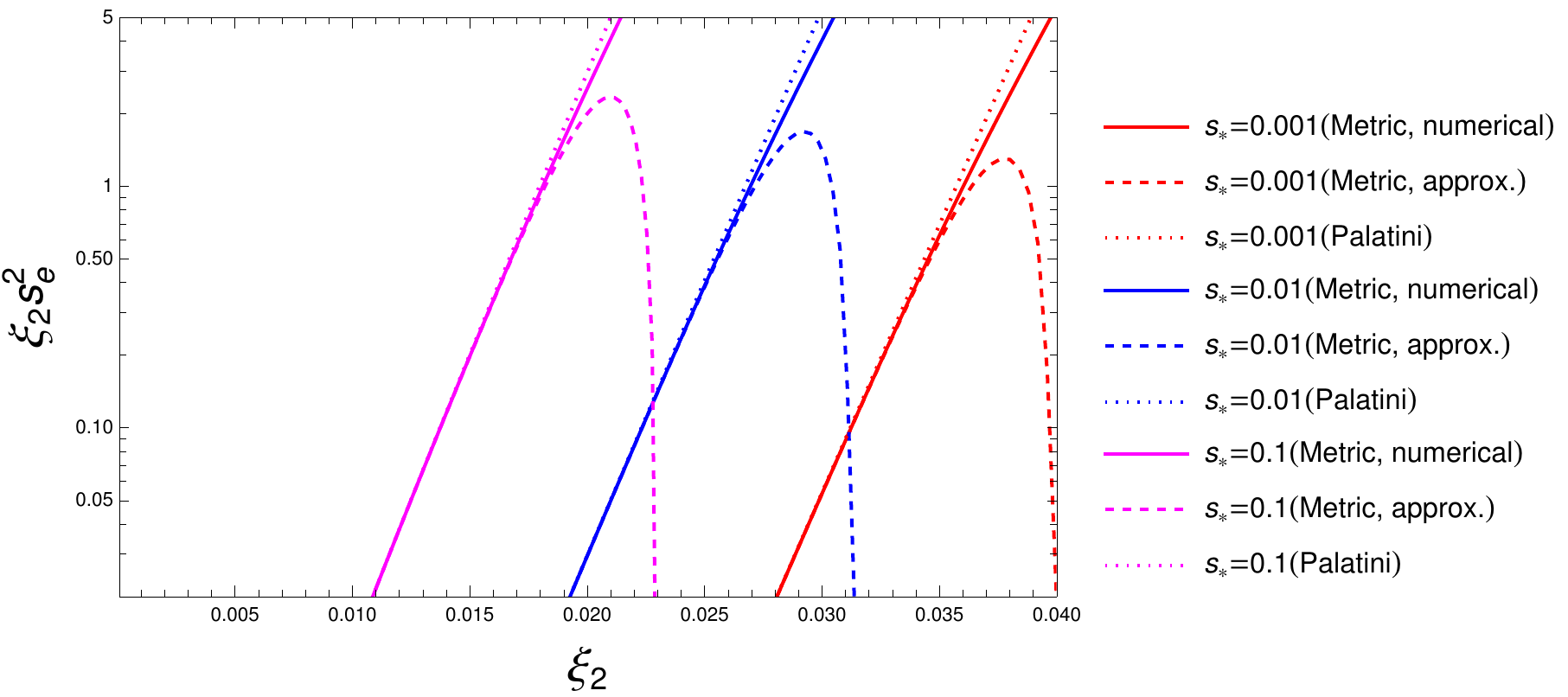}
\caption{Comparison between the analytical solution for $s_e$ in the Palatini formulation (dotted lines), the approximated analytical solution for $s_e$ in the metric formulation (dashed lines), and the numerical solution for $s_e$ in the metric formulation (solid lines), for three different $s_*$ values, 0.001 (red), 0.01 (blue), and 0.1 (magenta). We observe that the approximated analytical solution in the metric case starts to deviate from the numerical solution as the small nonminimal coupling assumption breaks down. We also see that the difference between the Palatini and the metric formulations is negligible in the small nonminimal coupling limit.}
\label{fig:send-comparison}
\end{figure}

Figure~\ref{fig:send-comparison} shows the comparison between the analytical solution for $s_e$ in the Palatini formulation \eqref{eqn:send-spivot-P}, the approximated analytical solution for $s_e$ in the metric formulation \eqref{eqn:send-spivot-M}, and the numerical solution for $s_e$ found by using $N=N_e$ with Eq.~\eqref{eqn:efoldings} in the metric formulation. We stress that the solution \eqref{eqn:send-spivot-P} in the Palatini case is exact.
Deviations between the approximated solution and the exact numerical solution in the metric case start to appear when the assumption of the small nonminimal coupling breaks down; we shall use $\xi_2 s_e^2 \leq 0.1$ as the condition for the small nonminimal coupling in this paper.
We also observe that the difference between the Palatini case and the metric case is negligible in the small nonminimal coupling limit; deviations become visible when the nonminimal coupling gets larger.

In order to discuss effects of the assistant field $s$ on the original $\phi$-field inflation model in a general way, we remain as agnostic as possible as to the form for the $\phi$-field potential, $V_{\rm J}(\phi)$. 
In the absence of the assistant field, the spectral index and the tensor-to-scalar ratio for the original $\phi$-field inflation model, denoted respectively by $n_s^{(0)}$ and $r^{(0)}$, are given by
\begin{align}
n_s^{(0)} = 1 - 6\epsilon^{(0)}_* + 2\eta^{(0)}_*
\,,\quad
r^{(0)} = 16\epsilon^{(0)}_*
\,,
\end{align}
under the slow-roll approximation, where the slow-roll parameters are defined as
\begin{align}
\label{eqn:SRparams0}
\epsilon^{(0)} \equiv 
\frac{1}{2}\left(
\frac{V_{{\rm J},\phi}}{V_{\rm J}}
\right)^2
\,,\quad
\eta^{(0)} \equiv
\frac{V_{{\rm J},\phi\phi}}{V_{\rm J}}
\,.
\end{align}
Comparing with the slow-roll parameters defined in Eq.~\eqref{eqn:SRparams}, we see that
\begin{align}
\epsilon^\phi = \epsilon^{(0)}e^{-2b}
\,,\quad
\eta^{\phi\phi} = \eta^{(0)}e^{-2b}
\,.
\end{align}
Writing $\epsilon^{(0)}_*$ and $\eta^{(0)}_*$ in terms of $n^{(0)}_s$ and $r^{(0)}$, we get
\begin{align}\label{eqn:epseta-nsr-phi}
\epsilon^\phi_* = \frac{r^{(0)}}{16}e^{-2b_*}
\,,\quad
\eta^{\phi\phi}_* = \frac{1}{2}\left(
n^{(0)}_s - 1 + \frac{3}{8}r^{(0)}
\right)e^{-2b_*}
\,.
\end{align}
We may thus substitute Eq.~\eqref{eqn:epseta-nsr-phi} into the observables, Eqs.~\eqref{eqn:scalarSI}, \eqref{eqn:TTSratio}, and \eqref{eqn:fNLlocal}, to get rid of the information on the $\phi$ field at the pivot scale.
There still remains, however, dependence on the $\phi$ field at the end of inflation through the $\phi$-field slow-roll parameters $\epsilon^\phi_e$ and $\eta^{\phi\phi}_e$ or, equivalently, $\epsilon^{(0)}_e$ and $\eta^{(0)}_e$.
In the current work, we consider two classes.

%%%%%%%%%%%%%%%%%%%%%%%%%%%%%%%%%%%%%%%%%%
\subsection{Class I: End of inflation via slow-roll violations}
\label{subsec:classI}
%%%%%%%%%%%%%%%%%%%%%%%%%%%%%%%%%%%%%%%%%%
The first class we consider is the case where end of inflation is set by $\epsilon_e \approx 1$, {\it i.e.}, violation of slow roll.
Examples include the chaotic inflation model with a power-law potential \cite{Linde:1983gd} and the loop inflation model with a logarithmic correction \cite{Dvali:1994ms}.
In this case, one may replace $\epsilon^\phi_e$ by $1 - \epsilon^\sigma_e$. 
Utilising Eq.~\eqref{eqn:epseta-nsr-phi}, one may express $n_s$ and $r$ in terms of $\xi_2$, $s_*$, $s_e$, $n_s^{(0)}$, and $r^{(0)}$.
The spectral index is given by
\begin{align}
n_s^{({\rm I})} &= 1
-\frac{(1+\xi_2 s_*^2)r^{(0)}}{8}
-\frac{16\xi_2^2 s_*^2}{1+\xi_2 s_*^2 (1 + 6 \kappa \xi_2)}
-\frac{32 (1 + \xi_2 s_e^2)^2
\left(1 + \xi_2 s_e^2 (1 + 6 \kappa \xi_2)\right)^2}{A_{\rm I}}
\nonumber\\
&\quad
+\frac{4 \xi_2^3  (1 + \xi_2 s_*^2)
\left(1-\xi_2^2 s_*^4 (1 + 6 \kappa \xi_2)\right)r^{(0)}}
{3 A_{\rm I} \left(1 + \xi_2 s_*^2 (1 + 6 \kappa \xi_2)\right)^3}
\nonumber\\
&\quad\quad\times 
\bigg[
\frac{\left(1 + \xi_2 s_*^2 (1 + 6 \kappa \xi_2)\right) B_{\rm I}}{4 \xi_2 s_*}
-\frac{32 s_*}{r^{(0)}} \left(1 + \xi_2 s_e^2 (1 - (8 - 6 \kappa) \xi_2)\right)
\bigg]^2
\nonumber\\
&\quad
-\frac{8 (1 + \xi_2 s_*^2)^2}{A_{\rm I}}
\bigg[
\frac{2 \left(1 + \xi_2 s_e^2 (1 - (8 - 6 \kappa) \xi_2)\right)^2}{r^{(0)}}
\left(8 - 8 n_s^{(0)} - 3 r^{(0)}\right)
\nonumber\\
&\quad\quad 
-\frac{6 \xi_2 \left(1 + \xi_2 s_e^2 (1 - (8 - 6 \kappa) \xi_2)\right) B_{\rm I}}{1 + \xi_2 s_*^2}
+\frac{\xi_2 \left(1 - 3 \xi_2 s_*^2 - 4 \xi_2^2 s_*^4 - 24 \kappa \xi_2^3 s_*^4\right) B_{\rm I}^2}{s_*^2 (1 + \xi_2 s_*^2)^2 (1 + \xi_2 s_*^2 (1 + 6 \kappa \xi_2))}
\nonumber\\
&\quad\quad
-\frac{128 \xi_2^2 s_*^2 \left(1 + \xi_2 s_e^2 \left(1 - (8 - 6 \kappa) \xi_2\right)\right)^2}{r^{(0)} (1 + \xi_2 s_*^2) \left(1 + \xi_2 s_*^2 (1 + 6 \kappa \xi_2)\right)}
\bigg]\,,\label{eqn:ns-I}
\end{align}
where
\begin{align}
A_{\rm I} &=
\frac{128 (1 + \xi_2 s_*^2) \left(1 + \xi_2 s_e^2 \left(1 - (8 - 6 \kappa) \xi_2\right)\right)^2}{r^{(0)}}
+\frac{\left(1 + \xi_2 s_*^2 (1 + 6 \kappa \xi_2)\right)B_{\rm I}^2}{s_*^2}
\,,\\
B_{\rm I} &=
\left(1 + \xi_2 s_e^2\right)\left(s_e^2 - s_*^2\right)
+8\xi_2 s_e^2
+2\xi_2^2 s_e^2 \left(
4 s_*^2 + 3 \kappa \left(s_e^2 - s_*^2\right)
\right)
\,.
\end{align}
The same procedure can be repeated for the tensor-to-scalar ratio. We obtain
\begin{align}
r^{({\rm I})} &= 128\big[
1+\xi_2 s_e^2(1+6\kappa\xi_2)
\big]^2
\left(\frac{1+\xi_2s_e^2}{1+\xi_2s_*^2}\right)^2
\nonumber\\
&\quad\times
\Bigg[
\frac{128}{r^{(0)}(1+\xi_2s_*^2)}
\big[
1+\xi_2s_e^2(1-8\xi_2+6\kappa\xi_2)
\big]^2
\nonumber\\
&\quad\quad
+\frac{1+\xi_2s_*^2(1+6\kappa\xi_2)}{s_*^2(1+\xi_2s_*^2)^2}
\big[
(1+\xi_2s_e^2(1+6\kappa\xi_2))(s_*^2-s_e^2)-8\xi_2s_e^2(1+\xi_2s_*^2)
\big]^2
\Bigg]^{-1}
\,.\label{eqn:r-I}
\end{align}
The expressions \eqref{eqn:ns-I} and \eqref{eqn:r-I} are exact so long as the slow-roll approximation holds. 
Together with Eq.~\eqref{eqn:send-spivot-P} or Eq.~\eqref{eqn:send-spivot-M}, we can determine the spectral index as well as the tensor-to-scalar ratio with a set of input parameters $\{\kappa,\xi_2,s_*,n_s^{(0)},r^{(0)}\}$\footnote{
To be precise, both $n_s^{(0)}$ and $r^{(0)}$ depend on the assistant field, namely $\xi_2$ and $s_*$, as the $\phi$-field value at the end of inflation changes when we take into account the dynamics of the assistant field. The dependence is, however, negligible as the small nonmiminal coupling limit is taken.
}.

The local-type nonlinearity parameter for Class I is given by
\begin{align}
f_{\rm NL, (I)}^{({\rm local})} &=
-\frac{10 (1 + \xi_2 s_*^2)^3 (1 + \xi_2 s_e^2)}{3 A_{\rm I}^2\left(1 + \xi_2 s_e^2 (1 + 6 \kappa \xi_2)\right)^2} \Bigg\{
\frac{B_{\rm I}^3 \left(1 + \xi_2 s_e^2 (1 + 6 \kappa \xi_2)\right)^3}{s_*^4 (1 + \xi_2 s_*^2)^2}
\nonumber\\
&\quad 
+\frac{256 \left(1 + \xi_2 s_e^2 (1 + 6 \kappa \xi_2)\right)^3 \left(1 + \xi_2 s_e^2 \left(1 - (8 - 6 \kappa) \xi_2\right)\right)^3}{(r^{(0)})^2 (1 + \xi_2 s_*^2)}\left(8 - 8 n_s^{(0)} - r^{(0)}\right)
\nonumber\\
&\quad 
+\frac{2 B_{\rm I}^2 \left(1 + \xi_2 s_*^2 (1 + 6 \kappa \xi_2)\right) \left(1 + \xi_2 s_e^2 (1 + 6 \kappa \xi_2)\right)^3 \left(1 + \xi_2 s_e^2 \left(1 - (8 - 6 \kappa) \xi_2\right)\right)}{s_*^2 (1 + \xi_2 s_*^2)^2}
\nonumber\\
&\quad 
-\frac{512 \left(1 + \xi_2 s_e^2 \left(1 - (8 - 6 \kappa) \xi_2\right)\right)}{(r^{(0)})^2 s_*^4 (1 + \xi_2 s_*^2)(1 + \xi_2 s_e^2)}
\nonumber\\
&\quad \times \Bigg[
8 \xi_2 s_*^2 s_e \left(1 + \xi_2 s_e^2 \left(1 - (8 - 6 \kappa) \xi_2\right)\right) - \frac{1}{16} B_{\rm I} r^{(0)} s_e \left(1 + \xi_2 s_*^2 (1 + 6 \kappa \xi_2)\right)
\Bigg]^2
\nonumber\\
&\quad \times \Bigg[
\left(1 + \xi_2 s_e^2 \left(1 - (8 - 6 \kappa) \xi_2\right)\right)^2
\nonumber\\
&\quad\quad
+ 8 \xi_2 \bigg[
\frac{\left(1 + 5 \xi_2 s_e^2 + 4 \xi_2^2 s_e^4 (1 + 6 \kappa \xi_2)\right) \left(1 + \xi_2 s_e^2 \left(1 - (8 - 6 \kappa) \xi_2\right)\right)}{1 + \xi_2 s_e^2 (1 + 6 \kappa \xi_2)}
\nonumber\\
&\quad\quad 
-2 \eta_e^{(0)} \xi_2 s_e^2 (1 + \xi_2 s_e^2) \left(1 + \xi_2 s_e^2 (1 + 6 \kappa \xi_2)\right)
\bigg]
\Bigg]
\Bigg\}
\,.
\label{eqn:fNL-I}
\end{align}
Compared to the tensor-to-scalar ratio and the spectral index, the nonlinearity parameter contains one more parameter, namely $\eta_e^{(0)}$. From Eq.~\eqref{eqn:fNL-I}, we note that the prefactor associated with $\eta_e^{(0)}$ is naturally small in the small nonminimal coupling limit. Consequently, the dependence on $\eta_e^{(0)}$ is negligible. Thus, when presenting model-independent results, we shall consider $\eta_e^{(0)}=0.01$. Different choices of $\eta_e^{(0)}$ do not lead to any visible difference.

\begin{figure}[t!]
\centering
\includegraphics[width=0.48\linewidth]{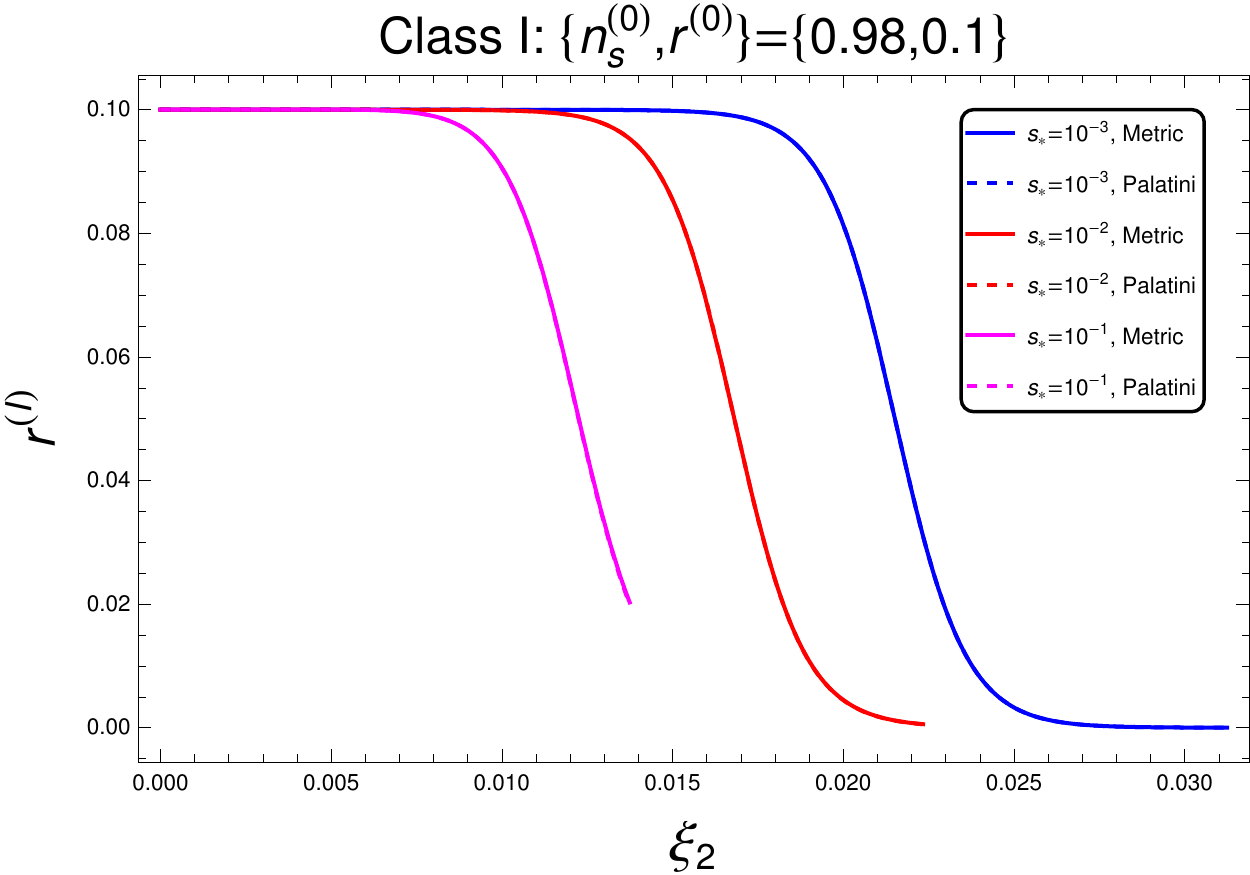}
\includegraphics[width=0.48\linewidth]{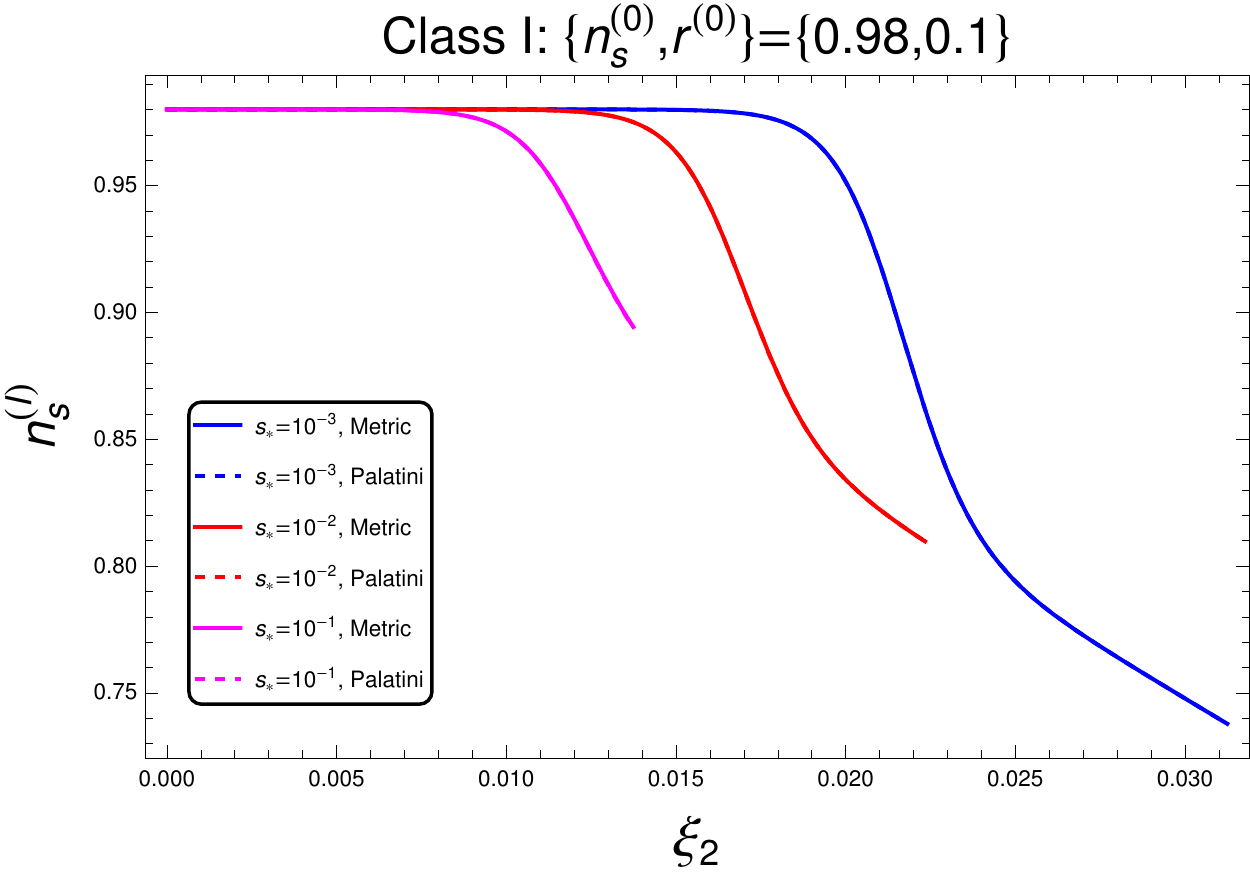}
\\
\includegraphics[width=0.48\linewidth]{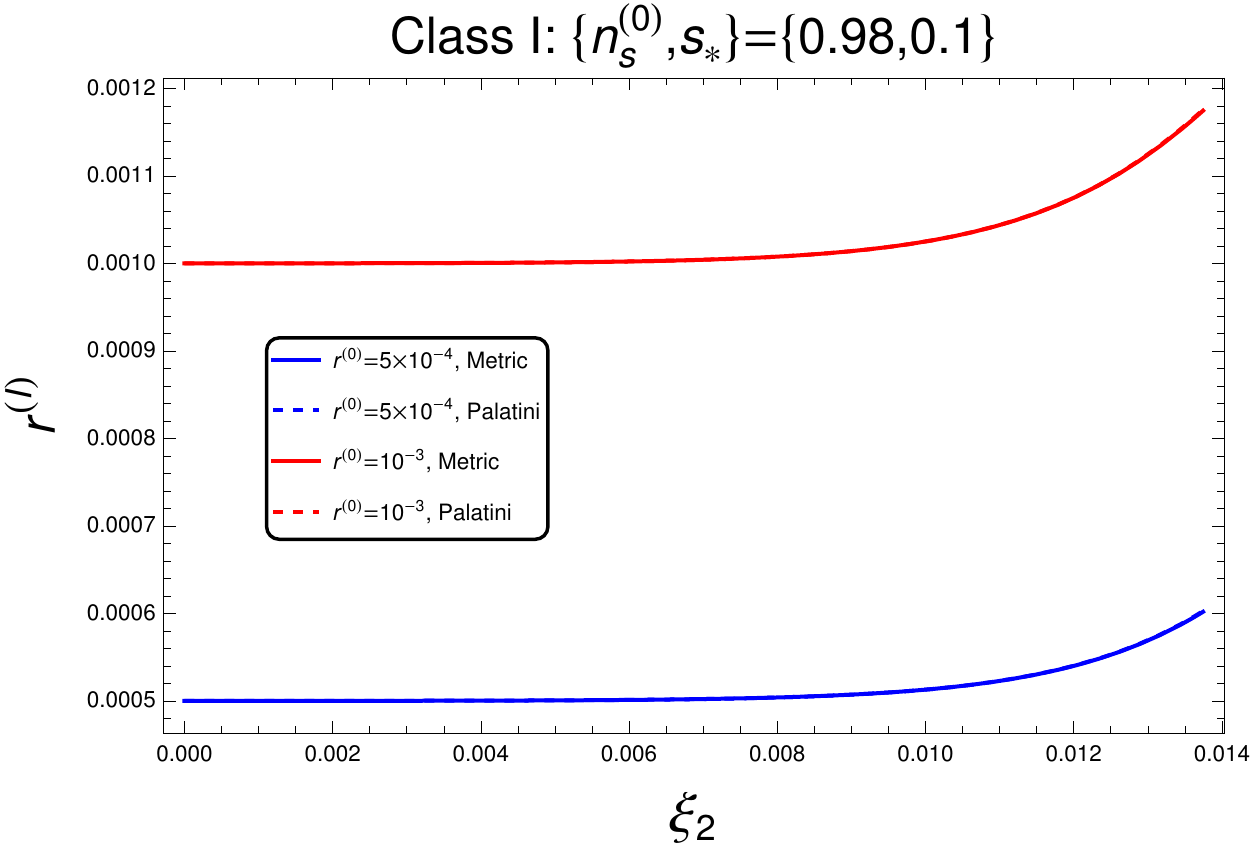}
\includegraphics[width=0.48\linewidth]{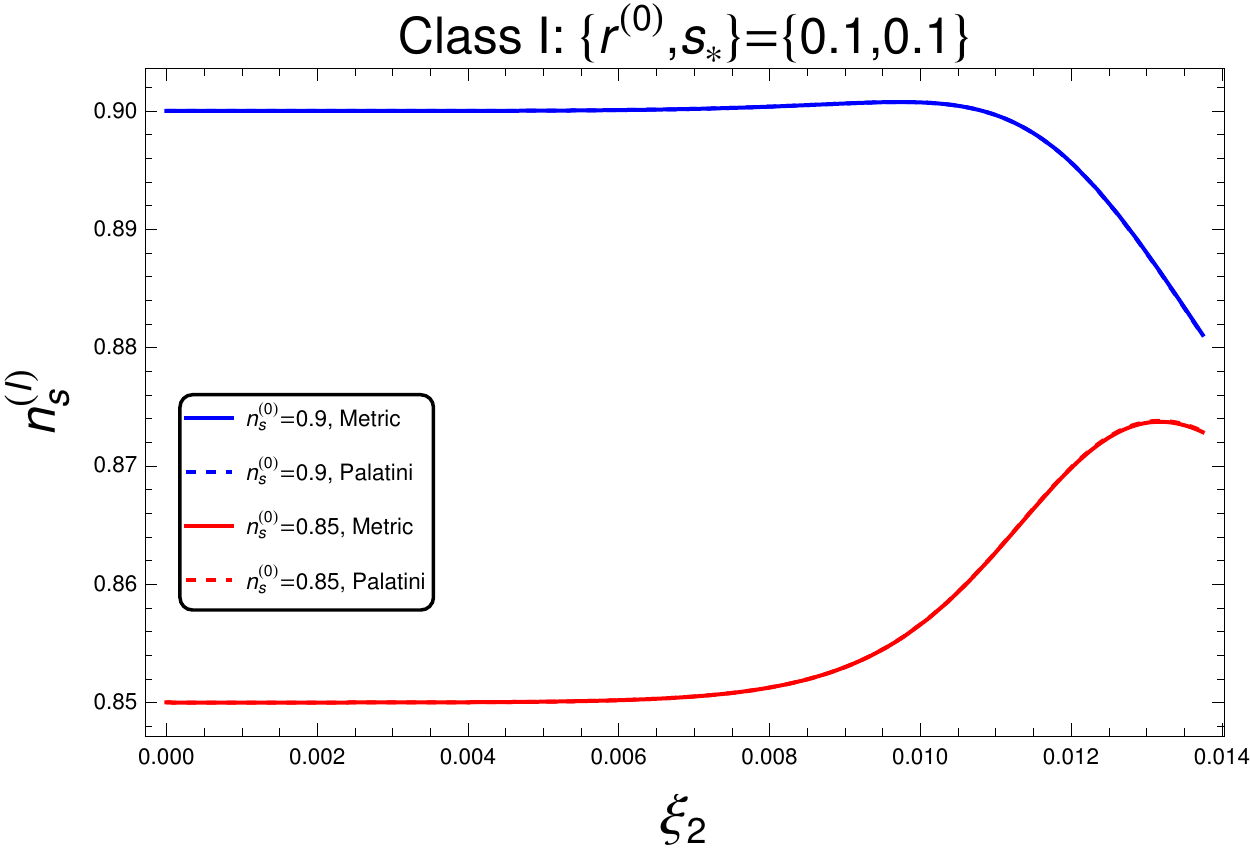}
\caption{In the upper panel, we show the dependence of the spectral index $n_s^{({\rm I})}$ (right panel) and the tensor-to-scalar ratio $r^{({\rm I})}$ (left panel) on the nonminimal coupling parameter $\xi_2$ for three different values of $s_*$, 0.001 (blue), 0.01 (red), and 0.1 (magenta), with the choice of $\{n_s^{(0)},r^{(0)}\}=\{0.98, 0.1\}$. The maximum values of $\xi_2$ are chosen in such a way that $\xi_2 s_e^2 = 0.1$ so that the condition of small nonminimal couplings is met.
We observe that both the spectral index and the tensor-to-scalar ratio decrease as the nonminimal coupling parameter increases; this behaviour is the general tendency as shown in Fig.~\ref{fig:nsr-I}.
In the lower panel, we see the behaviour of the tensor-to-scalar ratio for $r^{(0)}=5\times 10^{-4}$ (blue) and $10^{-3}$ (red) with the choice of $\{n_s^{(0)},s_*\}=\{0.98,0.1\}$ (left panel) and the behaviour of the spectral index for $n_s^{(0)}=0.9$ (blue) and $0.85$ (red) with $\{r^{(0)},s_*\}=\{0.1,0.1\}$ (right panel). When $n_s^{(0)}$ or $r^{(0)}$ takes a rather small value with a relatively large value of $s_*$, we observe the increasing behaviour. However, the change of tensor-to-scalar ratio is minuscule, and the increase of the spectral index is not large enough to go inside the latest Planck-BK bounds. 
In both the upper and lower panels, we see that the difference between the Palatini formulation (dashed) and the metric formulation (solid) is negligible as expected.
}
\label{fig:ns-r-I}
\end{figure}

\begin{figure}[t!]
\centering
\includegraphics[width=0.8\linewidth]{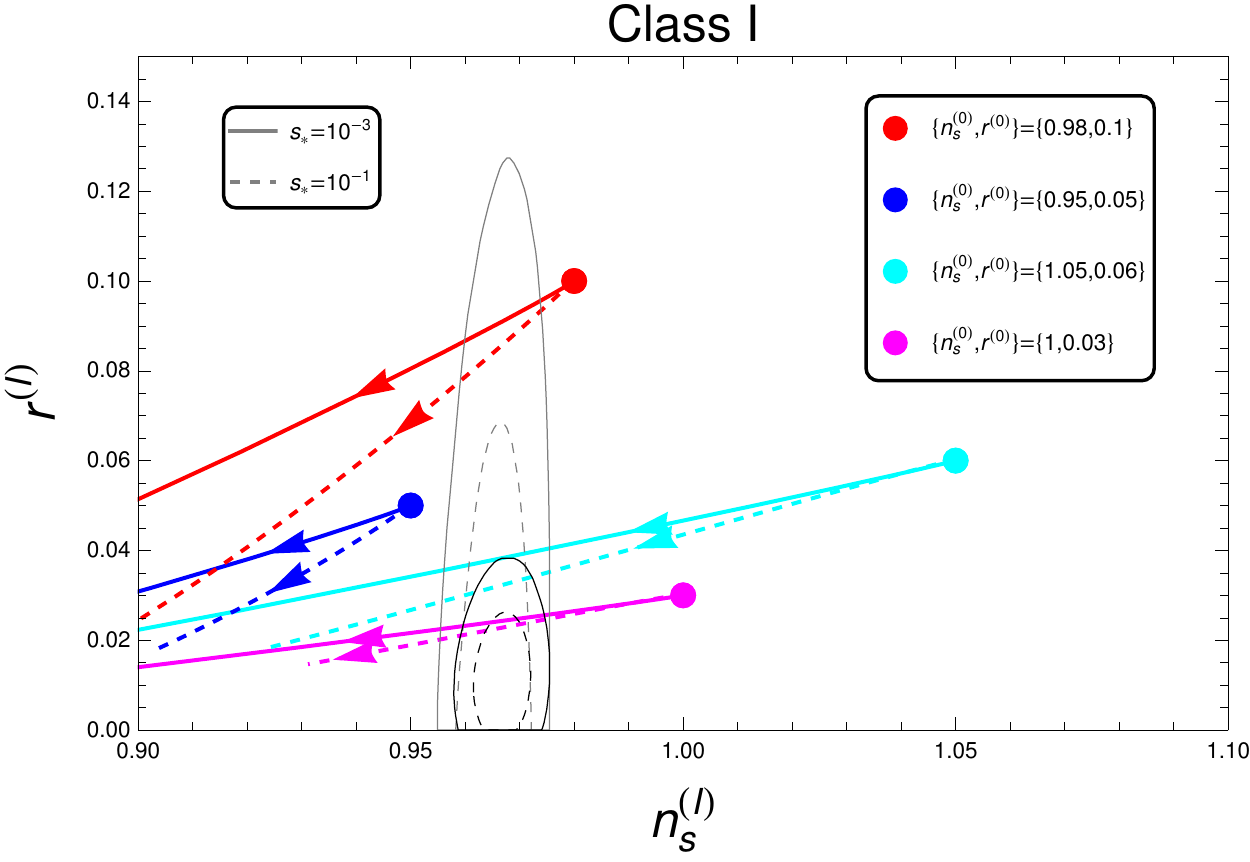}
\caption{Behaviour of the spectral index $n_s^{({\rm I})}$ and the tensor-to-scalar ratio $r^{({\rm I})}$ for $\{n_s^{(0)},r^{(0)}\}=\{0.98,0.1\}$ (red), $\{0.95, 0.05\}$ (blue), $\{1.05,0.06\}$ (cyan), and $\{1,0.03\}$ (magenta) with $s_*=10^{-3}$ (solid) and $10^{-1}$ (dashed). Only the metric cases are presented here as there is little difference between the Palatini and the metric formulations. The latest Planck-BK 1-sigma (2-sigma) bound is depicted as the black solid (dashed) line, while the Planck-only 1-sigma (2-sigma) bound is presented with the grey solid (dashed) line. Both the spectral index and the tensor-to-scalar ratio decrease as the nonminimal coupling parameter $\xi_2$ takes a larger value as indicated by arrows; the maximum value of $\xi_2$ is chosen such that $\xi_2 s_e^2 = 0.1$, {\it i.e.}, until the small nonminimal coupling condition is valid. Thus, one may save some otherwise ruled-out models; models that originally predict large spectral index and/or tensor-to-scalar ratio can be brought to the observationally-favoured region.}
\label{fig:nsr-I}
\end{figure}

The upper panel of Fig.~\ref{fig:ns-r-I} shows the dependence of the spectral index $n_s^{({\rm I})}$ (right panel) and the tensor-to-scalar ratio $r^{({\rm I})}$ (left panel) on $\xi_2$ for three different $s_*$ values with the choice of $\{n_s^{(0)},r^{(0)}\}=\{0.98, 0.1\}$. 
The ranges of $\xi_2$ are chosen such that $\xi_2 s_e^2$ satisfies the condition $\xi_2 s_e^2 \leq 0.1$ in order that the analytic formulae work well.
One may observe that both the spectral index and the tensor-to-scalar ratio decrease as the nonminimal coupling parameter $\xi_2$ increases.
The lower panel of Fig.~\ref{fig:ns-r-I} presents the behaviour of the tensor-to-scalar ratio (left panel) and the spectral index (right panel) for rather small values of $r^{(0)}$ and $n_s^{(0)}$, respectively, with a relatively large value of $s_*=0.1$. In this case, we observe the opposite behaviour; both the spectral index and the tensor-to-scalar ratio increase as the nonminimal coupling parameter $\xi_2$ increases. However, the change of the tensor-to-scalar ratio is minuscule, and the increase of the spectral index is not large enough to be allowed by the latest Planck-BK bounds.
For all the cases presented, the difference between the Palatini and metric formulations is negligible as the nonminimal coupling is small.
For most of the parameter space, the general tendency is that both the spectral index as well as the tensor-to-scalar ratio become lowered as we increase $\xi_2$.
Therefore, we may bring models that originally predict large $\{n_s^{(0)}, r^{(0)}\}$ to the observationally-favoured region as demonstrated in Fig.~\ref{fig:nsr-I} on the $n_s$--$r$ plane.

\begin{figure}[t!]
\centering
\includegraphics[scale=0.57]{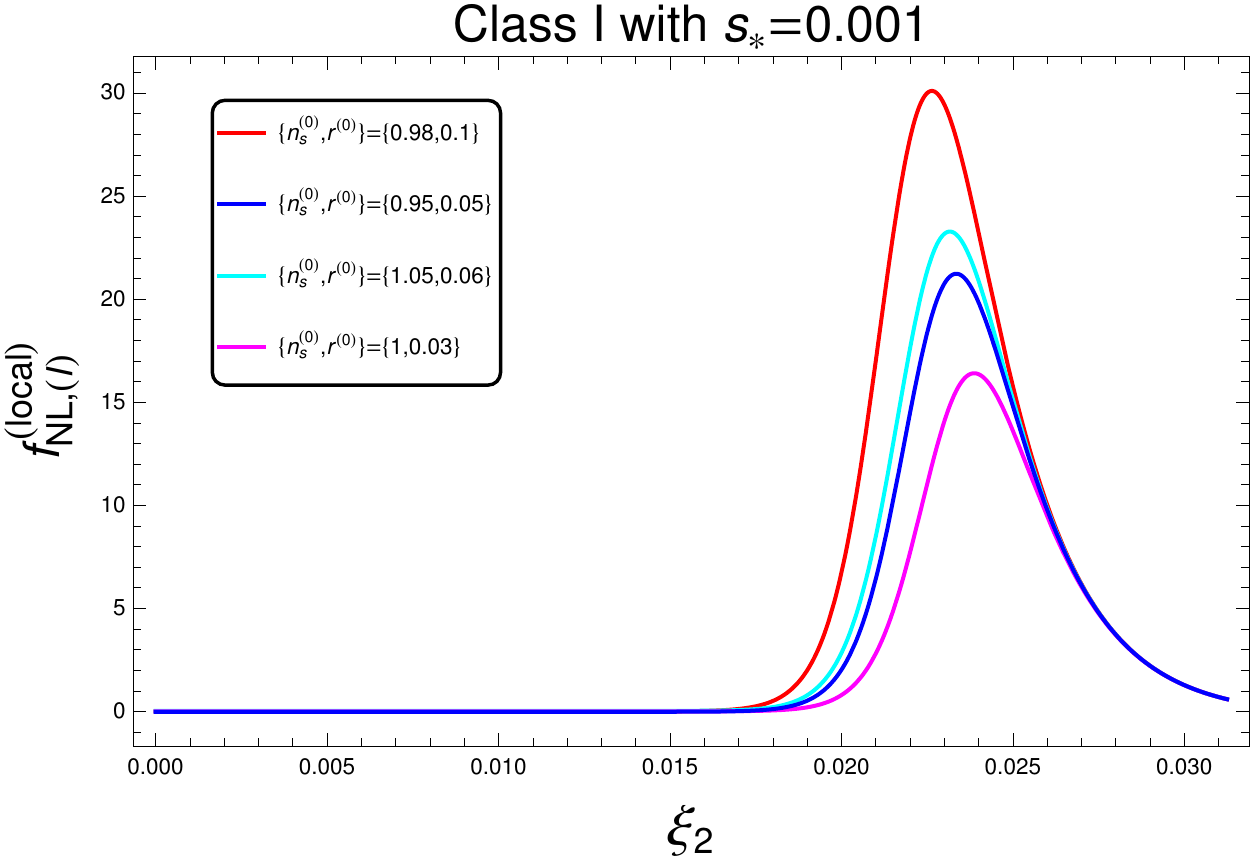}
\includegraphics[scale=0.59]{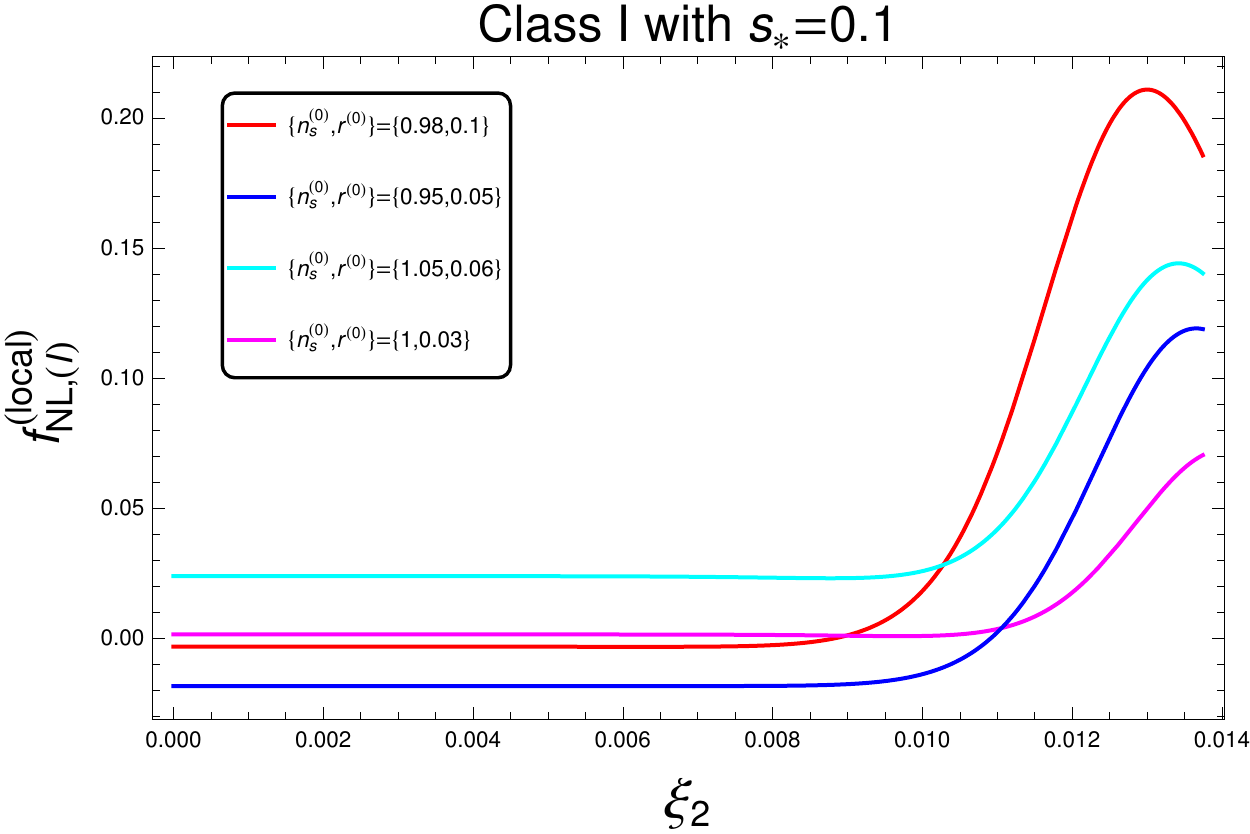}
\\
\includegraphics[scale=0.585]{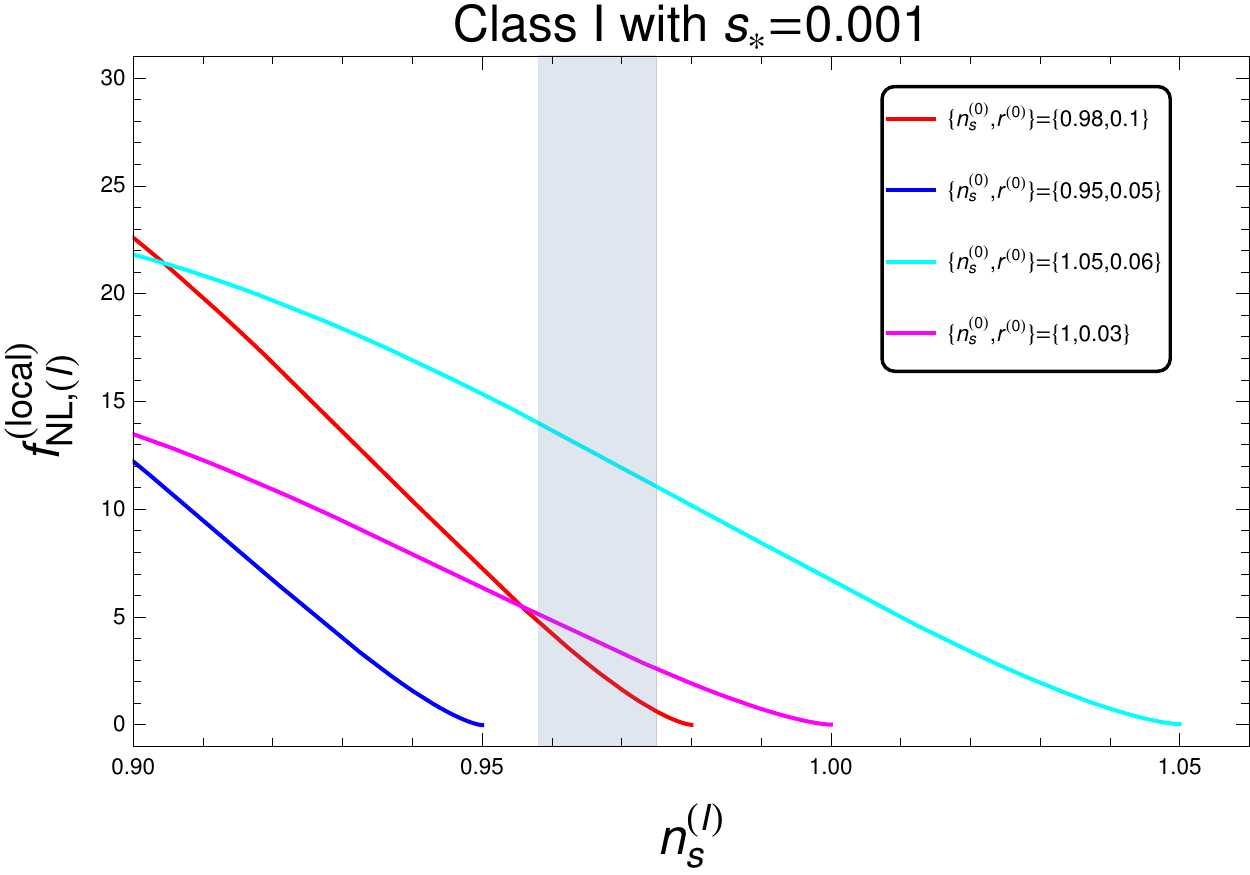}
\includegraphics[scale=0.59]{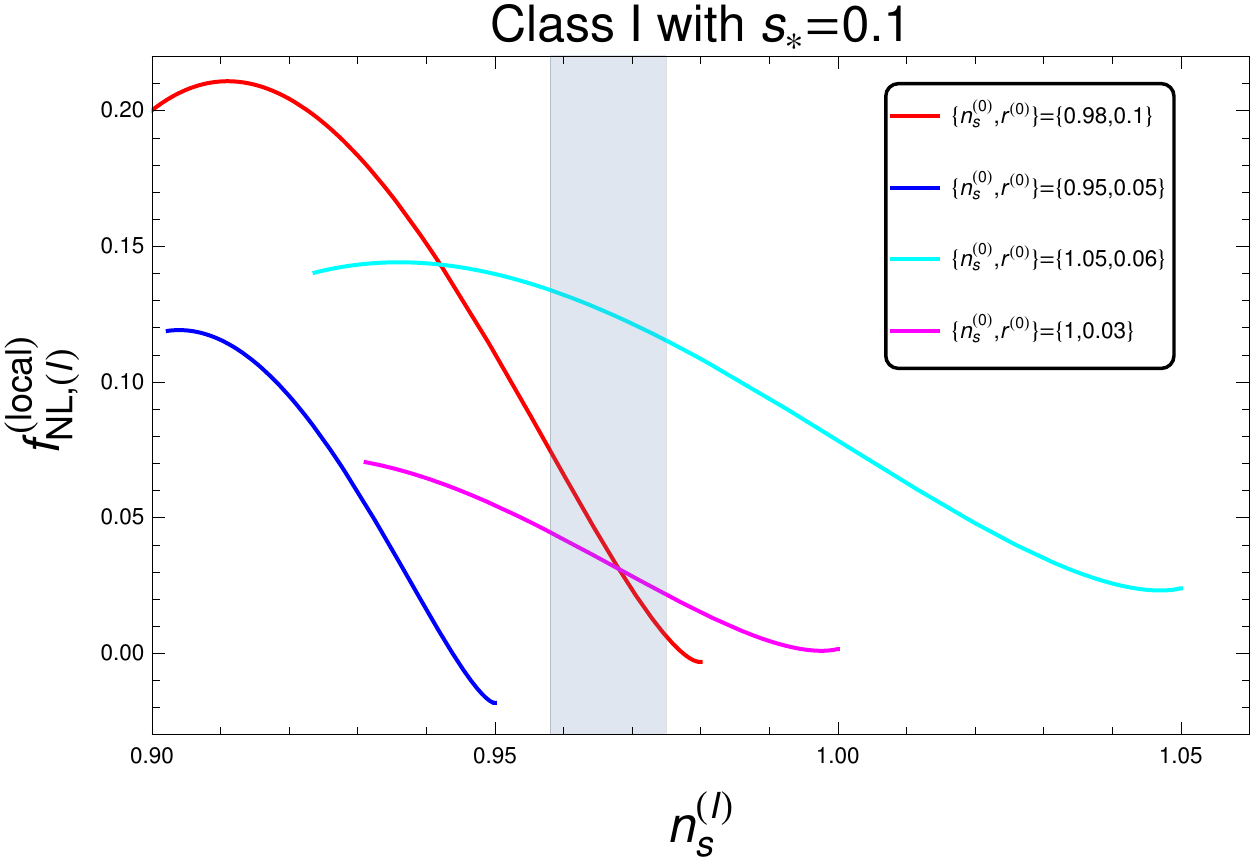}
\caption{
In the upper panel, the evolution of the nonlinearity parameter in terms of the nonminimal coupling parameter is shown for Class I. Two values of $s_*$ are considered, 0.001 (left panel) and 0.1 (right panel) with various choices of $\{n_s^{(0)},r^{(0)}\}=\{0.98,0.1\}$ (red), $\{0.95,0.05\}$ (blue), $\{1.05,0.06\}$ (cyan), and $\{1.0,0.03\}$ (magenta). 
We observe that the nonlinearity parameter initially shows an increasing behaviour.
In the lower panel, we present the predictions in the $f_{\rm NL, (I)}^{({\rm local})}$--$n_s^{({\rm I})}$ plane. The shaded region corresponds to the latest bounds on the spectral index. In the case of $s_*=0.001$, the nonlinearity parameter for the $\{n_s^{(0)},r^{(0)}\}=\{1.05,0.06\}$ case becomes slightly larger than the Planck 2-sigma bound, $-11.1 < f_{\rm NL}^{({\rm local})} < 9.3$. In the case of $s_*=0.1$, however, all the constraints are safely satisfied.
For all cases, $\eta_e^{(0)}=0.01$ is chosen. However, the dependence on $\eta_e^{(0)}$ is weak as the prefactor is small in the small nonminimal coupling limit. Only the metric formulation is presented as there is little difference between the metric and the Palatini formulations.
}
\label{fig:fNL-I}
\end{figure}

The behaviour of the nonlinearity parameter as we increase the nonminimal coupling parameter $\xi_2$ is shown in the upper panel of Fig.~\ref{fig:fNL-I}. There is little difference between the metric and the Palatini formulations, and thus, only the metric formulation is presented.
Four different choices of $\{n_s^{(0)},r^{(0)}\}$ are considered, $\{0.98,0.1\}$, $\{0.95,0.05\}$, $\{1.05,0.06\}$, and $\{1.0,0.03\}$, and two values of $s_*$ are chosen, 0.001 and 0.1. We observe that $f_{\rm NL}^{({\rm local})}$ initially increases as we increase $\xi_2$, and then it decreases. Moreover, we note that the $f_{\rm NL}^{({\rm local})}$ value becomes larger when the $s_*$ value is small which is an expected behaviour from the analytical expression \eqref{eqn:fNL-I}.
In the lower panel of Fig.~\ref{fig:fNL-I}, the predictions are shown in the $f_{\rm NL}^{({\rm local})}$--$n_s$ plane together with the observationally-allowed bound for the spectral index. For the case of $s_*=0.001$, the nonlinearity parameter for the $\{n_s^{(0)},r^{(0)}\}=\{1.05,0.06\}$ case becomes slightly larger than the Planck 2-sigma bound \cite{Planck:2019kim}, $-11.1 < f_{\rm NL}^{({\rm local})} < 9.3$. In the case of $s_*=0.1$, however, all the constraints are safely satisfied.
It is interesting to note that constraints on $f_{\rm NL}^{\rm (local)}$ can be improved as $|f_{\rm NL}^{\rm (local)}| < {\cal O}(0.1) -{\cal O}(1)$ in future galaxy surveys and 21 cm line of neutral hydrogen observations \cite{Camera:2013kpa,Ferramacho:2014pua,Raccanelli:2014kga,Yamauchi:2014ioa,Camera:2014bwa,dePutter:2014lna,Munoz:2015eqa,Yamauchi:2015mja,Sekiguchi:2018kqe}. Hence, the currently allowed parameter range can even further be probed by such future observations of non-Gaussianity.

%%%%%%%%%%%%%%%%%%%%%%%%%%%%%%%%%%%%%%%%%%
\subsection{Class II: End of inflation via a separate sector}
\label{subsec:classII}
%%%%%%%%%%%%%%%%%%%%%%%%%%%%%%%%%%%%%%%%%%

The second class we consider includes scenarios where end of inflation is provided by a separate sector other than the $\phi$ and $s$ fields.
Examples include the power-law inflation with an exponential potential \cite{Lucchin:1984yf} and the hybrid inflation model \cite{Cortes:2009ej}.
In the case of power-law inflation, the slow-roll parameters in the original $\phi$-field model become constant. Thus, inflation does not end via the standard slow-roll violation, and a separate end-of-inflation mechanism is needed. In the hybrid inflation model, the end of inflation is achieved by the so-called waterfall field\footnote{
There exists a parameter space where end of inflation via the breaking of the slow-roll condition is possible. This case falls into Class I discussed in Sec.~\ref{subsec:classI}.
}.
In the presence of the assistant field $s$, one may use the $s$ field to end inflation via the slow-roll violation, $\epsilon_e \approx 1$, in which case Class I applies\footnote{
When the assistant field is responsible for end of inflation, $s_*$ (or $\xi_2$) becomes a given quantity rather than a free, input parameter.
}. The violation of slow-roll via a nonminimal coupling can also be realised in some single-field nonminimal inflation models \cite{Takahashi:2020car}.
In this subsection, we do not assume that the assistant field is responsible for end of inflation. Instead, we assume that there is a separate mechanism that ends inflation without substantially modifying the observables.

As we do not have the condition $\epsilon_e \approx 1$, we are no longer able to replace $\epsilon_e^\phi$. In other words, information on the $\phi$ field at the end of inflation or, equivalently, $\epsilon_e^{(0)}$, is required. One exception is when the $\phi$-field slow-roll parameters are approximately constant, as in the power-law inflation case; in this special case, we have $\epsilon_e^{(0)} \approx \epsilon_*^{(0)} = r^{(0)}/16$, and thus, we do not need any information on the $\phi$-field value at the end of inflation.
We aim to present a general analysis without relying on a specific model. Therefore, we remain agnostic with respect to the original $\phi$-field model and treat $\epsilon_e^{(0)}$ as an independent parameter.
Choosing a specific model for the $\phi$ field would correspond to choosing a specific value for $\epsilon_e^{(0)}$.

From Eq.~\eqref{eqn:scalarSI}, we obtain the spectral index as follows:
\begin{align}
n_s^{({\rm II})} &= 1 - \frac{(1 + \xi_2 s_*^2)r^{(0)}}{8}
-\frac{16\xi_2^2 s_*^2}{1 + \xi_2 s_*^2 (1 + 6 \kappa \xi_2)}
\nonumber\\
&\quad
-\frac{32 (1 + \xi_2 s_e^2)^2}{A_{\rm II} \left(1 + \xi_2 s_e^2 (1 + 6 \kappa \xi_2)\right)^2}
\Big[
8 \xi_2^2 s_e^2
+ \epsilon^{(0)}_e (1 + \xi_2 s_e^2) \left(1 + \xi_2 s_e^2 (1 + 6 \kappa \xi_2)\right)
\Big]^2
\nonumber\\
&\quad
+\frac{r^{(0)}\xi_2 (1 + \xi_2 s_*^2) (1 + \xi_2 s_e^2)^2 \left(1 - \xi_2^2 s_*^4 (1 + 6 \kappa \xi_2)\right)}{12 A_{\rm II} s_*^2 \left(1 + \xi_2 s_*^2 (1 + 6 \kappa \xi_2)\right)}
\nonumber\\
&\quad\quad\times
\left[
\frac{B_{\rm II}}{1 + \xi_2 s_e^2 (1 + 6 \kappa \xi_2)}
-\frac{128 \epsilon^{(0)}_e \xi_2 s_*^2}{r^{(0)} \left(1 + \xi_2 s_*^2 (1 + 6 \kappa \xi_2)\right)}
\right]^2
\nonumber\\
&\quad 
-\frac{16 (1 + \xi_2 s_*^2)^2}{A_{\rm II}}
\Bigg[
\frac{(1 + \xi_2 s_e^2)^2}{r^{(0)}}
\left(\epsilon^{(0)}_e\right)^2 \left(8 - 8n_s^{(0)} - 3r^{(0)}\right)
\nonumber\\
&\quad\quad
-\frac{3 B_{\rm II} \epsilon^{(0)}_e \xi_2 (1 + \xi_2 s_e^2)^2}{(1 + \xi_2 s_*^2) \left(1 + \xi_2 s_e^2 (1 + 6 \kappa \xi_2)\right)}
-\frac{64 \xi_2^2 s_*^2 (1 + \xi_2 s_e^2)^2}{r^{(0)} (1 + \xi_2 s_*^2) \left(1 + \xi_2 s_*^2 (1 + 6 \kappa \xi_2)\right)}
\left(\epsilon^{(0)}_e\right)^2
\nonumber\\
&\quad\quad
+\frac{B_{\rm II}^2 \xi_2 (1 + \xi_2 s_e^2)^2 
\left(1 - \xi_2 s_*^2 (3 + 4 \xi_2 s_*^2 (1 + 6 \kappa \xi_2))\right)}{2 s_*^2 (1 + \xi_2 s_*^2)^2 
\left(1 + \xi_2 s_*^2 (1 + 6 \kappa \xi_2)\right) 
\left(1 + \xi_2 s_e^2 (1 + 6 \kappa \xi_2)\right)^2}
\Bigg]
\,,\label{eqn:ns-II}
\end{align}
where
\begin{align}
A_{\rm II} &= 
\frac{128 (1 + \xi_2 s_*^2) (1 + \xi_2 s_e^2)^2}{r^{(0)}}
\left(\epsilon^{(0)}_e\right)^2
+\frac{(1 + \xi_2 s_e^2)^2 \left(1 + \xi_2 s_*^2 (1 + 6 \kappa \xi_2)\right) B_{\rm II}^2}{s_*^2 \left(1 + \xi_2 s_e^2 (1 + 6 \kappa \xi_2)\right)^2}
\,,\\
B_{\rm II} &=
8 \xi_2 s_e^2 + \epsilon^{(0)}_e \left(s_e^2 - s_*^2\right)
\left(1 + \xi_2 s_e^2 (1 + 6 \kappa \xi_2)\right)
\,.
\end{align}
Similarly, we find the tensor-to-scalar ratio as
\begin{align}
r^{({\rm II})} &=
\frac{128 (1 + \xi_2 s_e^2)^2
\left(
8 \xi _2^2 s_e^2
+\epsilon^{(0)}_e (1 + \xi_2 s_e^2) 
\left(1 + \xi_2 s_e^2 (1 + 6 \kappa \xi_2)\right)
\right)^2}{\left(
1 + \xi_2 s_e^2 (1 + 6 \kappa \xi_2)
\right)^2}
\nonumber\\
&\quad\times
\Bigg[
\frac{(1 + \xi_2 s_e^2)^2 \left(1 + \xi_2 s_*^2 (1 + 6 \kappa \xi_2)\right)}{s_*^2 \left(1 + \xi_2 s_e^2 (1 + 6 \kappa \xi_2)\right)^2}
\Big[
8 \xi_2 s_e^2 + \epsilon^{(0)}_e (s_e^2 - s_*^2) \left(1 + \xi_2 s_e^2 (1 + 6 \kappa \xi_2)\right)
\Big]^2
\nonumber\\
&\quad\quad
+\frac{128 
(1 + \xi_2 s_*^2)
(1 + \xi_2 s_e^2)^2}{r^{(0)}}
\left(\epsilon^{(0)}_e\right)^2
\Bigg]^{-1}
\,.\label{eqn:r-II}
\end{align}
Note that these expressions are exact so long as the slow-roll approximation holds.
Together with the solution for $s_e$, Eq.~\eqref{eqn:send-spivot-P} or Eq.~\eqref{eqn:send-spivot-M}, the spectral index \eqref{eqn:ns-II} as well as the tensor-to-scalar ratio \eqref{eqn:r-II} can be computed with a set of input parameters $\{\kappa,\xi_2,s_*,\epsilon_e^{(0)},n_s^{(0)},r^{(0)}\}$.
Class II is more general than Class I in the sense that the expressions for Class I, Eq.~\eqref{eqn:ns-I} and Eq.~\eqref{eqn:r-I}, can be obtained by taking $\epsilon_e^{(0)} = (1-\epsilon_e^\sigma)e^{2b_e}$.
The special case of the constant slow-roll parameter can be captured by setting $\epsilon_e^{(0)} = r^{(0)}/16$.

The local-type nonlinearity parameter for Class II is given by
\begin{align}
f_{\rm NL, (II)}^{({\rm local})} &=
-\frac{10 (1 + \xi_2 s_*^2) (1 + \xi_2 s_e^2)}{3 A_{\rm II}^2}
\left(
\epsilon_e^{(0)} (1 + \xi_2 s_e^2)
+\frac{8 \xi_2^2 s_e^2}{1 + \xi_2 s_e^2 (1 + 6 \kappa \xi_2)}
\right)
\nonumber\\
&\quad \times 
\Bigg[ 
\frac{256 (\epsilon_e^{(0)})^3 (1 + \xi_2 s_*^2) (1 + \xi_2 s_e^2)^3}{(r^{(0)})^2}\left(8 - 8 n_s^{(0)} - r^{(0)}\right)
\nonumber\\
&\quad\quad 
+ \frac{2 B_{\rm II}^2 \epsilon_e^{(0)} (1 + \xi_2 s_e^2)^3
\left(1 + \xi_2 s_*^2 (1 + 6 \kappa \xi_2) \right)}{s_*^2 \left(1 + \xi_2 s_e^2 (1 + 6 \kappa \xi_2)\right)^2}
+\frac{B_{\rm II}^3 (1 + \xi_2 s_e^2)^3}{s_*^4 \left(1 + \xi_2 s_e^2 (1 + 6 \kappa \xi_2)\right)^3}
\Bigg]
\nonumber\\
&\quad 
+\frac{20 \epsilon_e^{(0)} s_e^2 (1 + \xi_2 s_*^2)^2 (1 + \xi_2 s_e^2)^4}{3 A_{\rm II}^2 \left(1 + \xi_2 s_e^2 (1 + 6 \kappa \xi_2)\right)^3}
\left(
\epsilon_e^{(0)} (1 + \xi_2 s_e^2)
+\frac{8 \xi_2^2 s_e^2}{1 + \xi_2 s_e^2 (1 + 6 \kappa \xi_2)}
\right)^{-1}
\nonumber\\
&\quad\times 
\left(
\frac{B_{\rm II} \left(1 + \xi_2 s_*^2 (1 + 6 \kappa \xi_2)\right)}{s_*^2 \left(1 + \xi_2 s_e^2 (1 + 6 \kappa \xi_2)\right)}
-\frac{128 \xi_2 \epsilon_e^{(0)}}{r^{(0)}}
\right)^2
\Big[
(\epsilon_e^{(0)})^2 (1 + \xi_2 s_e^2) \left(1 + \xi_2 s_e^2 (1 + 6 \kappa \xi_2)\right)^2
\nonumber\\
&\quad\quad 
+8 \xi_2 \epsilon_e^{(0)} \left(1 + 5 \xi_2 s_e^2 + 4 \xi_2^2 s_e^4 (1 + 6 \kappa \xi_2)\right)
-16 \xi_2^2 s_e^2 \eta_e^{(0)} \left(1 + \xi_2 s_e^2 (1 + 6 \kappa \xi_2)\right)
\Big]
\,.
\label{eqn:fNL-II}
\end{align}
Similar to the case of Class I, the nonlinearity parameter contains one more parameter compared to the tensor-to-scalar ratio and the spectral index, namely $\eta_e^{(0)}$. As the prefactor associated with $\eta_e^{(0)}$ is naturally small in the small nonminimal coupling limit, the dependence on $\eta_e^{(0)}$ is negligible. Thus, when presenting model-independent results, we shall consider $\eta_e^{(0)}=0.01$. Different choices of $\eta_e^{(0)}$ do not lead to any visible difference.

\begin{figure}[t!]
\centering
\includegraphics[width=0.48\linewidth]{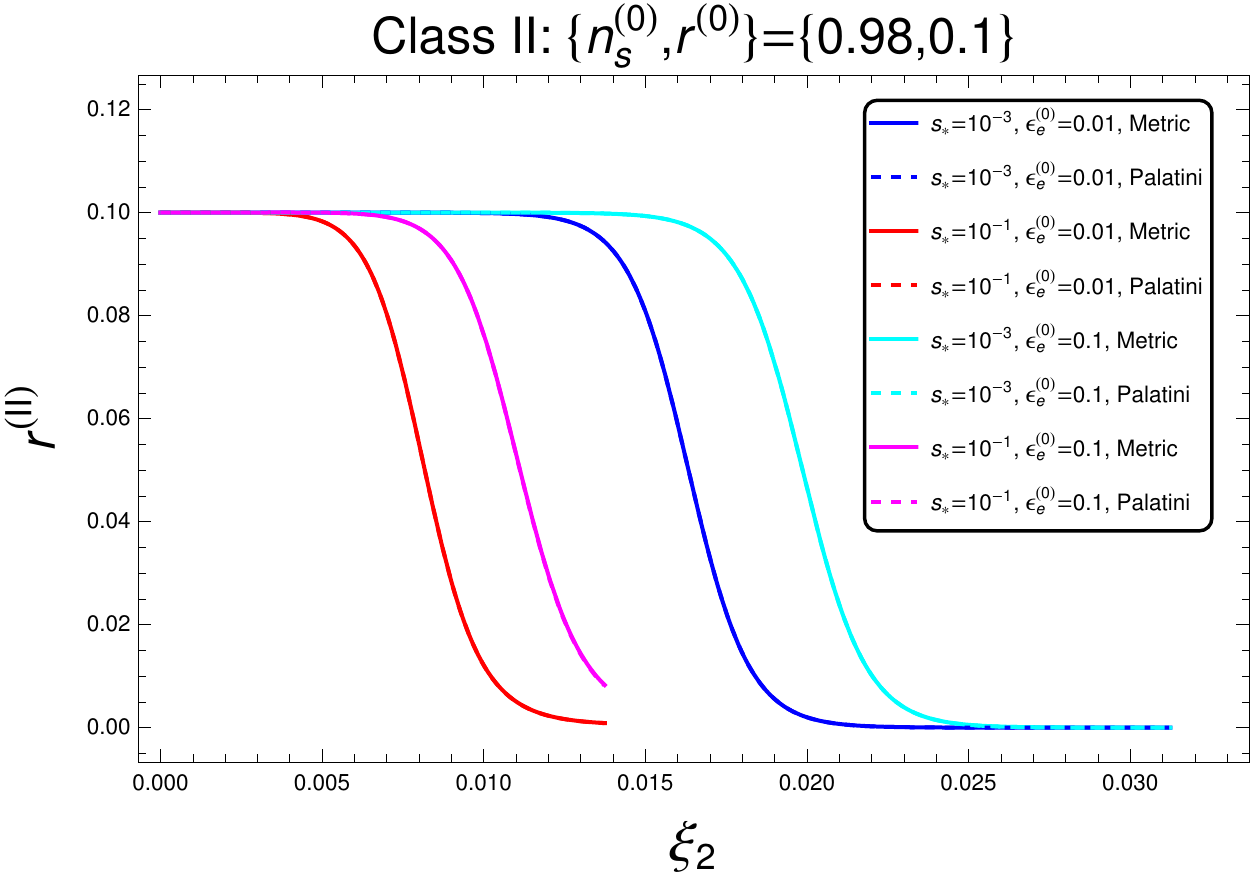}
\includegraphics[width=0.48\linewidth]{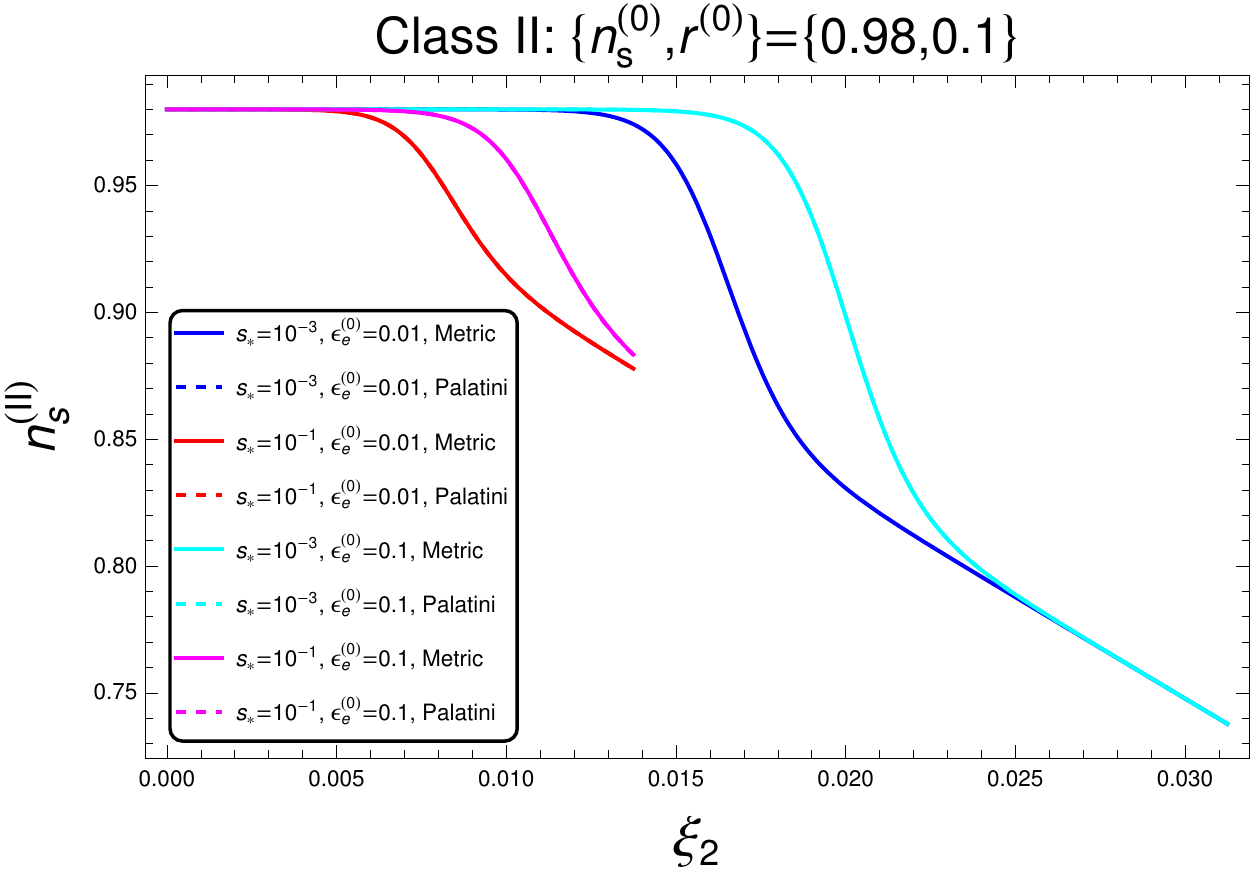}
\\
\includegraphics[width=0.48\linewidth]{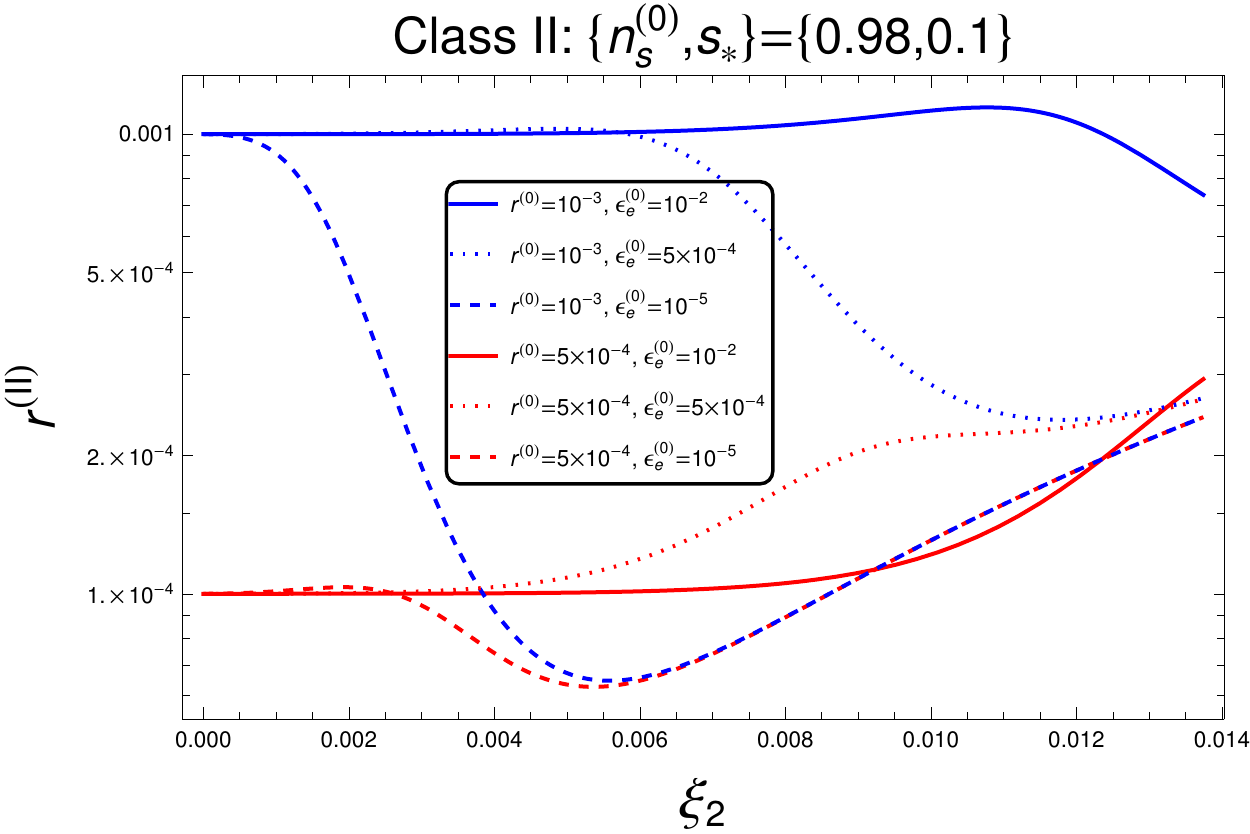}
\includegraphics[width=0.48\linewidth]{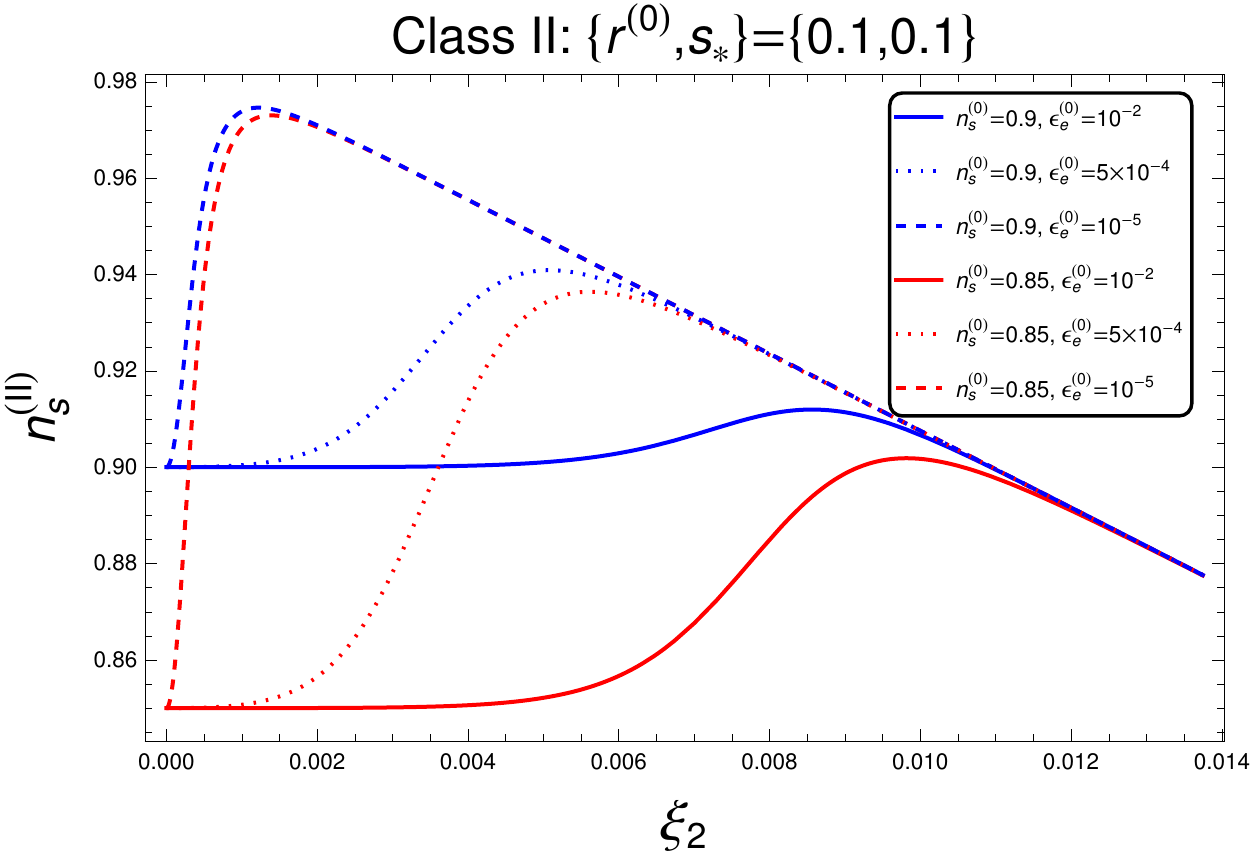}
\caption{In the upper panel, we show the dependence of the spectral index $n_s^{({\rm II})}$ and the tensor-to-scalar ratio $r^{({\rm II})}$ on the nonminimal coupling parameter $\xi_2$ for different choices of $\{s_*,e_e^{(0)}\}$ with $\{n_s^{(0)},r^{(0)}\}=\{0.98, 0.1\}$. 
The maximum values of $\xi_2$ are given via the small nonminimal coupling condition, $\xi_2 s_e^2 = 0.1$.
We observe that, similar to Class I, both the spectral index and the tensor-to-scalar ratio decrease as $\xi_2$ increases.
Also similar to Class I, the difference between the Palatini formulation (dashed) and the metric formulation (solid) is negligible.
In the lower panel, we see the behaviour of the tensor-to-scalar ratio for two different values of $r^{(0)}$ and three different values of $\epsilon_e^{(0)}$ with $\{n_s^{(0)},s_*\}=\{0.98,0.1\}$ (left panel) and the behaviour of the spectral index for different choices of $\{n_s^{(0)},\epsilon_e^{(0)}\}$ with $\{r^{(0)},s_*\}=\{0.1,0.1\}$ (right panel).
As the difference between the Palatini and the metric formulations is negligible, we only present the metric case in the lower panel.
When $\epsilon_e^{(0)}$ is chosen to be relatively large, the behaviour is similar to Class I.
On the other hand, when $\epsilon_e^{(0)}$ takes a very small value, the tensor-to-scalar ratio decreases initially and then increases again. Furthermore, unlike Class I, the increase of the spectral index is strong enough to go inside the allowed bounds.
The same behaviour is observed for different sets of $\{n_s^{(0)},r^{(0)}\}$; see also Fig.~\ref{fig:nsr-II}.}
\label{fig:ns-r-II}
\end{figure}

\begin{figure}[t!]
\centering
\includegraphics[width=0.49\linewidth]{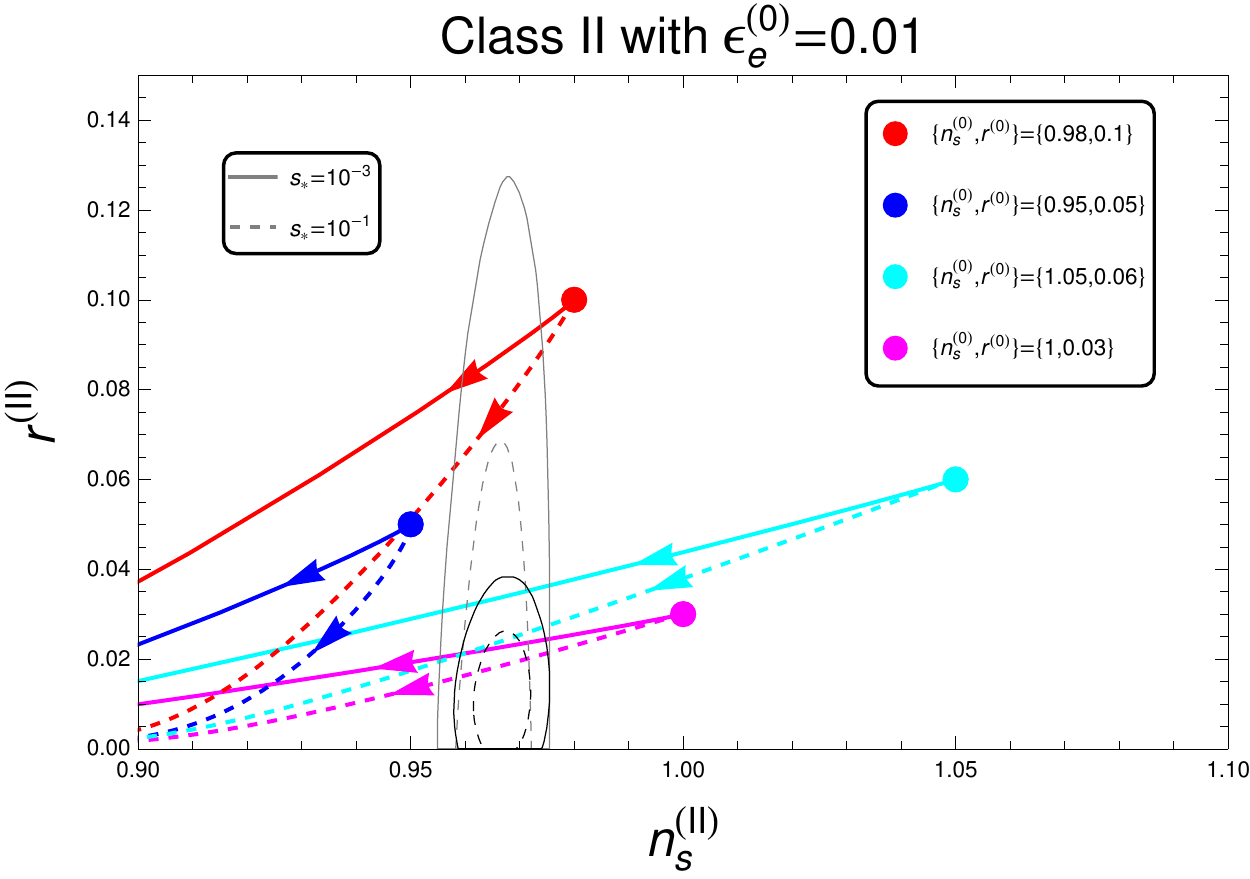}
\includegraphics[width=0.49\linewidth]{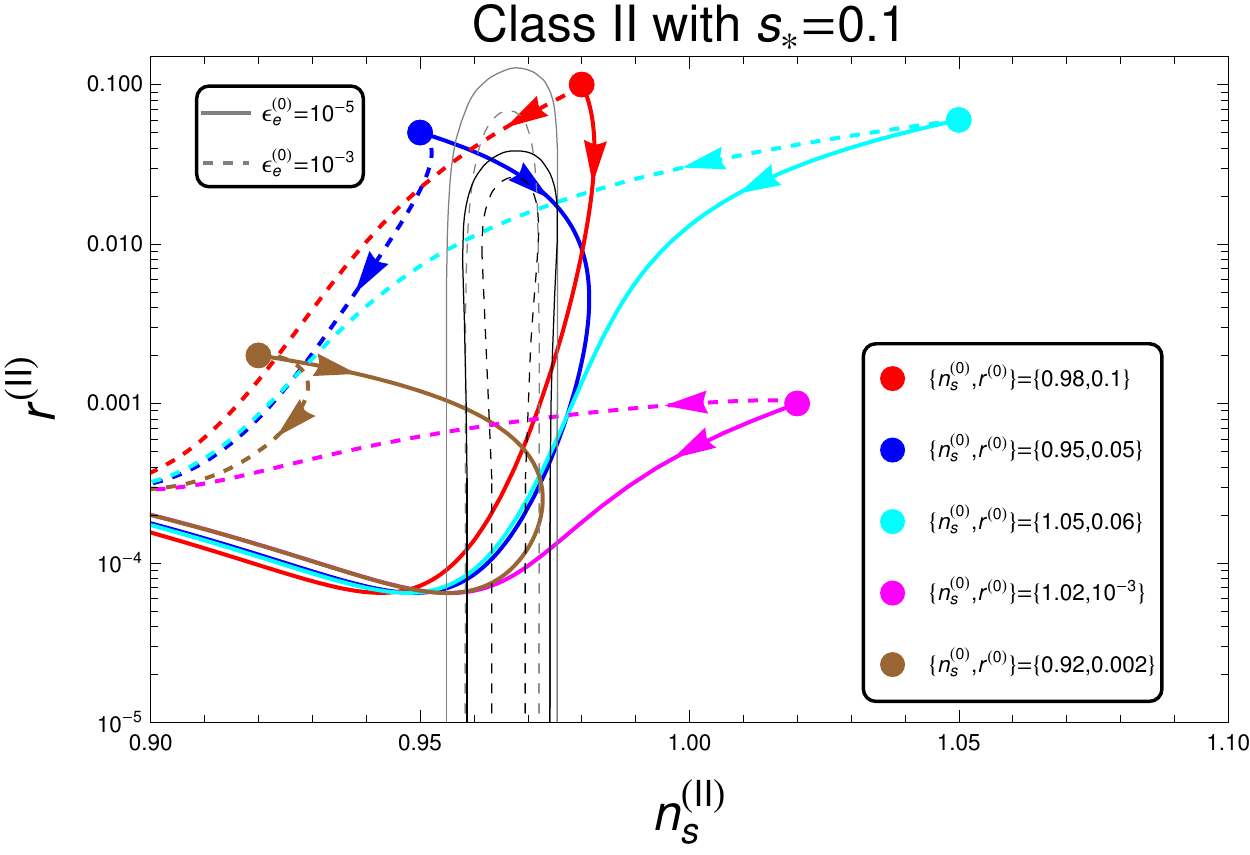}
\caption{{\it Left panel}: Behaviour of the spectral index $n_s^{({\rm II})}$ and the tensor-to-scalar ratio $r^{({\rm II})}$ for $\{n_s^{(0)},r^{(0)}\}=\{0.98,0.1\}$ (red), $\{0.95, 0.05\}$ (blue), $\{1.05,0.06\}$ (cyan), and $\{1,0.03\}$ (magenta) for $s_*=0.001$ (solid) and $0.1$ (dashed) with $\epsilon_e^{(0)}=0.01$.
{\it Right panel}: Behaviour of $n_s^{({\rm II})}$ and $r^{({\rm II})}$ for $\{n_s^{(0)},r^{(0)}\}=\{0.98,0.1\}$ (red), $\{0.95, 0.05\}$ (blue), $\{1.05,0.06\}$ (cyan), $\{1.02,0.001\}$ (magenta), and $\{0.92,0.002\}$ (brown) for $\epsilon_e^{(0)}=10^{-5}$ (solid) and $10^{-3}$ (dashed) with $s_*=0.1$.
As the difference between the Palatini and the metric formulations is negligible, only the metric cases are presented.
The latest Planck-BK 1-sigma (2-sigma) bound is depicted as the black solid (dashed) line, while the Planck-only 1-sigma (2-sigma) bound is presented with the grey solid (dashed) line.
When $\epsilon_e^{(0)}$ is relatively large, both the spectral index and the tensor-to-scalar ratio decrease as the nonminimal coupling parameter $\xi_2$ takes a larger value as indicated by arrows. Thus, similar to Class I, models that originally predict large spectral index and/or tensor-to-scalar ratio can be brought to the observationally-favoured region.
When $\epsilon_e^{(0)}$ takes a very small value, the spectral index initially shows an increasing behaviour for small $n_s^{(0)}$ cases, thereby entering the observationally-allowed bounds.
Thus, unlike Class I, models that originally predict small spectral index may as well become revived for small $\epsilon_e^{(0)}$ values.
}
\label{fig:nsr-II}
\end{figure}

In the upper panel of Fig.~\ref{fig:ns-r-II}, we present how the spectral index $n_s^{({\rm II})}$ and the tensor-to-scalar ratio $r^{({\rm II})}$ behave as we change $\xi_2$ for different sets of $\{s_*,\epsilon_e^{(0)}\}$ with a specific choice of $\{n_s^{(0)},r^{(0)}\}=\{0.98, 0.1\}$.
The maximum values of $\xi_2$ are chosen such that the small nonminimal coupling condition holds, {\it i.e.}, $\xi_2 s_e^2 = 0.1$.
The result is similar to Class I; both the spectral index and the tensor-to-scalar ratio decrease as the nonminimal coupling parameter $\xi_2$ increases. 
Furthermore, we see that the difference between the Palatini and metric formulations is negligible.
Thus, models that originally predict large $\{n_s^{(0)}, r^{(0)}\}$ and that belong to Class II can as well be brought to the observationally-favoured region.
In the lower panel of Fig.~\ref{fig:ns-r-II}, we show the behaviour of the tensor-to-scalar ratio (left panel) and the spectral index (right panel) for rather small $r^{(0)}$ and $n_s^{(0)}$ values, respectively, when $\epsilon_e^{(0)} = 10^{-2}$, $5\times 10^{-4}$, and $10^{-5}$.
When $\epsilon_e^{(0)}=10^{-2}$, the results are similar to Class I; the tensor-to-scalar ratio as well as the spectral index show an increasing behaviour, but the change is not significant.
On the other hand, when $\epsilon_e^{(0)}$ takes a very small value, the tensor-to-scalar ratio initially decreases and then increases again. The spectral index initially increases and then decreases again, which remains to be similar to Class I, but, unlike Class I, the increase of the spectral index is strong enough to bring the prediction to the allowed bounds. This is a stark difference between Class I and Class II: Not only can models that originally predict large spectral index and/or tensor-to-scalar ratio be brought to the observationally-favoured region, but models that predict small spectral index values may also be revived and become compatible with the latest Planck-BK bounds in the presence of the assistant field.
These behaviours are observed for different choices of $\{n_s^{(0)},r^{(0)}\}$ as demonstrated in Fig.~\ref{fig:nsr-II} on the $n_s$--$r$ plane.

\begin{figure}[t!]
\centering
\includegraphics[scale=0.565]{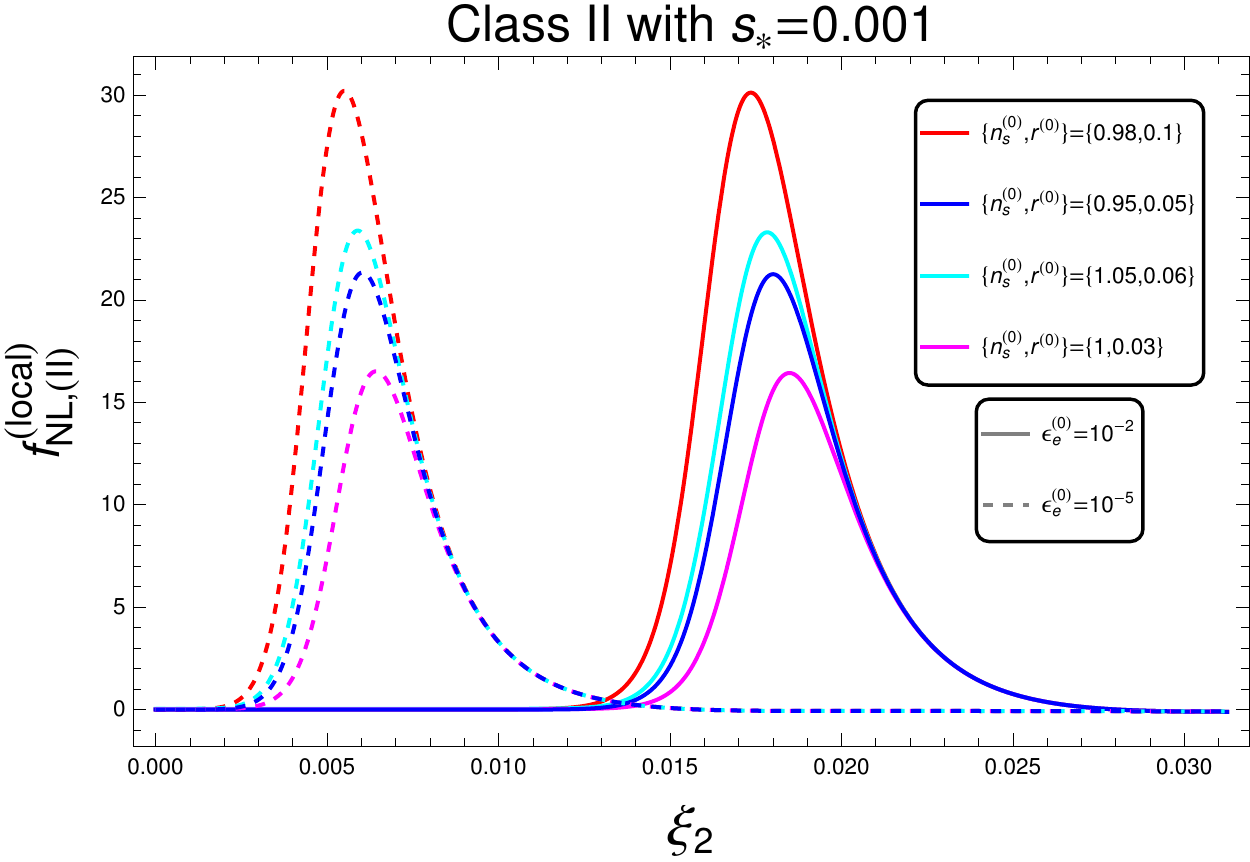}
\includegraphics[scale=0.59]{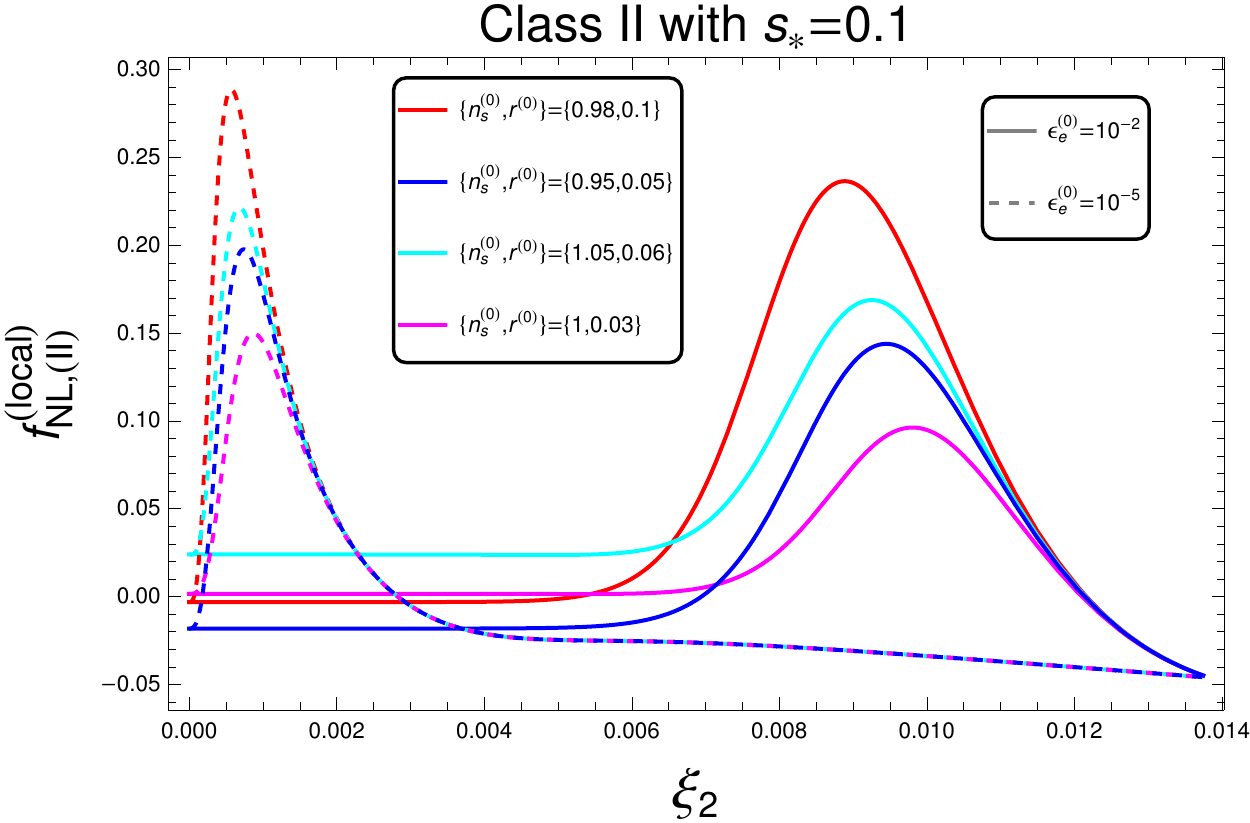}
\\
\includegraphics[scale=0.575]{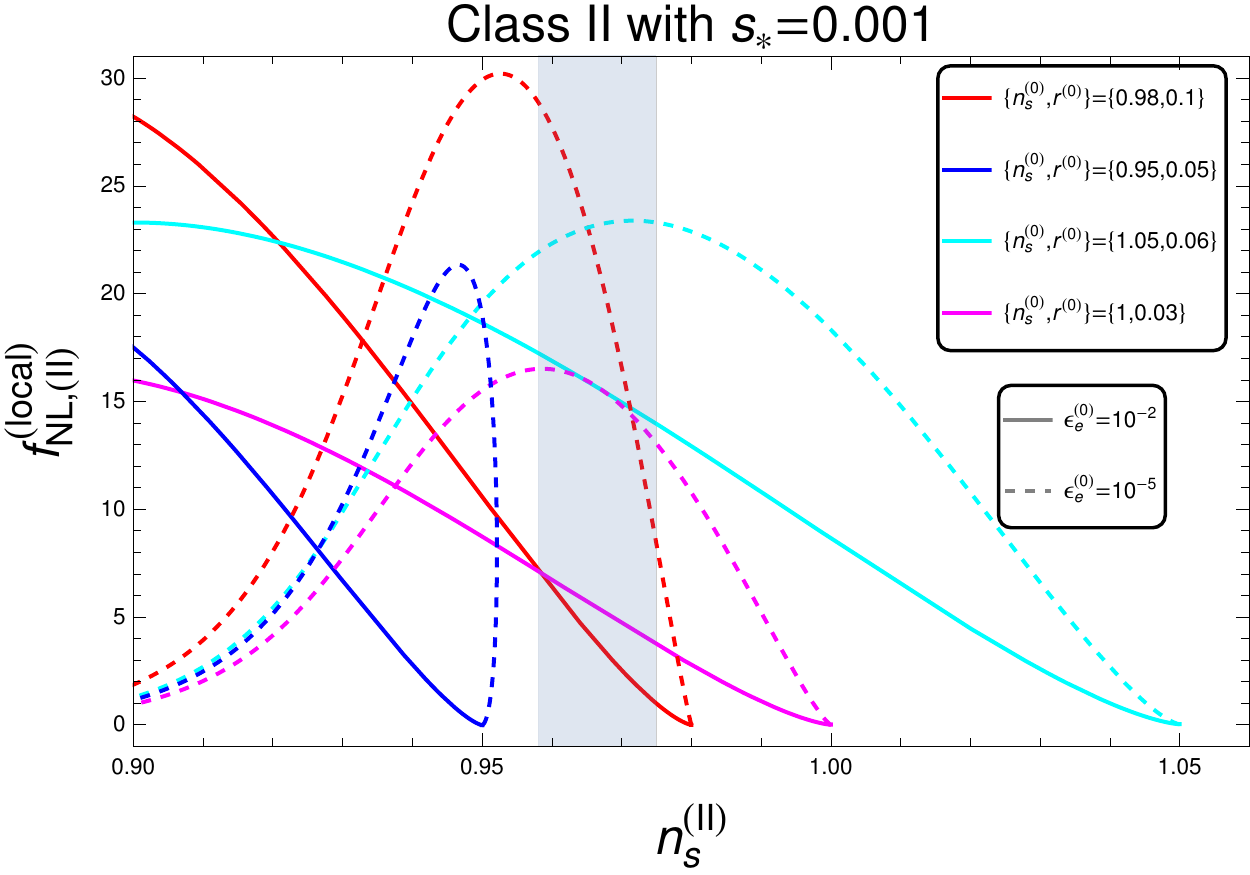}
\includegraphics[scale=0.59]{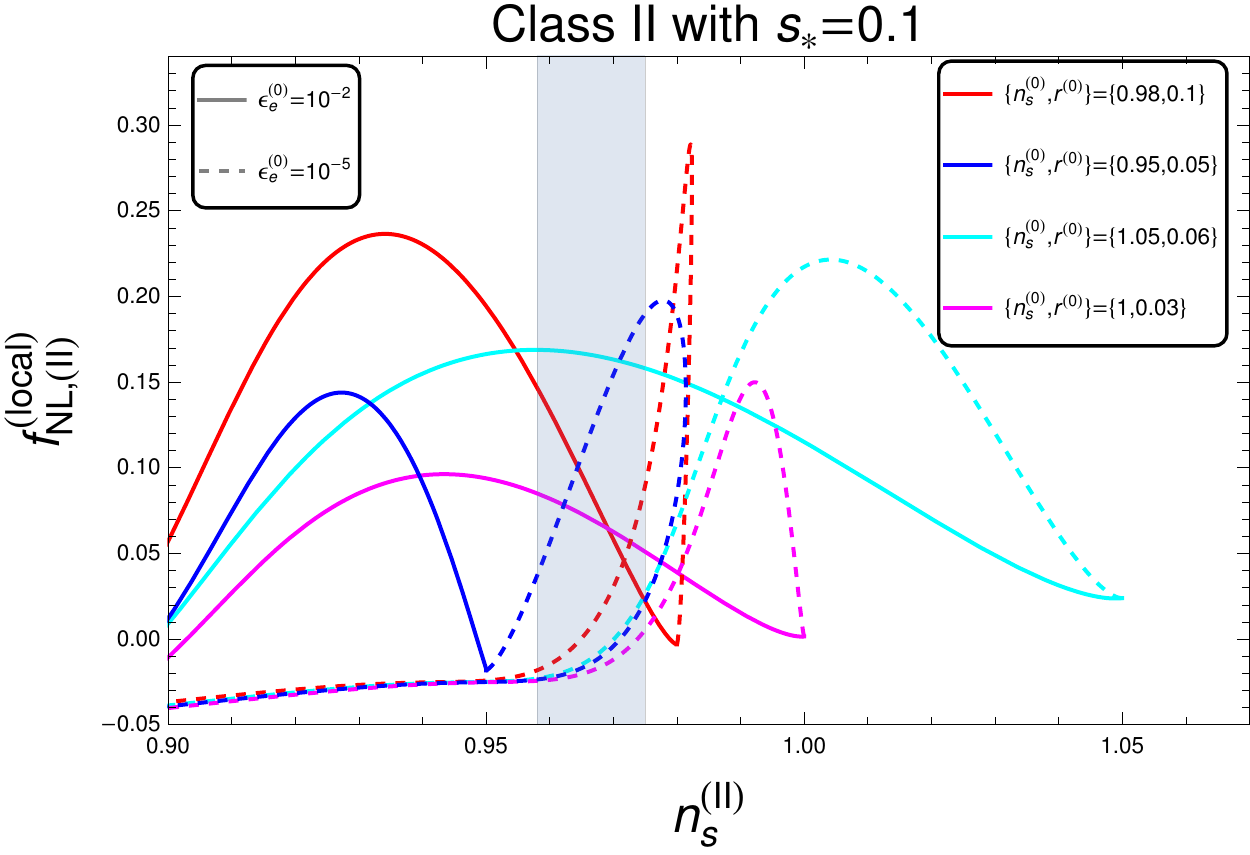}
\caption{
In the upper panel, the evolution of the nonlinearity parameter in terms of the nonminimal coupling parameter is shown for Class II. Two values of $s_*$ are considered, 0.001 (left panel) and 0.1 (right panel) with various choices of $\{n_s^{(0)},r^{(0)}\}=\{0.98,0.1\}$ (red), $\{0.95,0.05\}$ (blue), $\{1.05,0.06\}$ (cyan), and $\{1.0,0.03\}$ (magenta). Furthermore, two values of $\epsilon_e^{(0)}$ are considered, $10^{-2}$ (solid) and $10^{-5}$ (dashed). 
Similar to Class I, we observe that the nonlinearity parameter initially shows an increasing behaviour, and then it decreases.
In the lower panel, we present the predictions in the $f_{\rm NL, (II)}^{({\rm local})}$--$n_s^{({\rm II})}$ plane. The shaded region corresponds to the latest bounds on the spectral index. In the case of $s_*=0.001$ with $\epsilon_e^{(0)}=10^{-2}$, the nonlinearity parameter for the  $\{n_s^{(0)},r^{(0)}\}=\{1.05,0.06\}$ case becomes slightly larger than the Planck 2-sigma bound, $-11.1 < f_{\rm NL}^{({\rm local})} < 9.3$. When $\epsilon_e^{(0)}=10^{-5}$, only a narrow parameter range is allowed for the $\{n_s^{(0)},r^{(0)}\}=\{0.98,0.1\}$ case, while the rest cases are ruled out. In the case of $s_*=0.1$, however, all the constraints are safely satisfied for both $\epsilon_e^{(0)}=10^{-2}$ and $10^{-5}$.
For all cases, $\eta_e^{(0)}=0.01$ is chosen. However, the dependence on $\eta_e^{(0)}$ is weak as the prefactor is small in the small nonminimal coupling limit. Only the metric formulation is presented as there is little difference between the metric and the Palatini formulations.
}
\label{fig:fNL-II}
\end{figure}

The behaviour of the nonlinearity parameter as we increase the nonminimal coupling parameter $\xi_2$ is shown in the upper panel of Fig.~\ref{fig:fNL-II}. Only the metric formulation is presented as the difference between the metric and the Palatini formulations is negligible.
Four different choices of $\{n_s^{(0)},r^{(0)}\}$ are considered, $\{0.98,0.1\}$, $\{0.95,0.05\}$, $\{1.05,0.06\}$, and $\{1.0,0.03\}$, and two values of $s_*$ are chosen, 0.001 and 0.1. We additionally considered two scenarios of $\epsilon_e^{(0)}=10^{-2}$ and $\epsilon_e^{(0)}=10^{-5}$. Similar to Class I, we observe that $f_{\rm NL}^{({\rm local})}$ initially increases as we increase $\xi_2$, and then it decreases. Moreover, the $f_{\rm NL}^{({\rm local})}$ value tends to be larger when the $s_*$ value is small. In the lower panel of Fig.~\ref{fig:fNL-II}, the predictions are shown in the $f_{\rm NL}^{({\rm local})}$--$n_s$ plane. In the case of $s_*=0.001$ with $\epsilon_e^{(0)}=10^{-2}$, the nonlinearity parameter for the  $\{n_s^{(0)},r^{(0)}\}=\{1.05,0.06\}$ case becomes slightly larger than the Planck 2-sigma bound, $-11.1 < f_{\rm NL}^{({\rm local})} < 9.3$. When $\epsilon_e^{(0)}=10^{-5}$, only a narrow parameter range is allowed for the $\{n_s^{(0)},r^{(0)}\}=\{0.98,0.1\}$ case, while the rest cases are ruled out. In the case of $s_*=0.1$, however, all the constraints are safely satisfied for both $\epsilon_e^{(0)}=10^{-2}$ and $10^{-5}$.
The currently allowed parameter range can even further be probed by future observations of non-Gaussianity such as galaxy surveys and 21 cm line of neutral hydrogen observations \cite{Camera:2013kpa,Ferramacho:2014pua,Raccanelli:2014kga,Yamauchi:2014ioa,Camera:2014bwa,dePutter:2014lna,Munoz:2015eqa,Yamauchi:2015mja,Sekiguchi:2018kqe} as the constraints on $f_{\rm NL}^{\rm (local)}$ can be improved as $|f_{\rm NL}^{\rm (local)}| < {\cal O}(0.1) -{\cal O}(1)$.

%%%%%%%%%%%%%%%%%%%%%%%%%%%%%%%%%%%%%%%%%%
\section{Examples}
\label{sec:examples}
%%%%%%%%%%%%%%%%%%%%%%%%%%%%%%%%%%%%%%%%%%
As an application of our general analysis presented in Sec.~\ref{sec:genanalysis}, we consider three examples: the loop inflation model which belongs to Class I\footnote{
Chaotic inflation model is also categorised as Class I, and it has already been studied in Ref.~\cite{Hyun:2022uzc}.
} and the power-law inflation and hybrid inflation models which belong to Class II. These models are ruled out by the latest Planck-BK results. We show that, with the help of the assistant field, these three models may become compatible with the latest observations.
When presenting results, we only choose the metric formulation as there exists little difference between the metric and the Palatini formulations.

%%%%%%%%%%%%%%%%%%%%%%%%%%%%%%%%%%%%%%%%%%
\subsection{Loop inflation -- an example for Class I}
\label{subsec:loopinflation}
%%%%%%%%%%%%%%%%%%%%%%%%%%%%%%%%%%%%%%%%%%

%%
\begin{figure}[t!]
\centering
\includegraphics[width = 12.5cm]{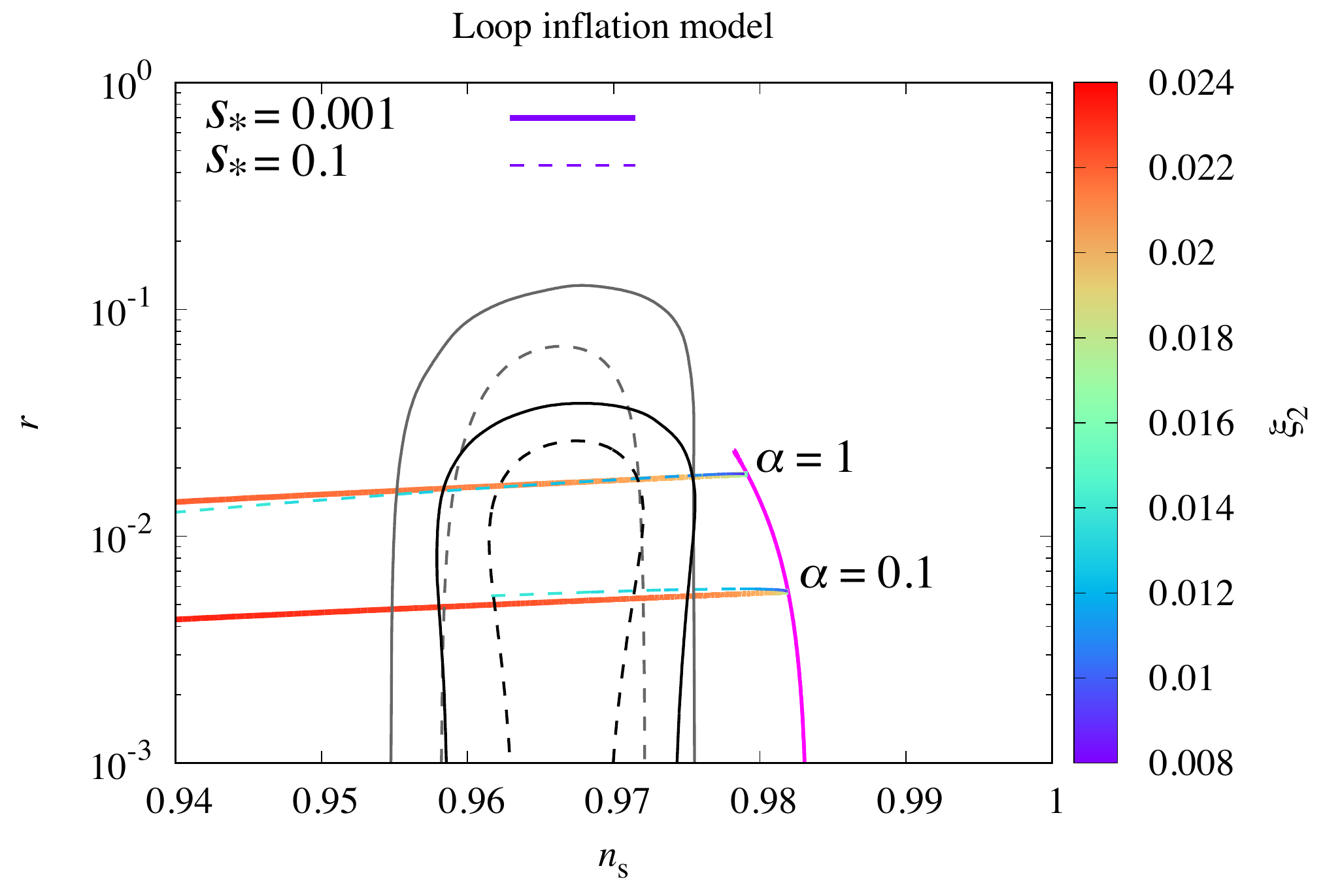}
\caption{
Predictions for the loop inflation model. The original model, depicted by the magenta line, is ruled out by the latest Planck \cite{Planck:2018jri} (grey) and Planck-BK \cite{BICEP:2021xfz} (black) bounds, due to the largeness of the spectral index. By varying the nonminimal coupling parameter $\xi_2$, the effect of the assistant field is presented for $\alpha=0.1$ and $\alpha=1$ with two values of $s_*$, $0.001$ (solid) and $0.1$ (dashed). The cut in the case of $\{\alpha,s_*\}=\{0.1,0.1\}$ is due to the fact that the maximum value of $\xi_2$ is chosen such that $\xi_2 s_e^2 = 0.1$.
In the presence of the assistant field, the spectral index gets suppressed. Thus, the model becomes compatible with the latest observational constraints.
}
\label{fig:LI_ns-r}
\end{figure}

Loop inflation is described by the following potential:
\begin{align}\label{eqn:LIpot}
V_{\rm J}(\phi)
= \Lambda^4 (1 + \alpha \log \phi)\,,
\end{align}
where $\alpha$ is the coefficient of one-loop correction term \cite{Dvali:1994ms}. As we are interested in phenomenological aspects of the model, we take $\alpha$ as a free parameter rather than focusing on its origin.
For the potential \eqref{eqn:LIpot}, we obtain the slow-roll parameters \eqref{eqn:SRparams0} as
\begin{align}
\epsilon^{(0)} 
= \frac{\alpha^2}{2\phi (1 + \alpha \log \phi)^{2}} \,, 
\quad
\eta^{(0)} 
= - \frac{\alpha}{\phi^2 (1 + \alpha \log \phi)} \,.
\end{align}
The field value of the inflaton $\phi$ at the end of inflation, $\phi_e$, characterised by $\epsilon^{(0)}(\phi=\phi_e) = 1$, is then given by
\begin{align}
\phi_e = \frac{1}{\sqrt{2}} \left[
W_0 \left(
\frac{e^{1/\alpha}}{\sqrt{2}}
\right)
\right]^{-1}
\,,
\end{align}
where $W_0$ is the zero-branch of Lambert's $W$ function, and we have assumed $\alpha > 0$. Since end of inflation may be achieved via the slow-roll violation, the loop inflation model belongs to Class I. Imposing 60 $e$-folds, {\it i.e.}, $60 = \int_e^* (V_{\rm J}/V_{{\rm J},\phi}) d\phi$, we obtain the field value at the pivot scale. We depict the original prediction for the spectral index $n_s^{(0)}$ and the tensor-to-scalar ratio $r^{(0)}$ in Fig.~\ref{fig:LI_ns-r} by varying $\alpha$ (magenta line). The original loop inflation model is clearly ruled out by the latest observational constraints. 

\begin{figure}
\centering
\includegraphics[width=7.5cm]{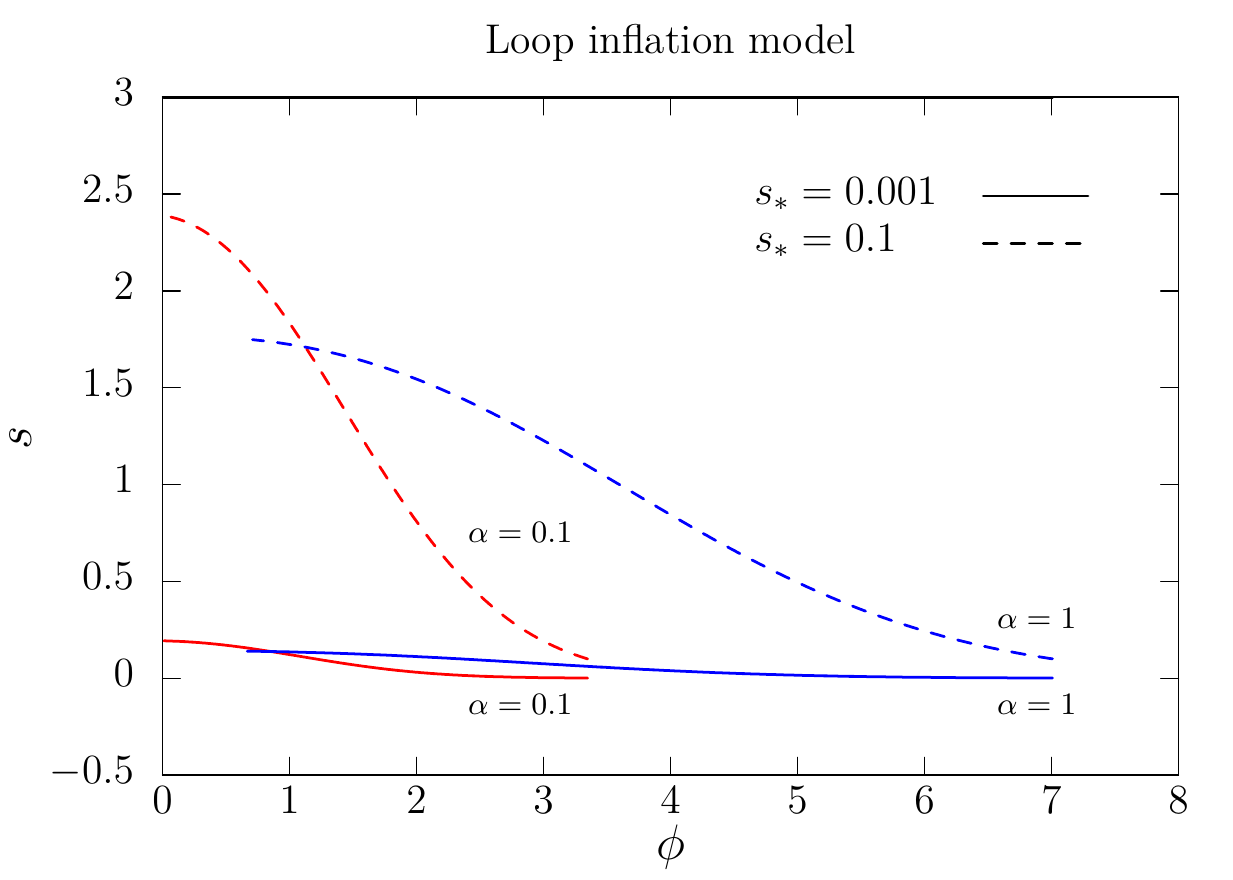}
\caption{Field trajectories for the loop inflation model. Two choices of $s_*$, 0.001 (solid) and 0.1 (dashed), are considered with $\alpha = 0.1$ (red) and $1$ (blue). For each case, we have chosen the value of $\xi_2$ in such a way that the spectral index becomes 0.965.}
\label{fig:LI_trajectory}
\end{figure}

Using the analytical expressions \eqref{eqn:ns-I} and \eqref{eqn:r-I}, the effect of the assistant field $s$ is presented in Fig.~\ref{fig:LI_ns-r} for $\alpha = 0.1$ and $\alpha = 1$ with two values of $s_*$, $0.001$ (solid line) and $0.1$ (dashed line). We see that the spectral index $n_s$ decreases as the nonminimal coupling parameter $\xi_2$ increases. Hence, we may rescue the original loop inflation model. For example, in the case of $\alpha = 0.1$, the model becomes compatible with the Planck-BK 2-sigma bounds for the range of $\xi_2 = (2.1 - 2.3) \times 10^{-2}$ for $s_* = 0.001$ and $\xi_2 = (1.2 - 1.4) \times 10^{-2}$ for $s_* = 0.1$. In the case of $\alpha = 1$, the corresponding ranges for the nonminimal coupling parameter are $\xi_2 = (1.9 - 2.1) \times 10^{-2}$ for $s_* = 0.001$ and $\xi_2 = (1.1 - 1.3) \times 10^{-2}$ for $s_* = 0.1$. For each case, we present in Fig. \ref{fig:LI_trajectory} the field trajectory during inflation that yields $n_s=0.965$.

\begin{figure}[t!]
\centering
\includegraphics[width = 15.6cm]{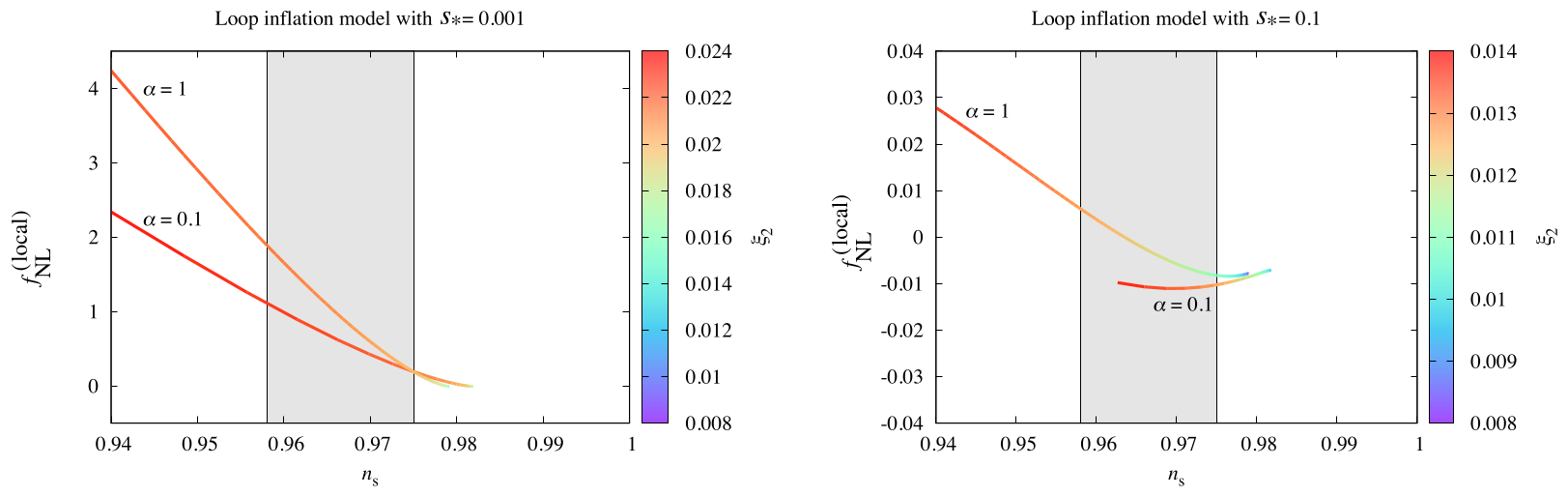}
\caption{
Nonlinearity parameter $f_{\rm NL}^{\rm (local)}$ in terms of the spectral index $n_s$ for the loop inflation model. Two choices of $s_*$ are considered, 0.001 (left) and 0.1 (right), and the cases with $\alpha=0.1$ and $1$ are shown in each panel.
The shaded region represents the Planck 2-sigma bound on the spectral index, $n_s = [0.955, 0.976]$. We note that, for both the $s_*=0.001$ and $s_*=0.1$ cases, the nonlinearity parameter is well within the Planck 2-sigma bound, $-11.1 < f_{\rm NL}^{({\rm local})} < 9.3$. We further observe that, for a relatively large value of $s_*$ such as the $s_* = 0.1$ case, the nonlinearity parameter tends to be tiny.
}
\label{fig:fNL-LI}
\end{figure}

Figure~\ref{fig:fNL-LI} shows the nonlinearity parameter $f_{\rm NL}^{(\rm local)}$, obtained by using Eq.~\eqref{eqn:fNL-I}, as a function of the spectral index $n_s$ for the loop inflation model. Similar to Fig.~\ref{fig:LI_ns-r}, two choices of $s_*$, 0.001 and 0.1, are considered. The shaded grey region indicates the Planck 2-sigma bound on the spectral index. We note that, for both the $s_*=0.001$ and $s_*=0.1$ cases, the nonlinearity parameter is well within the Planck 2-sigma bound, $-11.1 < f_{\rm NL}^{({\rm local})} < 9.3$. We further observe that, for a relatively large value of $s_*$ such as the $s_* = 0.1$ case, the nonlinearity parameter tends to be tiny.

\begin{figure}
\centering
\includegraphics[scale=0.6]{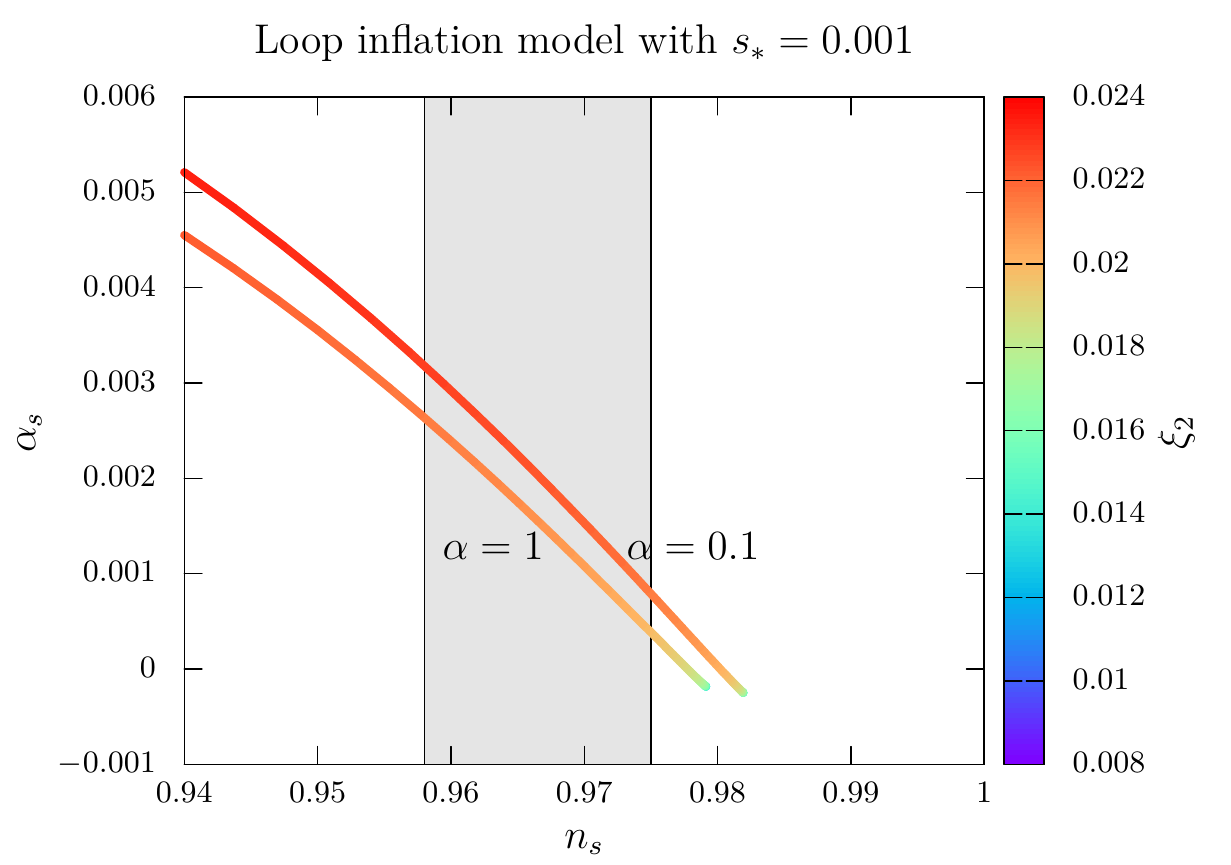}\;
\includegraphics[scale=0.6]{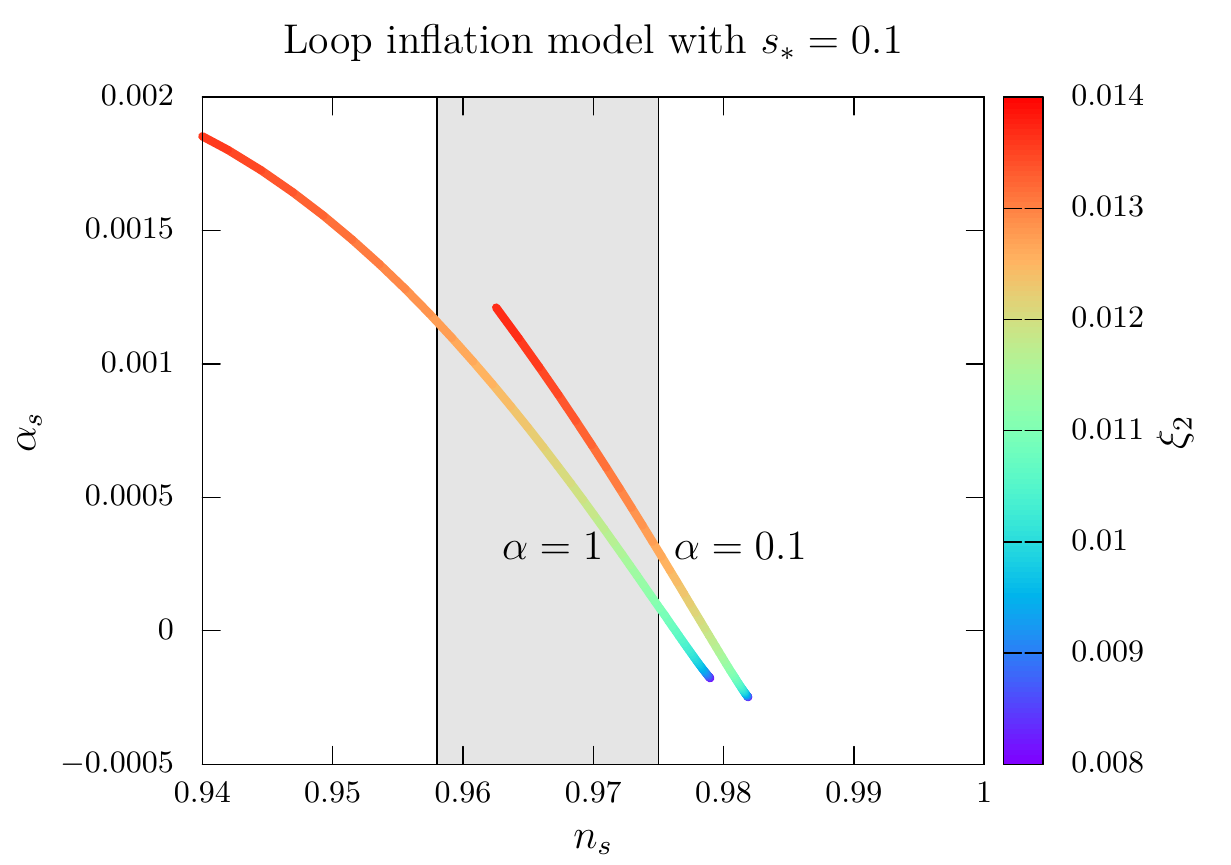}
\caption{Running of the spectral index $\alpha_s$ in terms of the spectral index $n_s$ for the loop inflation model. Similar to Fig.~\ref{fig:fNL-LI}, two choices of $s_*$ are considered with $\alpha=0.1$ and $1$ for both cases. We observe that $\alpha_s$ shows an increasing behaviour as $n_s$ decreases, and that a larger value of $s_*$ results in smaller $\alpha_s$. See also Appendix \ref{apdx:SIrunning}.}
\label{fig:alphas-LI}
\end{figure}

Finally, we present the prediction for the running of the spectral index $\alpha_s$, discussed in Appendix \ref{apdx:SIrunning}, in Fig.~\ref{fig:alphas-LI}. We have considered two choices of $s_*$ and two values for the parameter $\alpha$ as before. We observe that the running of the spectral index shows an increasing behaviour as the spectral index decreases. We further see that the running of the spectral index tends to be smaller when a larger value is considered for $s_*$.

%%%%%%%%%%%%%%%%%%%%%%%%%%%%%%%%%%%%%%%%%%
\subsection{Power-law inflation -- an example for Class II}
\label{subsec:powerlaw}
%%%%%%%%%%%%%%%%%%%%%%%%%%%%%%%%%%%%%%%%%%

%%
\begin{figure}[t!]
\centering
\includegraphics[width=12.5cm]{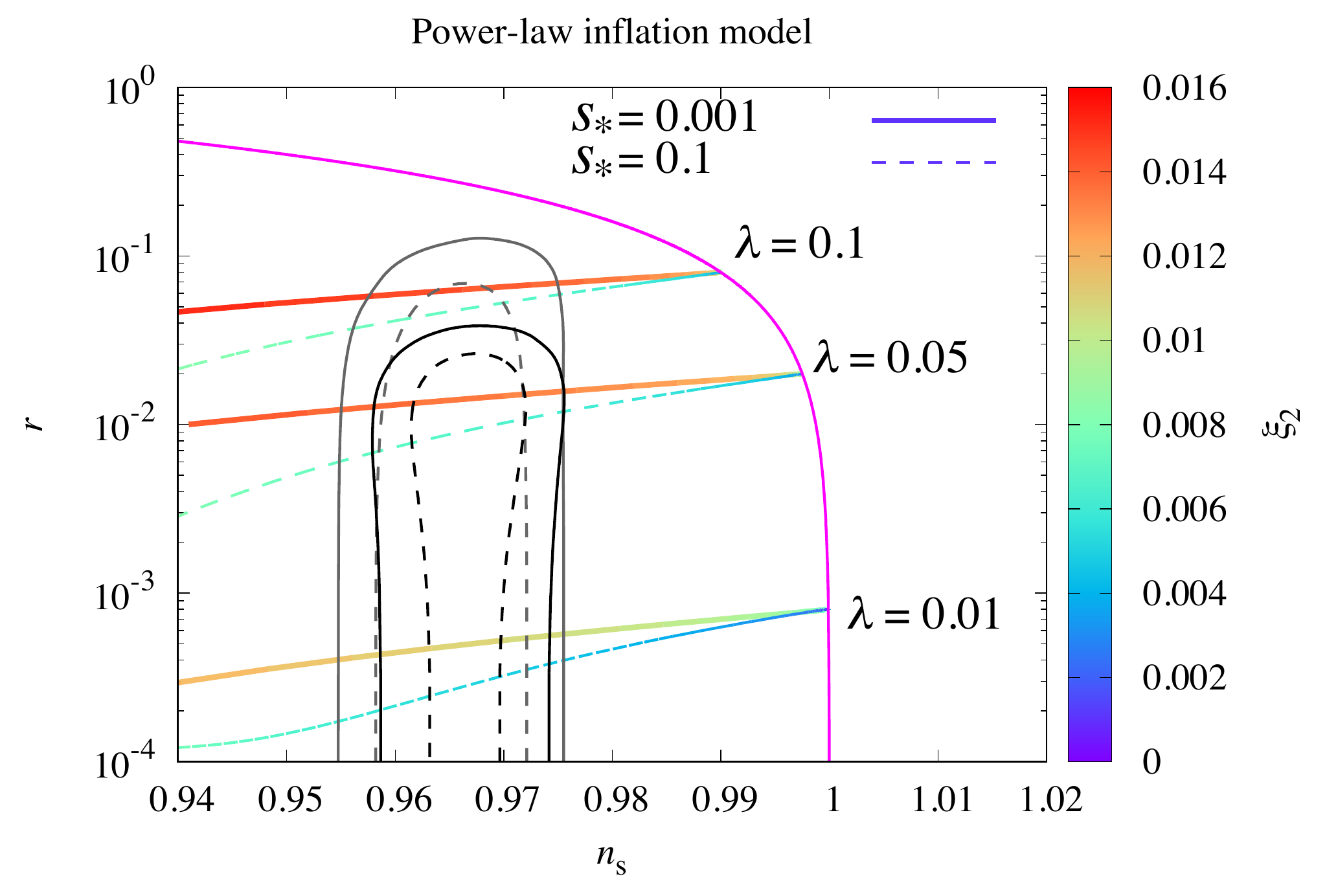}
\caption{Predictions for the power-law inflation model. The original model, depicted by the magenta line, is ruled out by the latest Planck (grey) and Planck-BK (black) bounds. By varying the nonminimal coupling parameter $\xi_2$, the effect of the assistant field is presented for $\lambda=0.1$, 0.05, and 0.01 with two choices of $s_*$, 0.001 (solid) and 0.1 (dashed). In the presence of the assistant field, the spectral index as well as the tensor-to-scalar ratio get suppressed. Thus, the power-law inflation model becomes compatible with the latest observational constraints.
}
\label{fig:PL_ns-r}
\end{figure}

Power-law inflation \cite{Lucchin:1984yf} is described by the potential
\begin{align}
V_{\rm J}(\phi) =
\Lambda^4 \exp(-\lambda\phi)
\,.
\end{align}
As the potential is given by an exponential function, the slow-roll parameters \eqref{eqn:SRparams0} become constants, 
\begin{align}
\epsilon^{(0)} =
\frac{1}{2}\lambda^2
\,,\quad 
\eta^{(0)} =
\lambda^2
\,.
\end{align}
Therefore, inflation does not end via slow-roll violations, calling for a separate sector for end of inflation. The power-law inflation model thus belongs to Class II with a special property of $\epsilon_e^{(0)} = \epsilon_*^{(0)}$.
The spectral index and the tensor-to-scalar ratio are given by
\begin{align}
n_s^{(0)} =
1 - \lambda^2
\,,\quad 
r^{(0)} =
8\lambda^2
\,.
\label{eqn: slow-roll parameters power-law}
\end{align}
The original predictions, $n_s^{(0)}$ and $r^{(0)}$, are depicted in Fig.~\ref{fig:PL_ns-r} as a magenta line for various values of $\lambda$. We see that the original model is ruled out by the latest Planck-BK bounds as either the spectral index or the tensor-to-scalar ratio is too large.

\begin{figure}
\centering
\includegraphics[width=7.5cm]{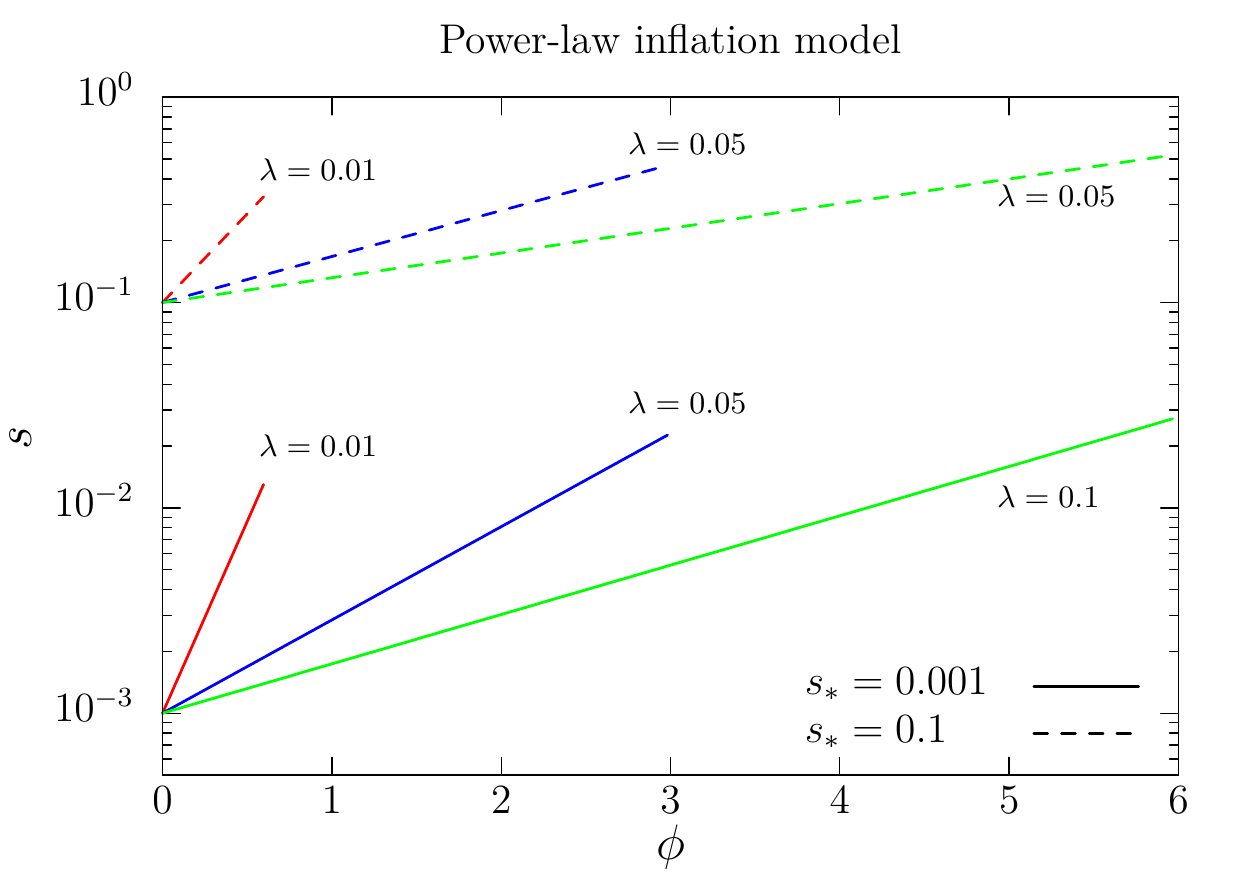}
\caption{Field trajectories for the power-law inflation model. Two choices of $s_*$, 0.001 (solid) and 0.1 (dashed), are considered with $\lambda = 0.01$ (red), $0.05$ (blue), and $1$ (green). For each case, we have chosen the value of $\xi_2$ in such a way that the spectral index becomes 0.965. For the demonstration of field trajectories, $\phi_*=0$ is chosen.}
\label{fig:PL_trajectory}
\end{figure}

From the general analysis performed in Sec.~\ref{subsec:classII}, one expects that the inclusion of the assistant field may bring the original predictions to the observationally-favoured region; see, {\it e.g.}, Fig.~\ref{fig:nsr-II}. 
Using the analytical expressions \eqref{eqn:ns-II} and \eqref{eqn:r-II}, the spectral index and the tensor-to-scalar ratio in the presence of the assistant field can be computed. In Fig.~\ref{fig:PL_ns-r}, we present the predictions in the $n_s$--$r$ plane for $\lambda=0.1$, 0.05, and 0.01 with two choices of $s_*$, 0.1 and 0.001, by varying the nonminimal coupling parameter $\xi_2$.
With the help of the assistant field, the power-law inflation model may thus become compatible again with the latest Planck-BK constraints.
For example, in the case of $\lambda=0.01$, the predictions for the spectral index and the tensor-to-scalar ratio enter the Planck-BK $2$-sigma bound for the range of $\xi_2=(1.05-1.11)\times 10^{-2}$ for $s_*=0.001$ and $\xi_2=(4.5 - 5.6)\times 10^{-3}$ for $s_*=0.1$. In the case of $\lambda=0.05$, the corresponding range for the nonminimal coupling parameter is $\xi_2=(1.29-1.36)\times 10^{-2}$ for $s_*=0.001$ and $\xi_2=(6.0 - 6.9)\times 10^{-3}$ for $s_*=0.1$.
When $\lambda =0.1$, although $n_s$ and $r$ get smaller by increasing $\xi_2$, and in particular, $n_s$ can be brought to the observationally allowed range, the suppression of $r$ is not enough to be allowed by the Planck-BK constraints.
Field trajectories for aforementioned parameter choices are shown in Fig. \ref{fig:PL_trajectory} with $\phi_*=0$ as an example.

\begin{figure}[t!]
\centering
\includegraphics[width = 15.6cm]{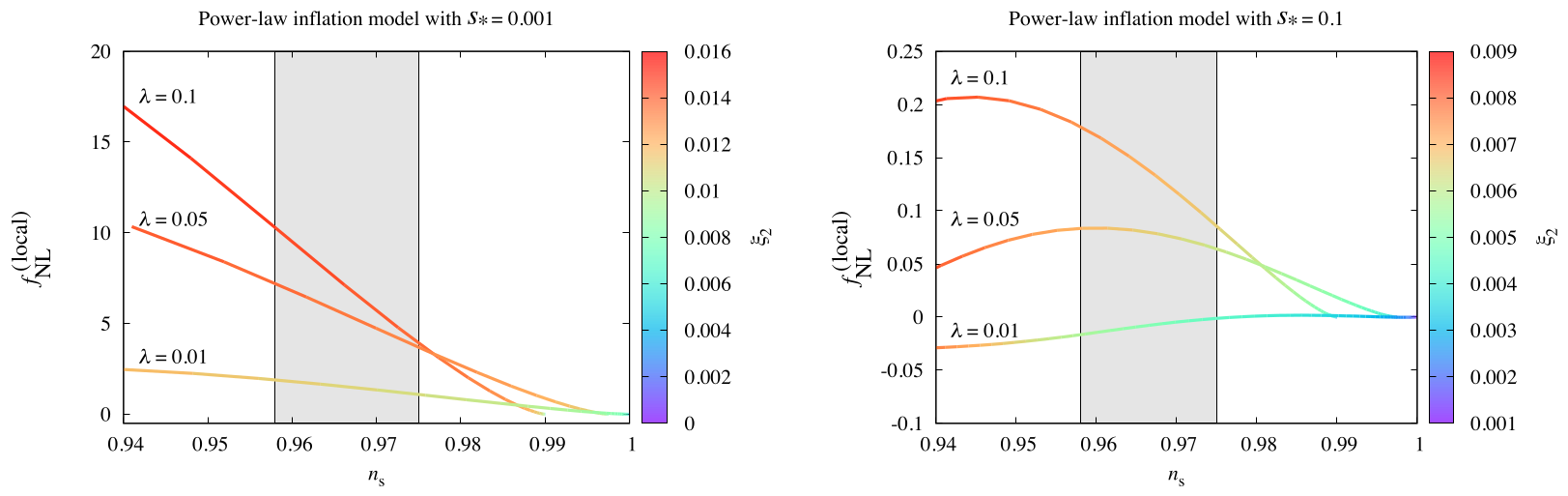}
\caption{
Nonlinearity parameter $f_{\rm NL}^{\rm (local)}$ in terms of the spectral index $n_s$ for the power-law inflation model. Two choices of $s_*$ are considered, 0.001 (left) and 0.1 (right). The shaded region represents the Planck 2-sigma bound on the spectral index, $n_s = [0.955, 0.976]$.
In the case of $\lambda = 0.1$ with $s_* = 0.001$, the nonlinearity parameter may become slightly larger than the Planck 2-sigma bound, $-11.1 < f_{\rm NL}^{({\rm local})} < 9.3$.
On the other hand, for a relatively large value of $s_*$ such as the $s_* = 0.1$ case, the nonlinearity parameter tends to be tiny.
}
\label{fig:fNL-PL}
\end{figure}

Figure~\ref{fig:fNL-PL} shows the nonlinearity parameter $f_{\rm NL}^{(\rm local)}$, obtained by using Eq.~\eqref{eqn:fNL-II}, as a function of the spectral index $n_s$ for the power-law inflation model. Similar to Figs.~\ref{fig:PL_ns-r}, two choices of $s_*$, 0.001 and 0.1, are considered. The shaded grey region indicates the Planck 2-sigma bound on the spectral index.
In the case of $\lambda = 0.1$ with $s_* = 0.001$, the nonlinearity parameter may become slightly larger than the Planck 2-sigma bound, $-11.1 < f_{\rm NL}^{({\rm local})} < 9.3$.
On the other hand, for a relatively large value of $s_*$ such as the $s_* = 0.1$ case, the nonlinearity parameter tends to be tiny.

\begin{figure}
\centering
\includegraphics[scale=0.6]{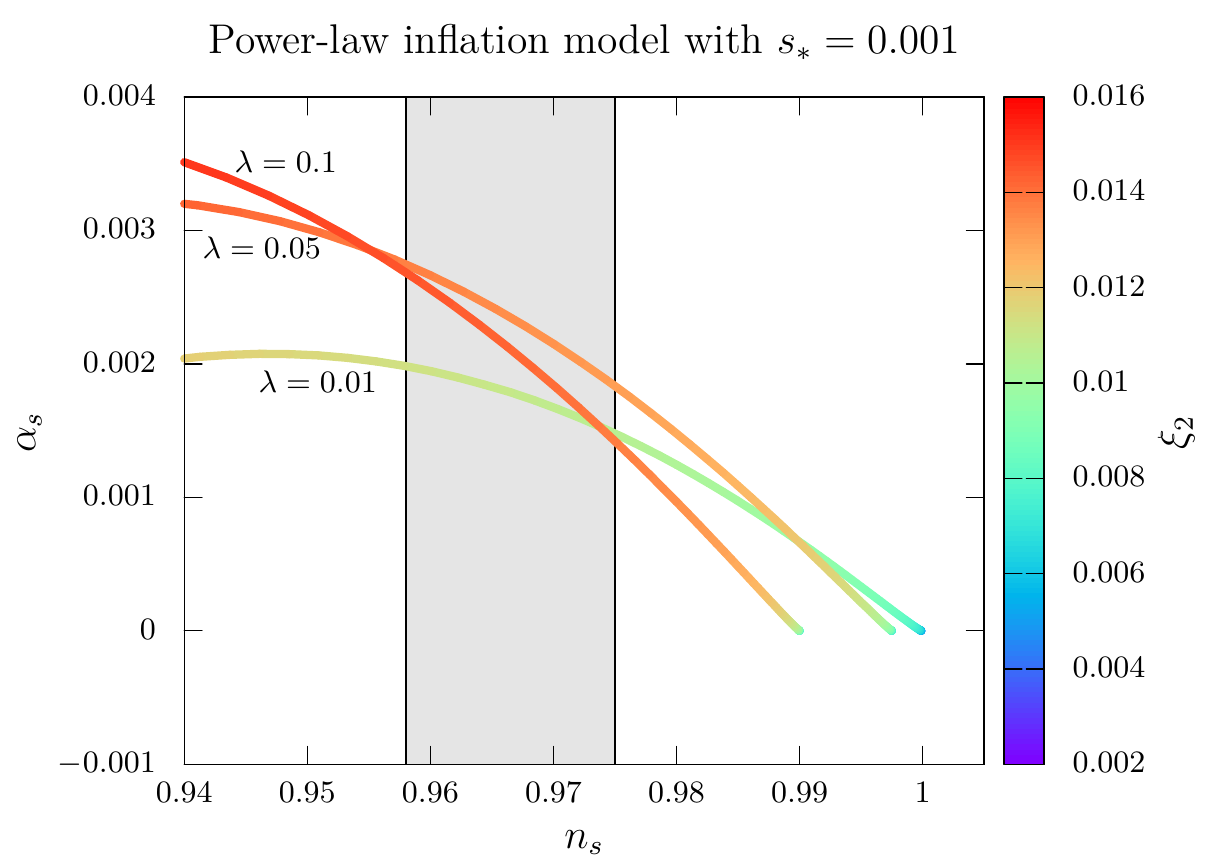}\;
\includegraphics[scale=0.6]{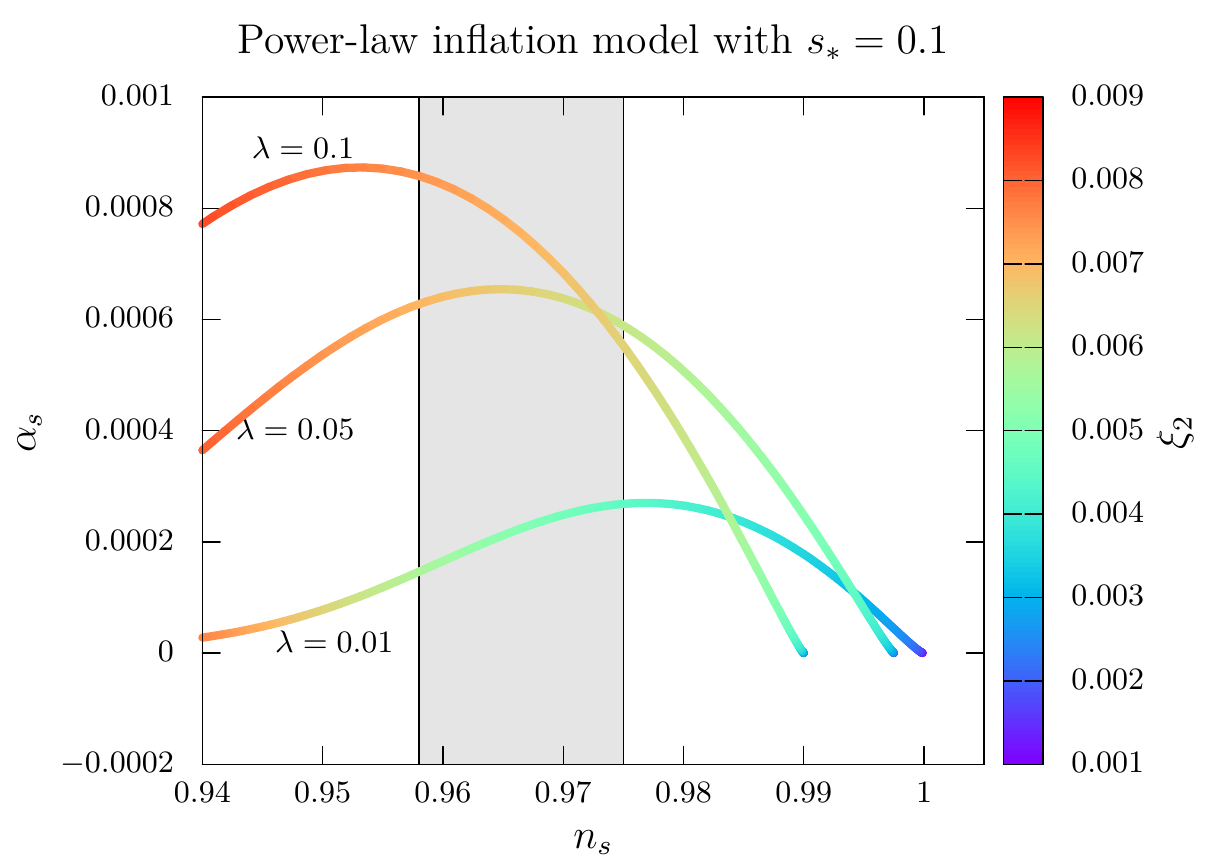}
\caption{Running of the spectral index $\alpha_s$ in terms of the spectral index $n_s$ for the power-law inflation model. Similar to Fig.~\ref{fig:fNL-PL}, two choices of $s_*$ are considered with three different values for $\lambda$. We observe that $\alpha_s$ initially increases and then decreases as $n_s$ decreases, and that a larger value of $s_*$ results in smaller $\alpha_s$. See also Appendix \ref{apdx:SIrunning}.}
\label{fig:alphas-PL}
\end{figure}

Finally, the running of the spectral index $\alpha_s$, discussed in Appendix \ref{apdx:SIrunning}, is presented in Fig.~\ref{fig:alphas-PL}. Again, two choices of $s_*$ and three values for the parameter $\lambda$ are considered. We observe that, as the spectral index decreases, the running of the spectral index first shows an increasing behaviour, and then it decreases. We further see that the running of the spectral index tends to be smaller when a larger value is considered for $s_*$. We note that this behaviour aligns with the general analysis presented in Appendix \ref{apdx:SIrunning}.

%%%%%%%%%%%%%%%%%%%%%%%%%%%%%%%%%%%%%%%%%%
\subsection{Hybrid inflation -- an example for Class II}
\label{subsec:hybrid}
%%%%%%%%%%%%%%%%%%%%%%%%%%%%%%%%%%%%%%%%%%

\begin{figure}[t!]
\centering
\includegraphics[width=15.6cm]{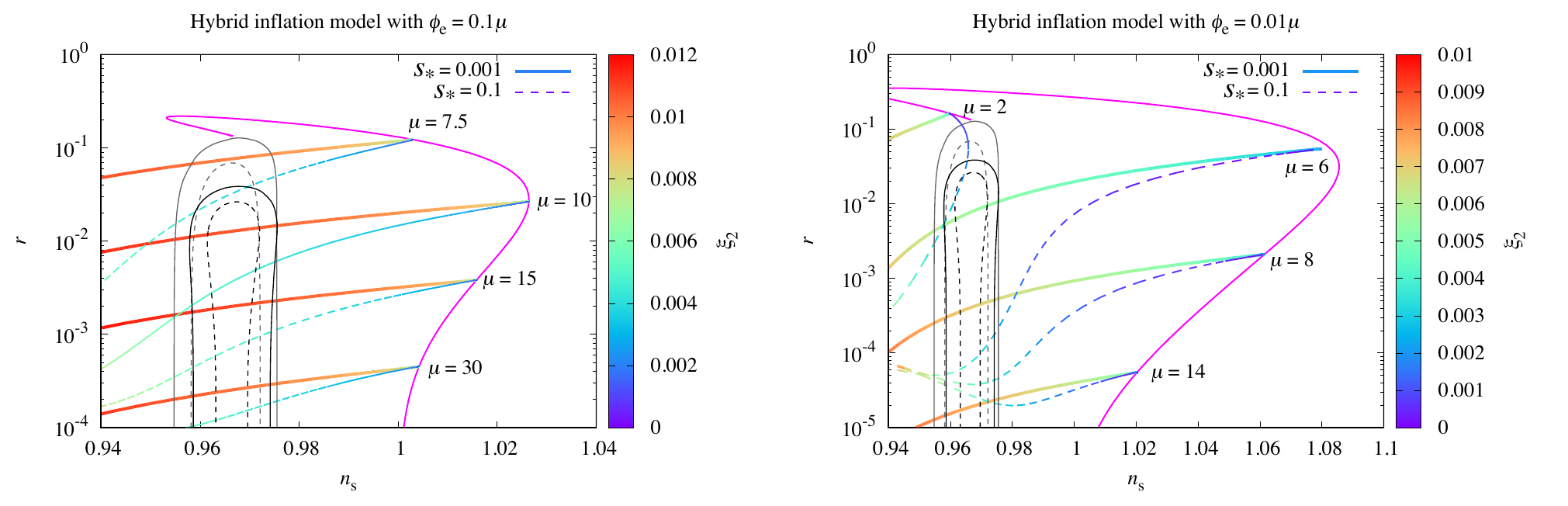}
\caption{Predictions for the hybrid inflation model. Two values of $\phi_e$ are considered: $\phi_e = 0.1\mu$ (left panel) and $\phi_e = 0.01\mu$ (right panel). The original model, depicted by the magenta lines, is ruled out by the latest Planck (grey) and Planck-BK (black) bounds. By varying the nonminimal coupling parameter $\xi_2$, the effect of the assistant field is presented for $\mu = \{7.5, 10, 15, 30\}$ when $\phi_e = 0.1\mu$ (left panel) and $\mu = \{2, 6, 8, 14\}$ when $\phi_e = 0.01\mu$ (right panel) with two choices of $s_*$, 0.001 (solid) and 0.1 (dashed). In the presence of the assistant field, the model may become compatible with the latest observational constraints.
} 
\label{fig:HI_ns-r}
\end{figure}

The scalar potential for the hybrid inflation model \cite{Cortes:2009ej} is given by
\begin{align}
V(\phi, \psi) 
= \frac{1}{2} m^2 \phi^2 
+ \frac{\lambda^\prime}{4}(\psi^2 - \Delta^2)^2
+ \frac{\lambda}{2} \phi^2 \psi^2
\,,
\end{align}
where $\psi$ is the waterfall field, $m$ is a mass of $\phi$, $\lambda$ and $\lambda^\prime$ are dimensionless coupling constants, and $\Delta$ is a constant parameter that has the dimension of mass. The inflationary trajectory mainly goes through the valley of the potential along $\psi = 0$. Thus, during inflation, the hybrid inflation model can effectively be described by a single-field model with the inflaton potential given by
\begin{align}
V_{\rm J}(\phi) = \Lambda^4 \left[1 + \left(\frac{\phi}{\mu}\right)^2 \right]
\,,
\end{align}
where $\Lambda = (\lambda^\prime/4)^{1/4} \Delta $ and $\mu = \sqrt{\lambda^\prime/2} \Delta^2/m$. The slow-roll parameters \eqref{eqn:SRparams0} are given by
\begin{align}
\epsilon^{(0)}
= \frac{2 (\phi/\mu)^2}{\mu^2[1+(\phi/\mu)^2]^2}
\,,\quad
\eta^{(0)}
= \frac{2}{\mu^2[1+(\phi/\mu)^2]}
\,.
\end{align}
We are interested in the parameter space where end of inflation is governed by the waterfall phase, not by slow-roll violations. We thus focus on the $\mu > 1/\sqrt{2}$ region so that the slow-roll parameters remain to be smaller than unity for any value of $\phi$. In this case, the model falls into Class II. We shall view $\phi_e$, the field value of the inflaton at the end of inflation, as a free parameter, together with $\mu$. Once $\phi_e$ and $\mu$ are fixed, we use 60 $e$-folds, {\it i.e.}, $60 = \int_e^* (V_{\rm J}/V_{{\rm J},\phi}) d\phi$, to find the field value at the pivot scale and thus the values of the spectral index and the tensor-to-scalar ratio.
The original predictions are shown in Fig.~\ref{fig:HI_ns-r} for $\phi_e = 0.1\mu$ and $\phi_e = 0.01\mu$ (magenta lines) by varying $\mu$.
We see that the original model is ruled out by the latest observational constraints.

\begin{figure}[t!]
\centering
\includegraphics[width=7.5cm]{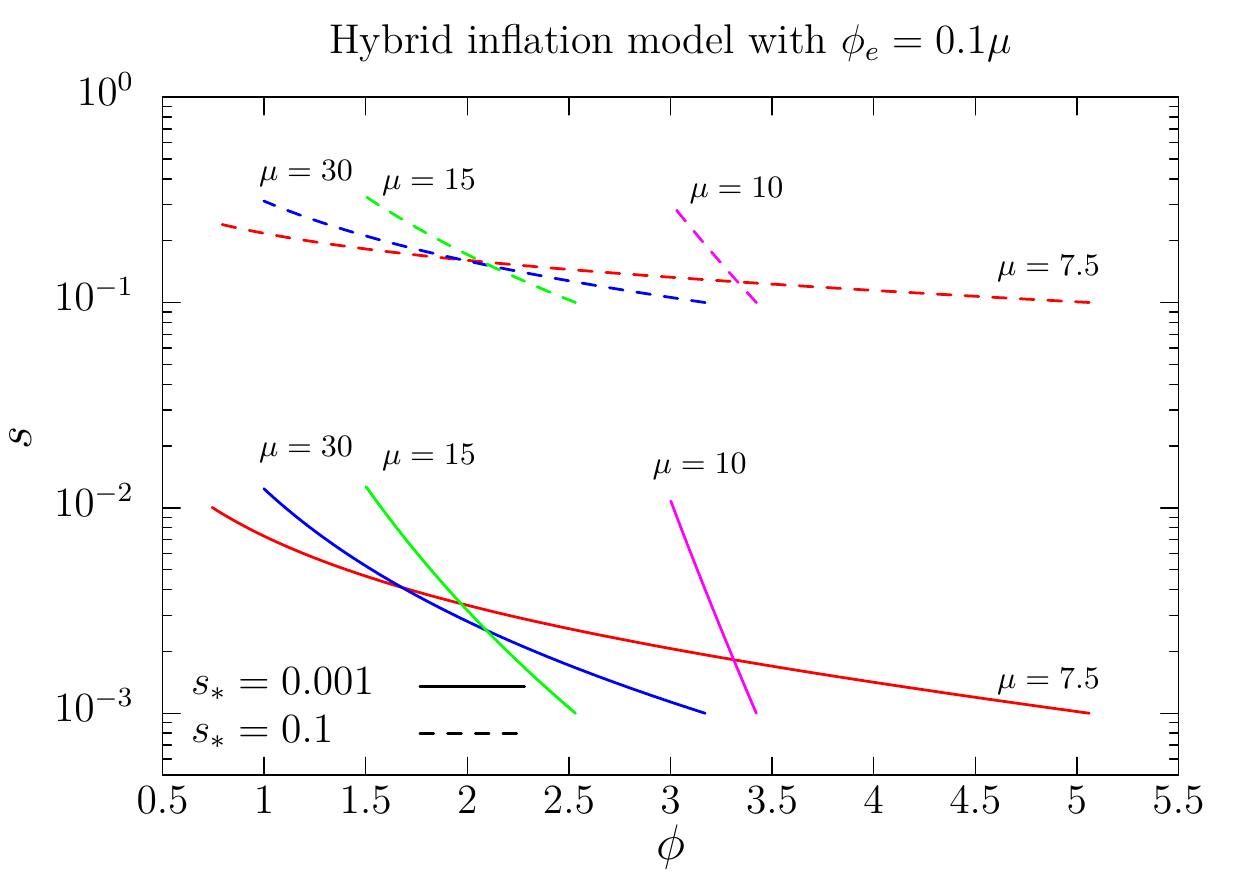}
\includegraphics[width=7.5cm]{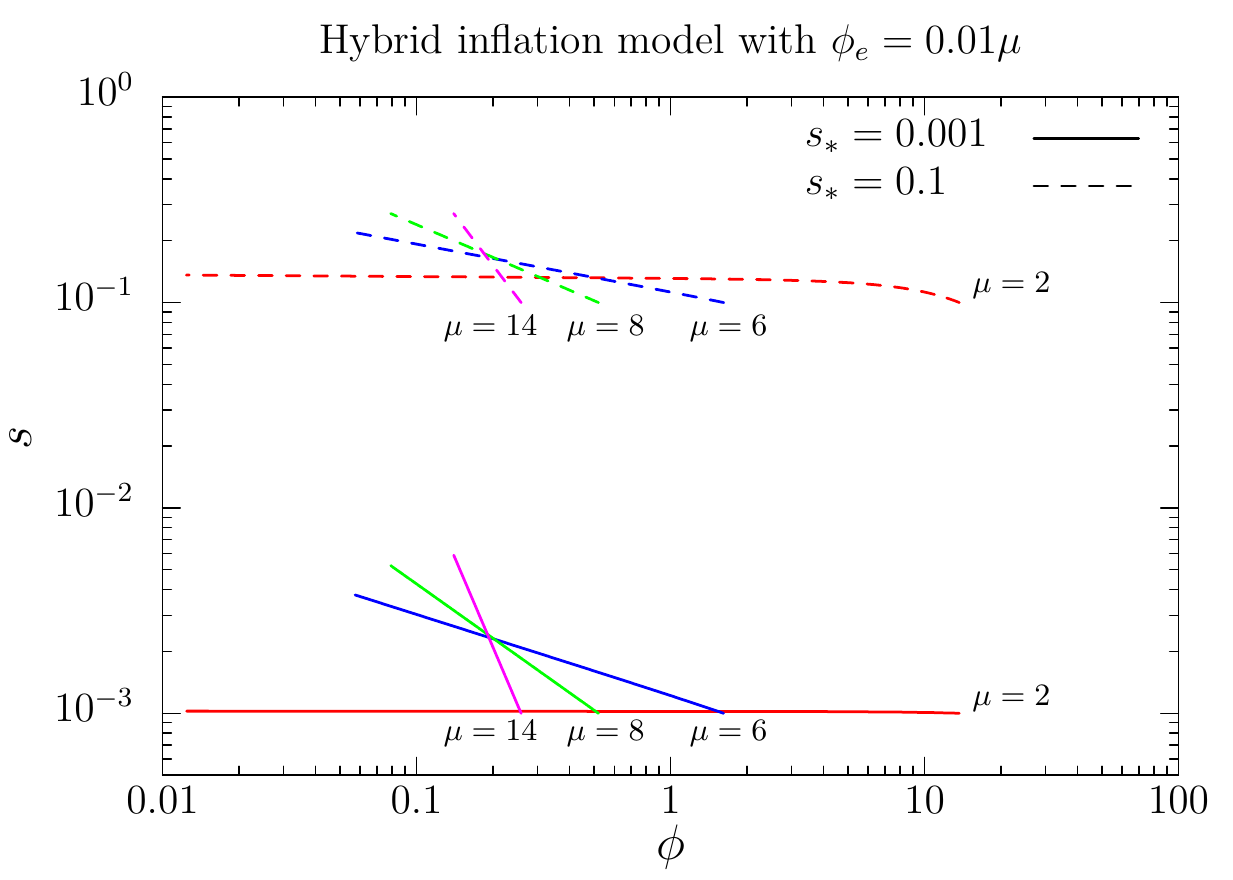}
\caption{Field trajectories for the hybrid inflation model. Two choices of $s_*$, 0.001 (solid) and 0.1 (dashed), are considered. The left panel shows the $\phi_e = 0.1 \mu$ case with $\mu = 7.5$ (red), $10$ (magenta), $15$ (green), and $30$ (blue), whereas the right panel shows the $\phi_e = 0.01\mu$ case with $\mu = 2$ (red), $6$ (blue), $8$ (green), and $14$ (magenta). For each case, we have chosen the value of $\xi_2$ in such a way that the spectral index becomes 0.965, except for $\{s_*,\phi_e,\mu\}=\{0.001,0.01\mu,2\}$ in which $n_s= 0.965$ cannot be reached for any values of $\xi_2$, and hence, $\xi_2=0.0001$ is chosen for illustration.
}
\label{fig:HI_trajectory}
\end{figure}

In Fig.~\ref{fig:HI_ns-r}, using the analytical expressions \eqref{eqn:ns-II} and \eqref{eqn:r-II}, the spectral index $n_s$ and the tensor-to-scalar ratio $r$ in the presence of the assistant field are presented for $\mu=\{7.5, 10, 15, 30\}$ with $s_*=\{0.001,0.1\}$ for the case of $\phi_e = 0.1\mu$ and $\mu=\{2, 6, 8, 14\}$ with $s_*=\{0.001,0.1\}$ for the case of $\phi_e = 0.01\mu$.
As expected, the inclusion of the assistant field brings the original predictions to the observationally-favoured region by suppressing the spectral index as well as the tensor-to-scalar ratio.
For example, in the case of $\phi_e = 0.1 \mu$ and $\mu = 10$, the prediction becomes compatible with the Planck-BK 2-sigma bounds for the range of $\xi_2 = (1.0 - 1.1) \times 10^{-2}$ for $s_* = 0.001$ and $\xi_2 = (4.3 - 5.3) \times 10^{-3}$ for $s_* = 0.1$. In the case of $\phi_e = 0.01 \mu$ and $\mu = 6$, the corresponding ranges for the nonminimal coupling parameter are $\xi_2 = (5.3 - 5.9) \times 10^{-3}$ and $\xi_2 = (2.2 - 4.3) \times 10^{-3}$ for $s_\ast = 0.001$ and $s_\ast = 0.1$, respectively.
Field trajectories for all the parameter choices are shown in Fig. \ref{fig:HI_trajectory}.

\begin{figure}[t!]
\centering
\includegraphics[width = 15.6cm]{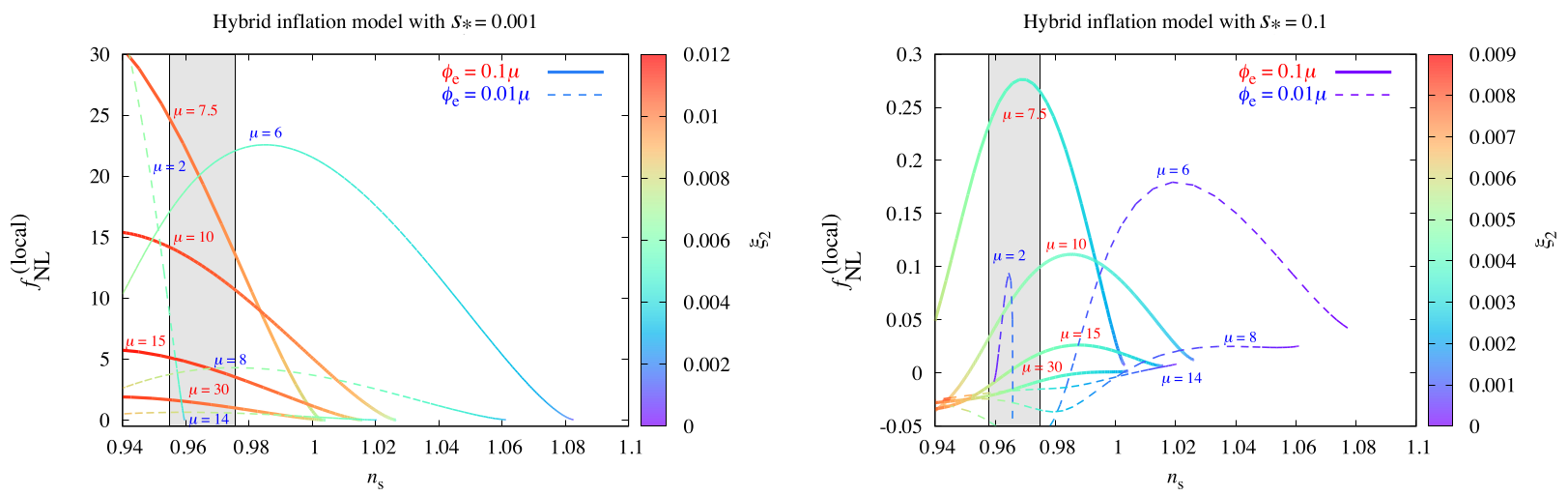}
\caption{
Nonlinearity parameter $f_{\rm NL}^{\rm (local)}$ in terms of the spectral index $n_s$ for the hybrid inflation model. Two choices of $s_*$ are considered, 0.001 (left) and 0.1 (right). In addition, $\phi_e = 0.1\mu$ (solid lines) and $\phi_e = 0.01\mu$ (dashed lines) are considered. The shaded region represents the Planck 2-sigma bound on the spectral index, $n_s = [0.955, 0.976]$. 
In the case of $s_* = 0.001$, the nonlinearity parameter may become large and incompatible with the Planck 2-sigma bound, $-11.1 < f_{\rm NL}^{({\rm local})} < 9.3$.
On the other hand, for a relatively large value of $s_*$ such as the $s_* = 0.1$ case, the nonlinearity parameter tends to be tiny.
}
\label{fig:fNL-HI}
\end{figure}

Figure~\ref{fig:fNL-HI} shows the nonlinearity parameter $f_{\rm NL}^{(\rm local)}$, obtained by using Eq.~\eqref{eqn:fNL-II}, as a function of the spectral index $n_s$ for the hybrid inflation model. Similar to Fig.~\ref{fig:HI_ns-r}, two choices of $s_*$, 0.001 and 0.1, are considered with $\phi_e = 0.1\mu$ and $\phi_e = 0.01\mu$. The shaded grey region indicates the Planck 2-sigma bound on the spectral index.
In the case of $s_* = 0.001$, the nonlinearity parameter may become large.
For example, in the cases of $\{\phi_e,\mu\}=\{0.1\mu,7.5\}$, $\{0.1\mu,10\}$, and $\{0.01\mu,6\}$, the nonlinearity parameter is incompatible with the Planck 2-sigma bound, $-11.1 < f_{\rm NL}^{({\rm local})} < 9.3$.
On the other hand, the nonlinearity parameter tends to be tiny in the $s_* = 0.1$ case.

\begin{figure}
\centering
\includegraphics[scale=0.6]{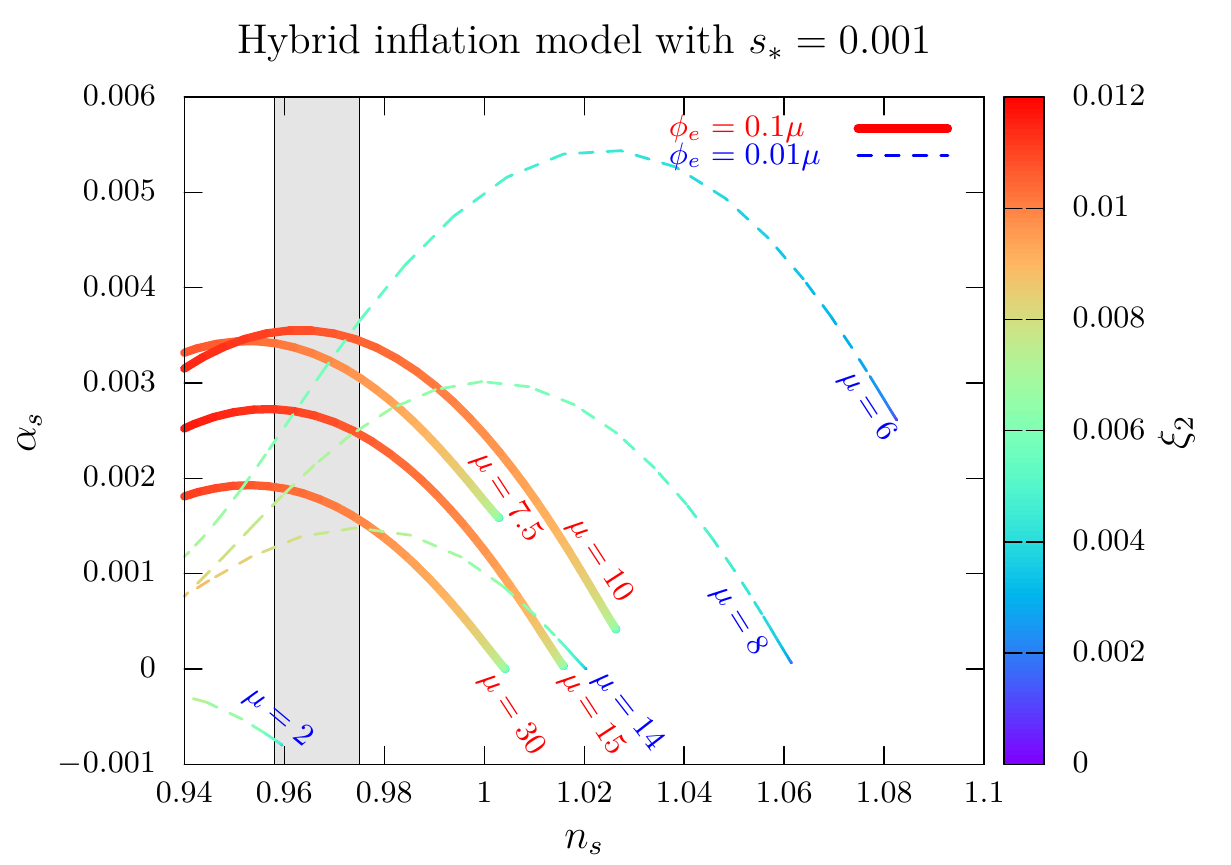}\;
\includegraphics[scale=0.6]{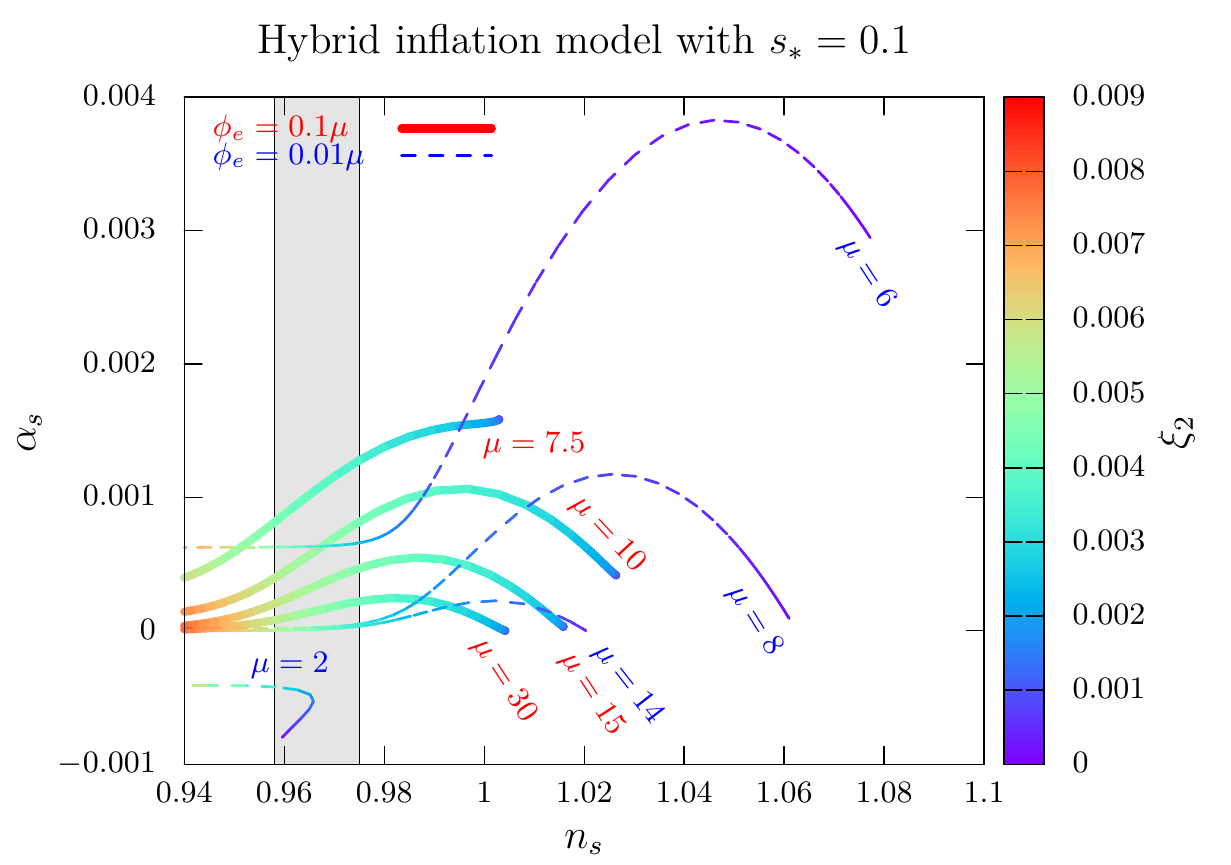}
\caption{Running of the spectral index $\alpha_s$ in terms of the spectral index $n_s$ for the hybrid inflation model. Similar to Fig.~\ref{fig:fNL-HI}, two choices of $s_*$ are considered with two different values for $\phi_e$. We observe that $\alpha_s$ initially increases and then decreases as $n_s$ decreases, and that a larger value of $s_*$ results in smaller $\alpha_s$. See also Appendix \ref{apdx:SIrunning}.}
\label{fig:alphas-HI}
\end{figure}

The running of the spectral index $\alpha_s$, discussed in Appendix \ref{apdx:SIrunning}, is presented in Fig.~\ref{fig:alphas-HI}. Similar to Fig.~\ref{fig:fNL-HI}, two choices of $s_*$ and two values for $\phi_e$ are considered. The general tendency of the behaviour of $\alpha_s$ remains the same as those for the previous two example models. In other words, the running of the spectral index shows an increasing behaviour before it decreases again, and it tends to take a smaller value when $s_*$ takes a larger value. For the detailed discussion of the running of the spectral index, see Appendix \ref{apdx:SIrunning}.

%%%%%%%%%%%%%%%%%%%%%%%%%%%%%%%%%%%%%%%%%%
\section{Conclusion}
\label{sec:conc}
%%%%%%%%%%%%%%%%%%%%%%%%%%%%%%%%%%%%%%%%%%
We have investigated the impact of the assistant field on single-field inflationary models. The assistant field is characterised by being nonminimally coupled to gravity, effectively massless, and having no direct coupling to the original inflaton field. Without specifying a potential for the inflaton field, we performed a general analysis and presented analytical expressions for inflationary observables such as the spectral index $n_s$, the tensor-to-scalar ratio $r$, and the local-type nonlinearity parameter $f_{\rm NL}^{\rm (local)}$, for both the metric and Palatini formulations.

For Class I, where the end of inflation is achieved through slow-roll violations, the assistant field reduces both $n_s$ and $r$, potentially reviving many single-field models that were previously ruled out by observations. For Class II, where the end of inflation is determined by a separate sector, a small $\epsilon_e^{(0)}$ value may bring a small $n_s$ into the observationally-favoured region. The compatibility of the nonlinearity parameter $f_{\rm NL}^{\rm (local)}$ with observations is also shown for both Class I and Class II.

Our results are demonstrated using three example models: loop inflation (Class I), power-law inflation, and hybrid inflation (Class II). These models were previously ruled out due to large values of $n_s$ and/or $r$. Our findings show that the presence of the assistant field can bring the predictions of $n_s$ and $r$ into the observationally-favoured region, making the models compatible with the current observational bounds.

It is worth noting that not all single-field models can be revived with the help of the assistant field. For instance, models that originally predict a small $n_s$ and belong to either Class I or Class II with a large $\epsilon_e^{(0)}$ cannot be revived. The impact of a massive assistant field on such models remains an open question and will be the subject of future study.

%%%%%%%%%%%%%%%%%%%%%%%%%%%%%%%%%%%%%%%%%%
\acknowledgments
%%%%%%%%%%%%%%%%%%%%%%%%%%%%%%%%%%%%%%%%%%
This work was supported by National Research Foundation grants funded by the Korean government (NRF-2021R1A4A2001897) and (NRF-2019R1A2C1089334) (S.C.P.), and JSPS KAKENHI Grants No.~19K03874 (T.T.).

\appendix

%%%%%%%%%%%%%%%%%%%%%%%%%%%%%%%%%%%%%%%%%%
\section{Running of the Spectral Index}
\label{apdx:SIrunning}
%%%%%%%%%%%%%%%%%%%%%%%%%%%%%%%%%%%%%%%%%%

The running of the scalar spectral index in a general multi-field inflation model is given by \cite{Gong:2014kpa}
\begin{align}
\alpha_s &\equiv \frac{dn_s}{d\ln k} \simeq
4\epsilon^2-2\frac{\dot{\epsilon}}{H}
+2\frac{N_{,i}N_{,j}}{G^{mn}N_{,m}N_{,n}}\left(
4\epsilon w^{ij} + 2w^i{}_k w^{jk} - w^{ij}{}_{;k}\frac{\dot{\varphi}^k_0}{H}
\right) - (n_s-1)^2
\,,\label{eqn:SIRunning}
\end{align}
evaluated at the horizon crossing, which we denote by the subscript $*$, where
\begin{align}
w_{ij} =
u_{(i;j)} + \frac{1}{3}R_{m(ij)n}\frac{\dot{\varphi}^m_0\dot{\varphi}^n_0}{H^2}
\,,\qquad
u_i =
-\frac{V_{,i}}{3H^2}
\,,\label{eqn:wandu}
\end{align}
with $\varphi^i_0$ being the background fields. For the two-field model under consideration, one finds that
\begin{align*}
\frac{\dot{\epsilon}}{H}\bigg\vert_* &\approx
-2\epsilon^\phi_*\left(
\eta^{\phi\phi}_*
-2\epsilon^\phi_*
\right)
-2\epsilon^\sigma_*\left(
\eta^{\sigma\sigma}_*
-2\epsilon^\sigma_*
\right)
+\epsilon^\phi_*\sqrt{\epsilon^\sigma_* \epsilon^b_*} s^b_* s^\sigma_*
\,,
\end{align*}
and
\begin{align*}
2\frac{N_{,i}N_{,j}}{G^{mn}N_{,m}N_{,n}}\left(
4\epsilon w^{ij} + 2w^i{}_k w^{jk} - w^{ij}{}_{;k}\frac{\dot{\varphi}^k}{H}
\right)
&\approx
\frac{2(N_{,\phi})^2A^{\phi\phi}
+ 2(N_{,\sigma})^2A^{\sigma\sigma}
+ 4N_{,\phi}N_{,\sigma}A^{\phi\sigma}}{
e^{-2b}(N_{,\phi})^2+(N_{,\sigma})^2
}
\,,
\end{align*}
where
\begin{align*}
A^{\phi\phi} &=
4\epsilon w^{\phi\phi}
+2e^{2b}(w^{\phi\phi})^2
+2(w^{\phi\sigma})^2
+e^{-b}s^\phi \sqrt{2\epsilon^\phi}w^{\phi\phi}{}_{;\phi}
+s^\sigma\sqrt{2\epsilon^\sigma}w^{\phi\phi}{}_{;\sigma}
\,,\\
A^{\sigma\sigma} &=
4\epsilon w^{\sigma\sigma}
+2e^{2b}(w^{\phi\sigma})^2
+2(w^{\sigma\sigma})^2
+e^{-b}s^\phi\sqrt{2\epsilon^\phi}w^{\sigma\sigma}{}_{;\phi}
+s^\sigma\sqrt{2\epsilon^\sigma}w^{\sigma\sigma}{}_{;\sigma}
\,,\\
A^{\phi\sigma} &=
4\epsilon w^{\phi\sigma}
+2e^{2b}w^{\phi\phi}w^{\phi\sigma}
+2w^{\phi\sigma}w^{\sigma\sigma}
+e^{-b}s^\phi\sqrt{2\epsilon^\phi}w^{\phi\sigma}{}_{;\phi}
+s^\sigma\sqrt{2\epsilon^\sigma}w^{\phi\sigma}{}_{;\sigma}
\,,
\end{align*}
and
\begin{align*}
N_{,\phi}\Big\vert_{*} &=
\frac{1}{\sqrt{2}}\frac{s^\phi_*}{\sqrt{\epsilon^\phi_*}}\frac{\epsilon^\phi_e}{\epsilon_e}e^{2b_e-b_*}
\,,\\
N_{,\sigma}\Big\vert_{*} &=
\frac{1}{\sqrt{2}}\frac{s^\sigma_*}{\sqrt{\epsilon^\sigma_*}}\left(
1-\frac{\epsilon^\phi_e}{\epsilon_e}e^{2b_e-2b_*}
\right)
\,.
\end{align*}
From the definition of $w_{ij}$ in Eq. \eqref{eqn:wandu}, we find that
\begin{align*}
w^{\phi\phi}\Big\vert_* &=
e^{-2b_*}\left[
2\epsilon^\phi_* - \eta^{\phi\phi}_* - \frac{1}{2}s^b_*s^\sigma_*\sqrt{\epsilon^b_*\epsilon^\sigma_*} + \frac{1}{24}\epsilon^\sigma_*\left(\eta^b_*+2\epsilon^b_*\right)
\right]
\,,\\
w^{\phi\sigma}\Big\vert_* &=
\frac{1}{2}e^{-b_*}s^\phi_*\sqrt{\epsilon^\phi_*}\left[
s^b_* \sqrt{\epsilon^b_*}
-\frac{1}{12} s^\sigma_* \sqrt{\epsilon^\sigma_*}\left(\eta^b_* + 2\epsilon^b_*\right)
\right]
\,,\\
w^{\sigma\sigma}\Big\vert_* &=
2\epsilon^\sigma_* - \eta^{\sigma\sigma}_* + \frac{1}{24}\epsilon^\phi_*\left(\eta^b_* + 2\epsilon^b_*\right)
\,,
\end{align*}
and
\begin{align*}
w^{\phi\phi}{}_{;\phi}\Big\vert_{*} &= 
\frac{1}{\sqrt{2\epsilon^\phi_*}}s^\phi_* e^{-b_*}\left[
-\xi^{\phi\phi\phi}_*
+6\eta^{\phi\phi}_*\epsilon^\phi_*
-8(\epsilon^\phi_*)^2
+\frac{1}{2}\epsilon^b_*\epsilon^\phi_*
-\frac{1}{24}s^b_* s^\sigma_* \sqrt{\epsilon^b_*\epsilon^\sigma_*}\epsilon^\phi_*\Big(\eta^b_*+2\epsilon^b_*\Big)
\right]
\,,\\
w^{\phi\phi}{}_{;\sigma}\Big\vert_{*} &= 
\frac{1}{\sqrt{2\epsilon^b_*}}e^{-2b_*}\bigg\{
\frac{1}{3}s^b_*\epsilon^\sigma_*\xi^b_*
+s^b_*\epsilon^b_*\Big(\eta^{\phi\phi}_*-2\epsilon^\phi_*\Big)
+\frac{1}{2}s^b_*\epsilon^b_*\left[
\frac{1}{12}\epsilon^\sigma_*\eta^b_*-\Big(\eta^{\phi\phi}_*-2\epsilon^\sigma_*\Big)
\right]
\nonumber\\
&\qquad\qquad\qquad
+\frac{1}{24}s^\sigma_*\sqrt{\epsilon^b_*\epsilon^\sigma_*}\left[
2\Big(\eta^b_*+2\epsilon^b_*\Big)\Big(\eta^{\sigma\sigma}_*-2\epsilon^\sigma_*\Big)-3\eta^b_*
\right]
\bigg\}
\,,\\
w^{\sigma\sigma}{}_{;\phi}\Big\vert_{*} &=
\frac{1}{24\sqrt{2}}s^\phi_* \sqrt{\epsilon^\phi_*} e^{b_*}\left[
\Big(\eta^b_*+2\epsilon^b_*\Big)\Big(2\eta^{\phi\phi}_*-4\epsilon^\phi_*+s^b_*s^\sigma_*\sqrt{\epsilon^b_*\epsilon^\sigma_*}\Big)-12\epsilon^b_*
\right]
\,,\\
w^{\sigma\sigma}{}_{;\sigma}\Big\vert_{*} &=
\frac{1}{\sqrt{2\epsilon^\sigma_*}}s^\sigma_*\left\{
-\xi^{\sigma\sigma\sigma}_*
+2\epsilon^\sigma_*\Big(3\eta^{\sigma\sigma}_*-4\epsilon^\sigma_*\Big)
-\frac{1}{12}s^b_*s^\sigma_*\sqrt{\frac{\epsilon^\sigma_*}{\epsilon^b_*}}\epsilon^\phi_*\left[\Big(\epsilon^b_*\Big)^2-4\xi^b_*\right]
\right\}
\,,\\
w^{\phi\sigma}{}_{;\phi}\Big\vert_{*} &=
\frac{1}{48\sqrt{2}}\bigg\{
s^b_*\sqrt{\epsilon^b_*}\left[
48\Big(\eta^{\phi\phi}_*-2\epsilon^\phi_*\Big)
-24\Big(\eta^{\sigma\sigma}_*-2\epsilon^\sigma_*\Big)
+\Big(\epsilon^\phi_*-\epsilon^\sigma_*\Big)\Big(\eta^b_*+2\epsilon^b_*\Big)
\right]
\nonumber\\
&\qquad\qquad\qquad
+2s^\sigma_*\sqrt{\epsilon^\sigma_*}\left[
6\sqrt{\epsilon^\sigma_*}-\Big(\eta^{\phi\phi}_*-2\epsilon^\phi_*\Big)\Big(\eta^b_*+2\epsilon^b_*\Big)
\right]
\bigg\}
\,,\\
w^{\phi\sigma}{}_{;\sigma}\Big\vert_{*} &=
\frac{1}{24\sqrt{2}}s^\phi_*\sqrt{\epsilon^\phi_*}e^{-b_*}\bigg\{
3\Big(\eta^b_*-2\epsilon^b_*\Big)
-\Big(\eta^b_*+2\epsilon^b_*\Big)\Big(\eta^{\sigma\sigma}_*-2\epsilon^\sigma_*\Big)
\nonumber\\
&\qquad\qquad\qquad\qquad\qquad
+\frac{1}{2}s^b_*s^\sigma_*\sqrt{\frac{\epsilon^\sigma_*}{\epsilon^b_*}}\left[2\Big(\epsilon^b_*\Big)^2-\epsilon^b_*\eta^b_*-16\xi^b_*\right]
\bigg\}
\,.
\end{align*}
Here, we have defined
\begin{align*}
\xi^{\phi\phi\phi} \equiv
\frac{V_{,\phi\phi\phi}V_{,\phi}}{V^2}e^{-4b}
\,,\quad
\xi^{\sigma\sigma\sigma} \equiv
\frac{V_{,\sigma\sigma\sigma}V_{,\sigma}}{V^2}
\,,\quad 
\xi^b \equiv 
8b_{,\sigma\sigma\sigma}b_{,\sigma}
\,.
\end{align*}
Combining all the individual terms, together with Eq.~\eqref{eqn:scalarSI}, into Eq.~\eqref{eqn:SIRunning} gives the slow-roll-approximated analytical expression for the running of the scalar spectral index. As the calculation is straightforward and the resultant expression is rather long, we do not present the final expression here.

\begin{figure}[t!]
\centering
\includegraphics[scale=0.6]{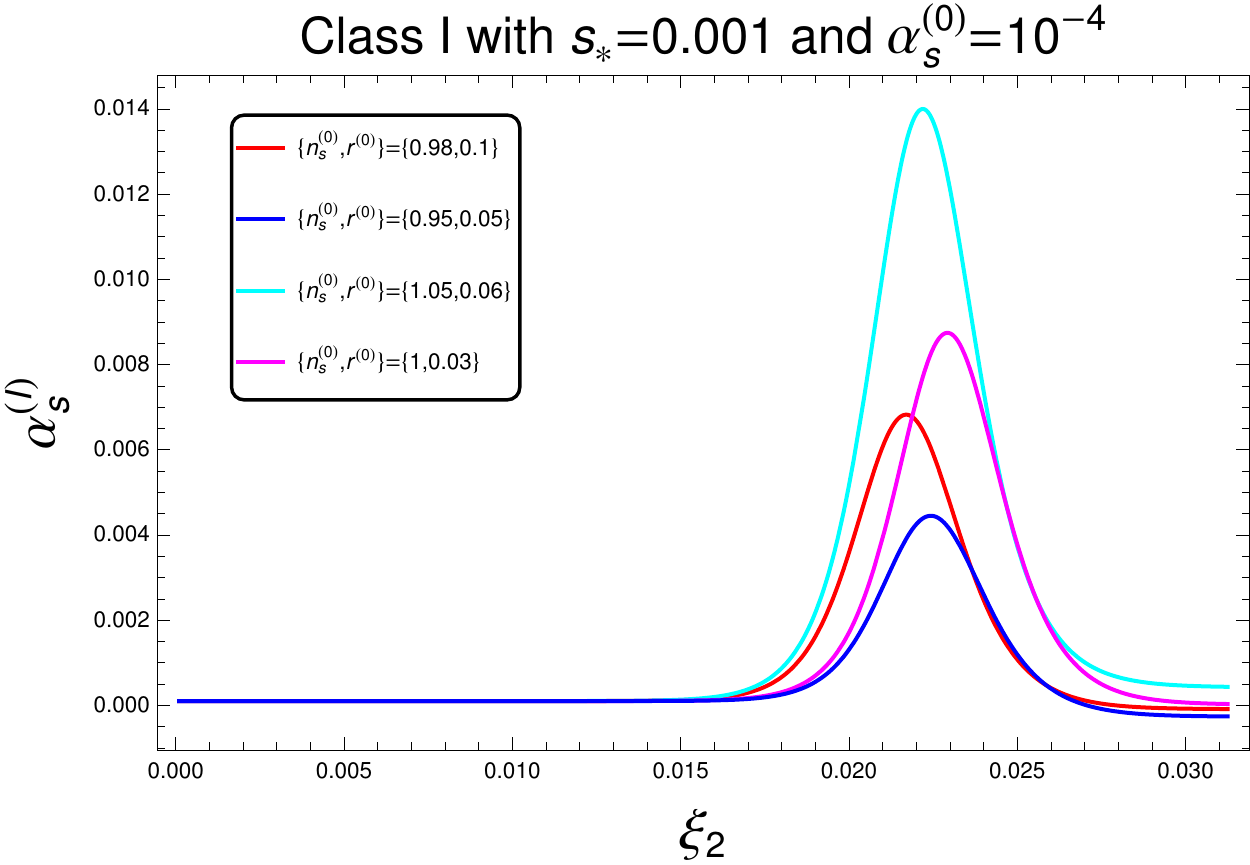}\;
\includegraphics[scale=0.6]{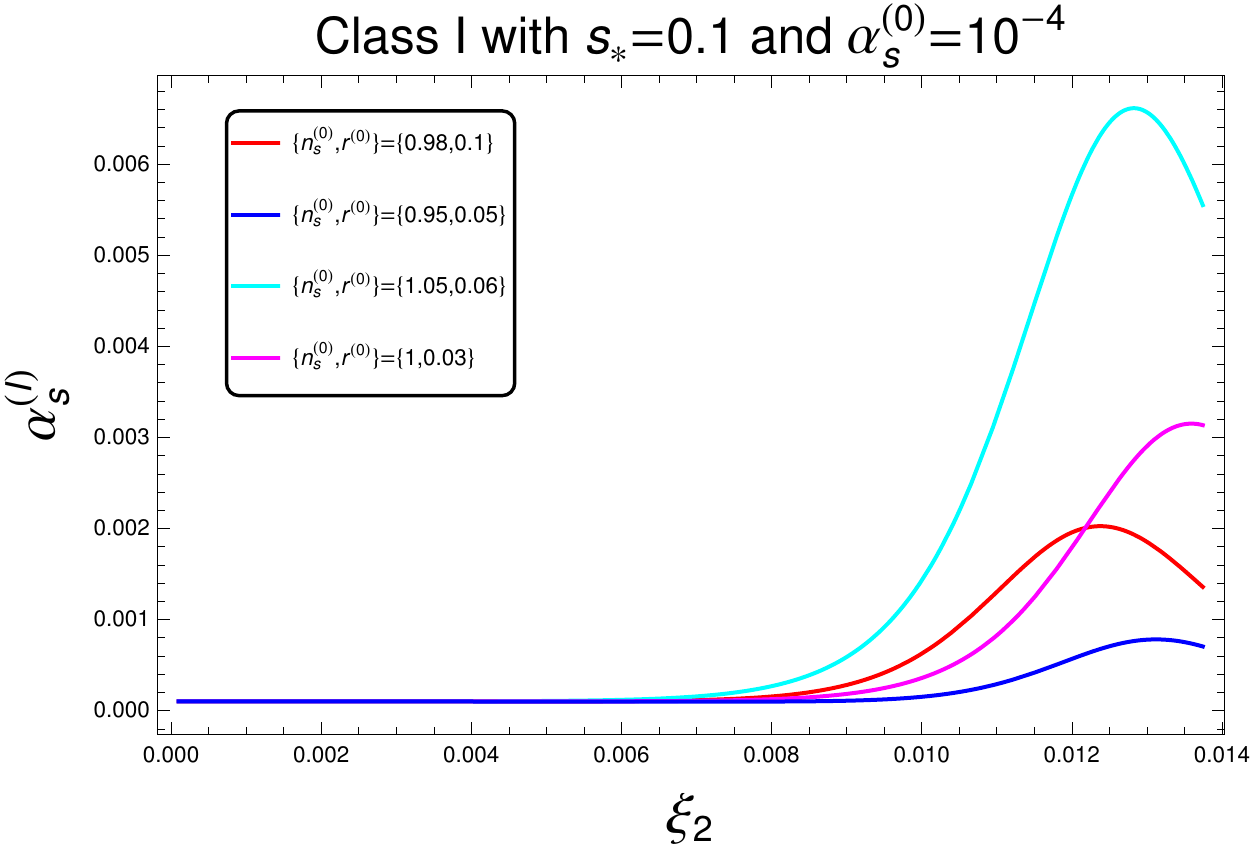}
\\
\includegraphics[scale=0.6]{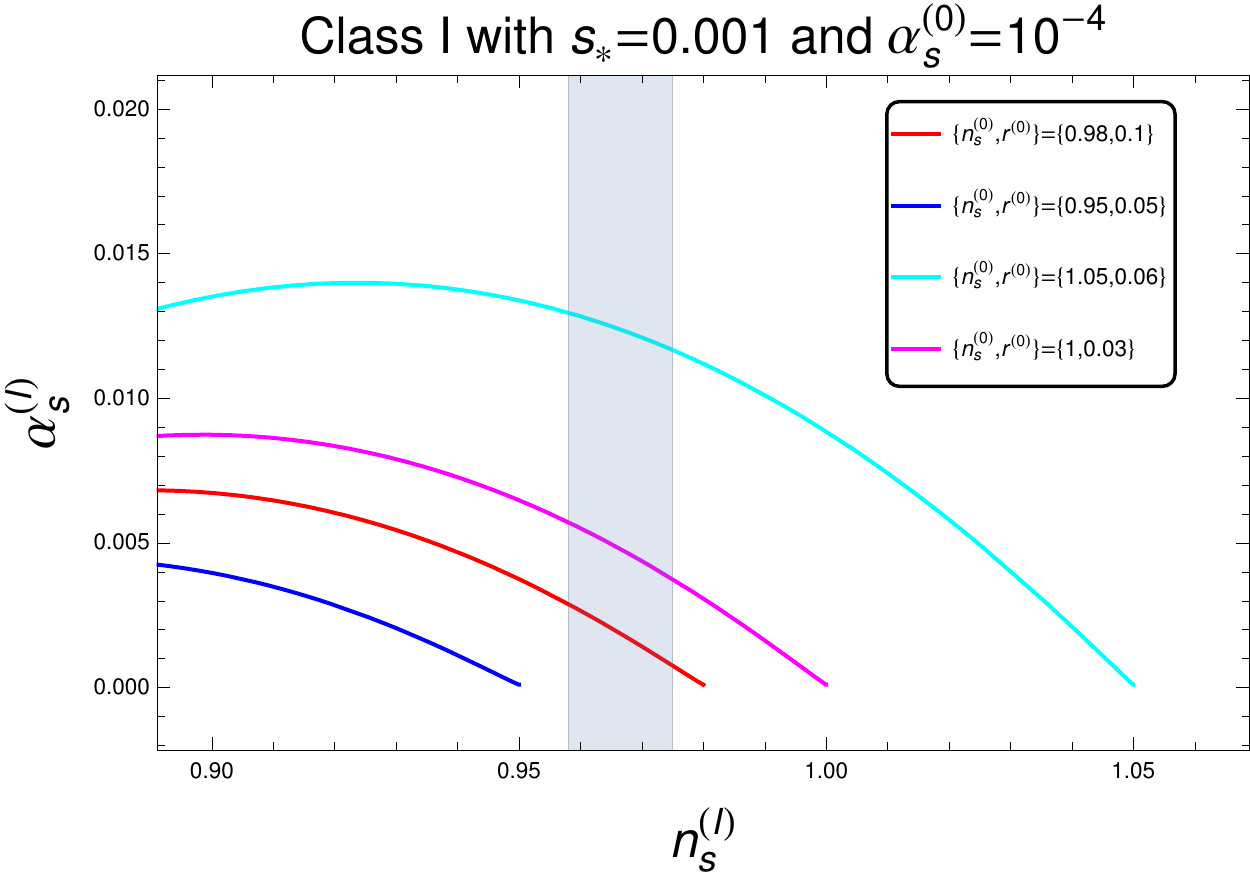}\;
\includegraphics[scale=0.6]{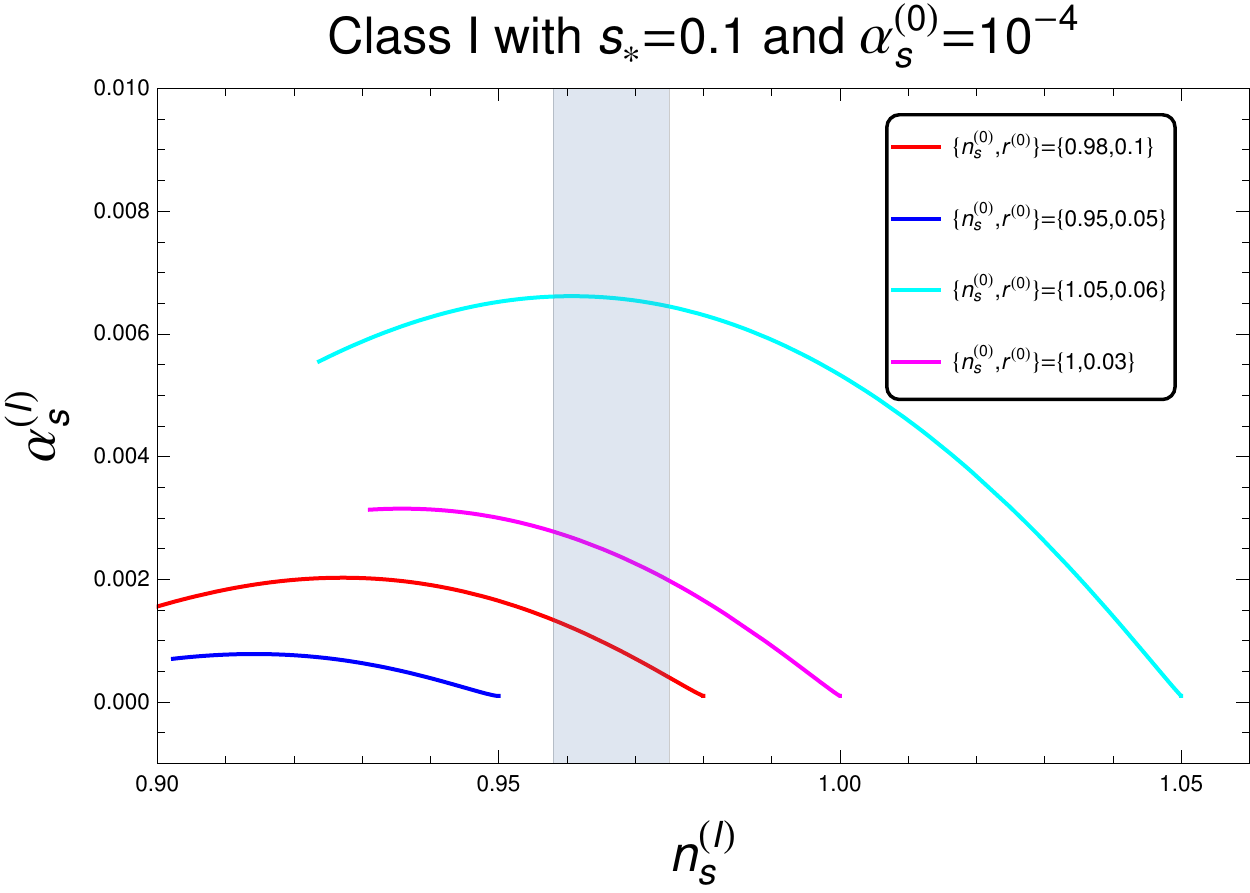}
\caption{In the upper panel, the evolution of the running of the spectral index is shown for Class~I in terms of the nonminimal coupling parameter $\xi_2$. Two values of $s_*$ are considered, 0.001 (left) and 0.1 (right) with various choices of $\{n_s^{(0)},r^{(0)}\}$. Furthermore, we have fixed $\alpha_s^{(0)}=10^{-4}$. In the lower panel, we present the predictions in the $\alpha_s^{({\rm I})}$--$n_s^{({\rm I})}$ plane. The shaded region corresponds to the latest bounds on the spectral index. We observe that the running first grows and then eventually decreases as $\xi_2$ increases. Only the metric formulation is considered.}
\label{fig:alphas-I}
\end{figure}
\begin{figure}[ht!]
\centering
\includegraphics[scale=0.6]{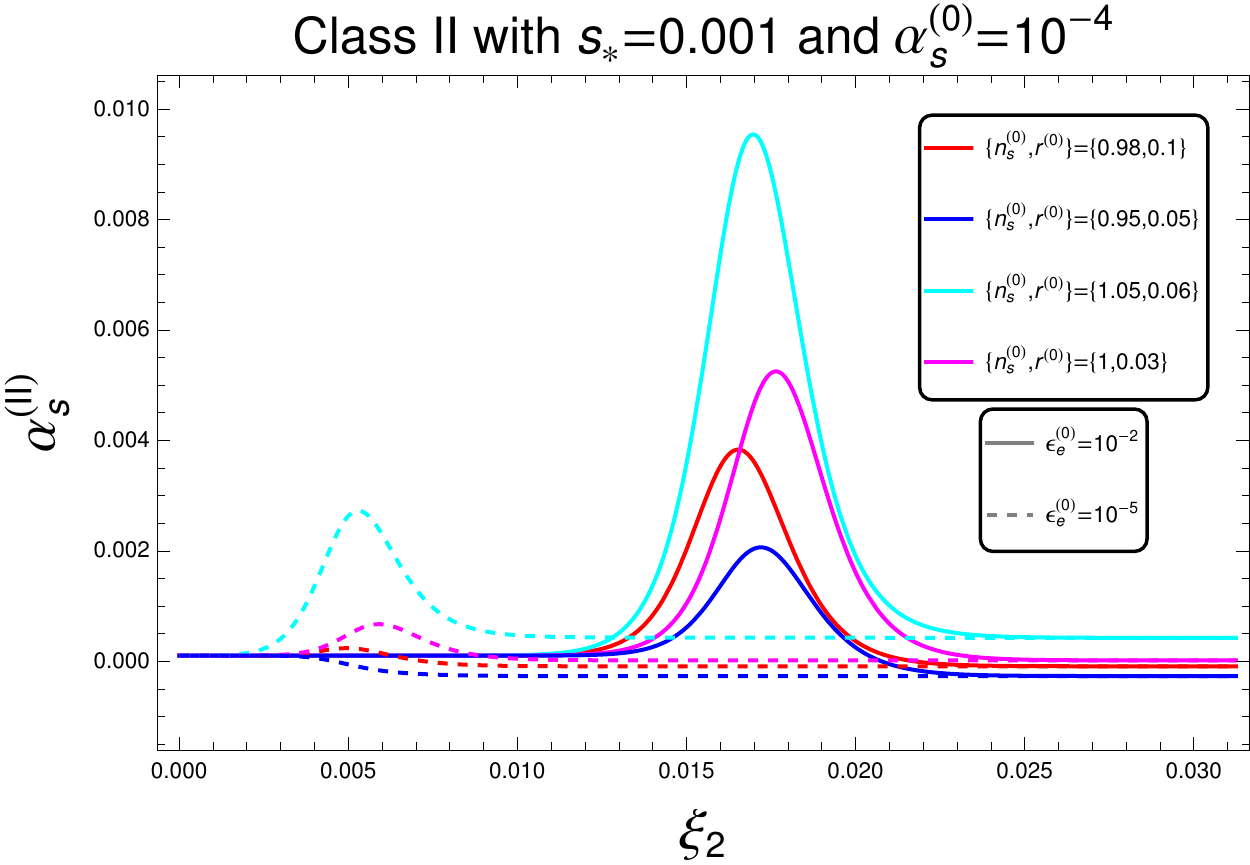}\;
\includegraphics[scale=0.6]{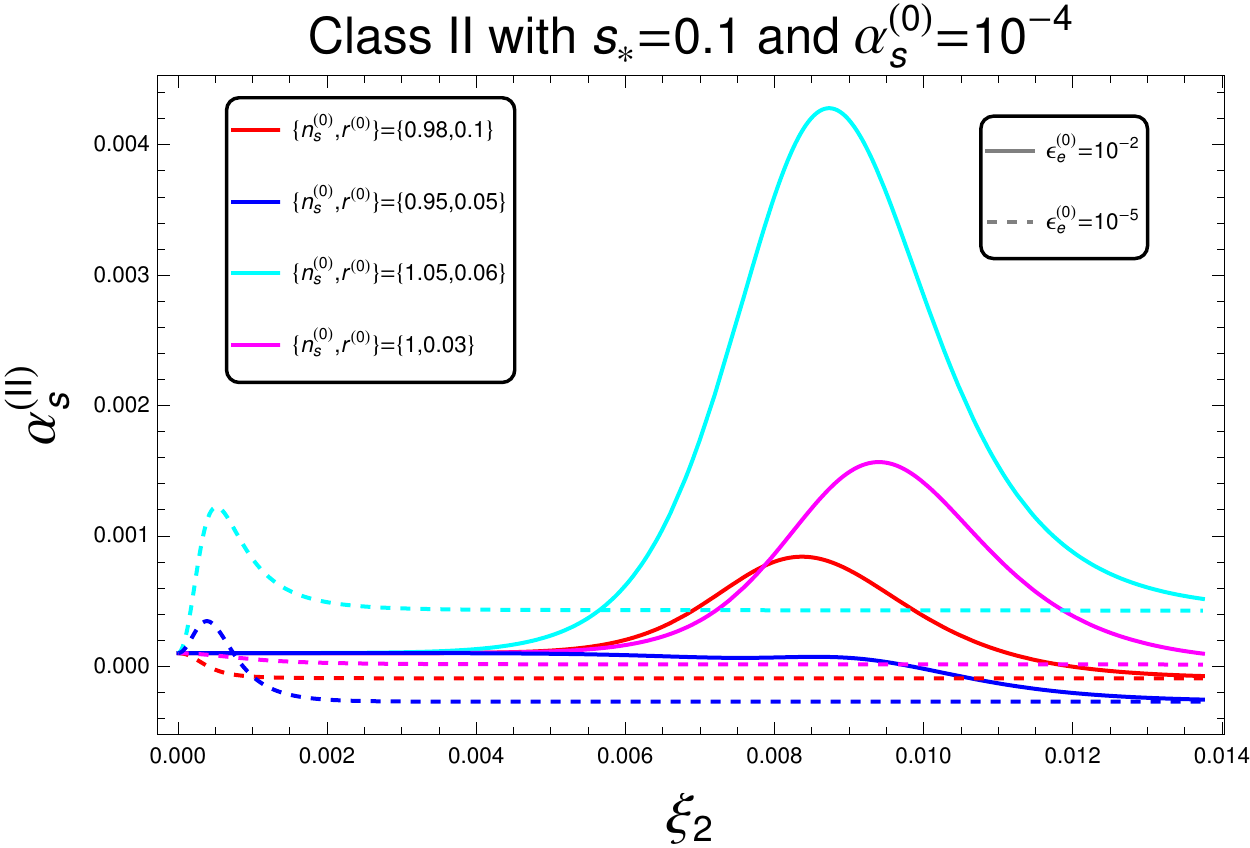}
\\
\includegraphics[scale=0.6]{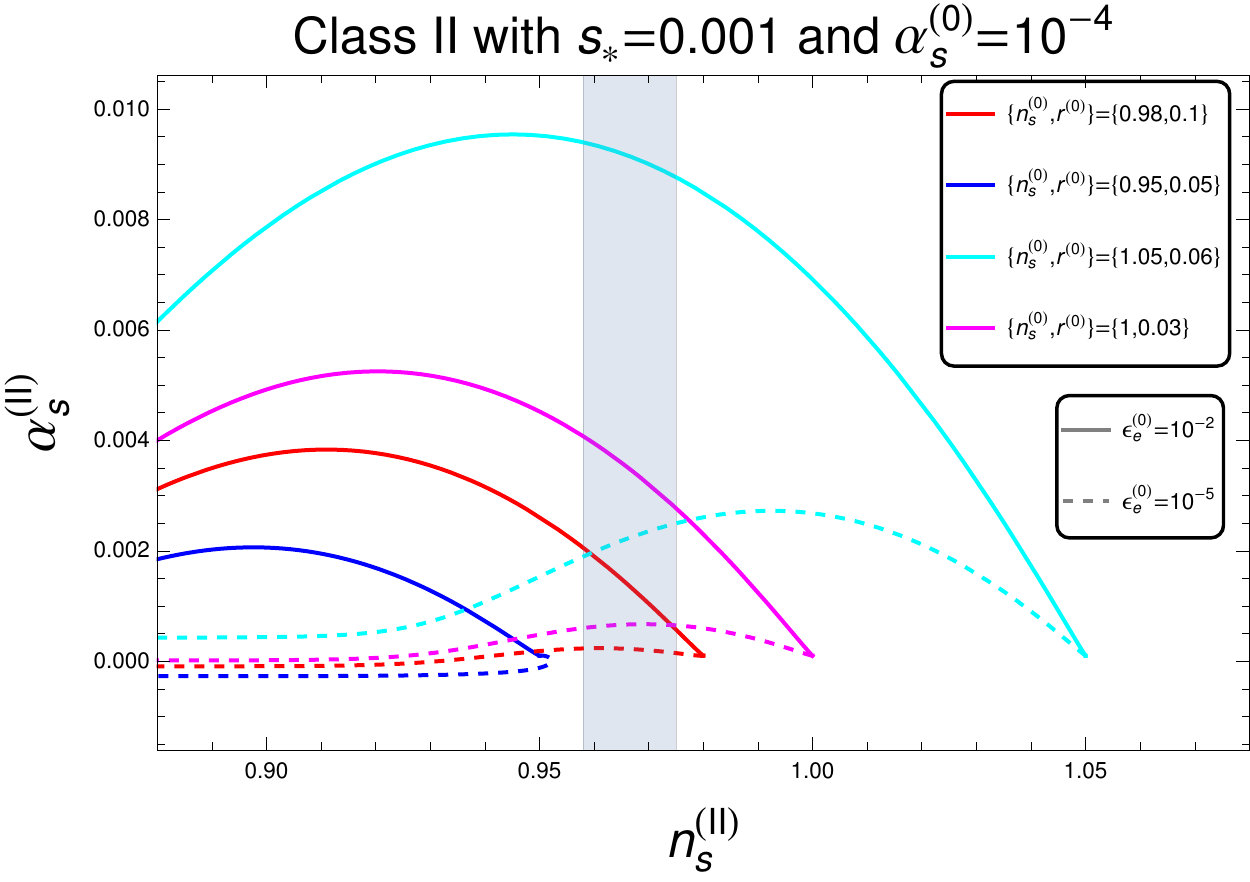}\;
\includegraphics[scale=0.6]{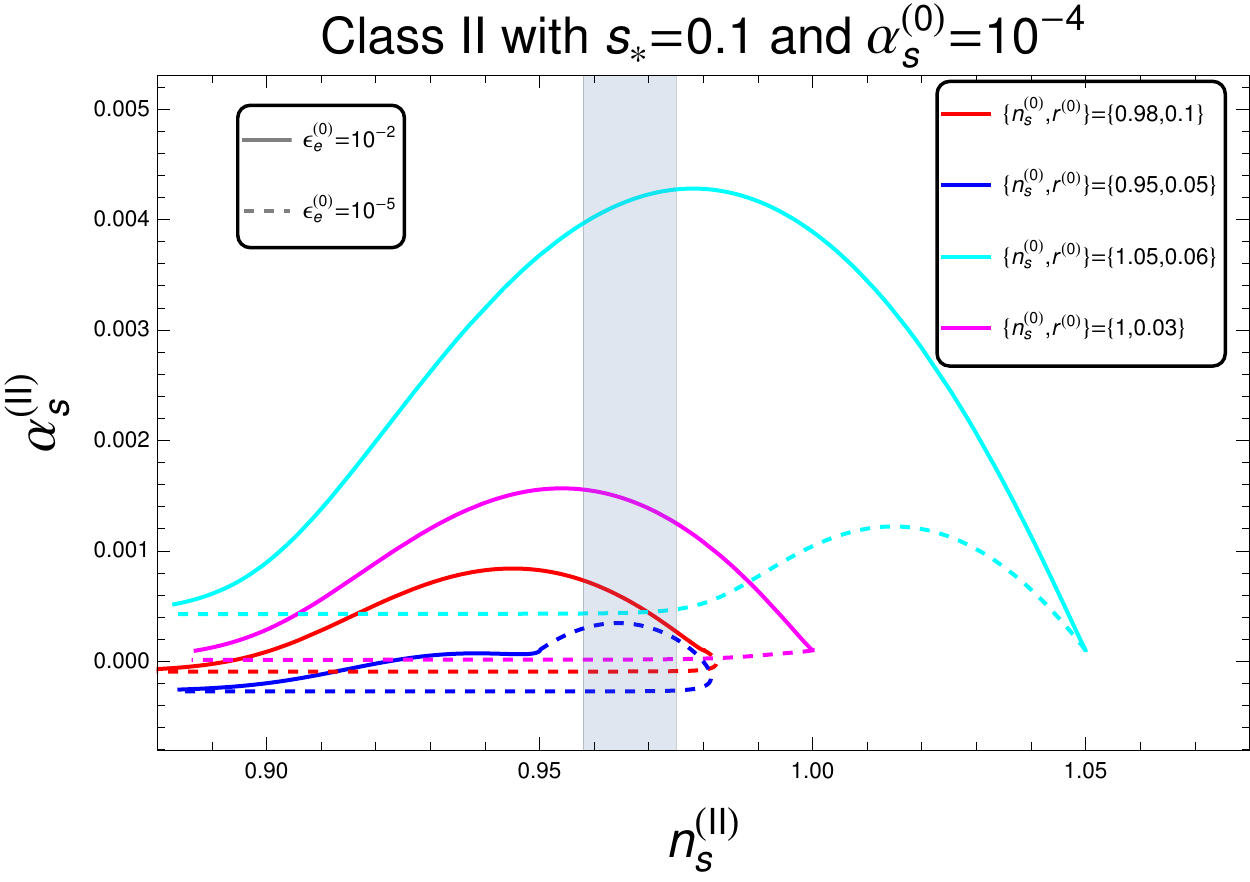}
\caption{In the upper panel, the evolution of the running of the spectral index is shown for Class~II in terms of the nonminimal coupling parameter $\xi_2$. Two values of $s_*$ are considered, 0.001 (left) and 0.1 (right) with various choices of $\{n_s^{(0)},r^{(0)}\}$. Furthermore, two different values, $10^{-2}$ and $10^{-5}$, are chosen for $\epsilon_e^{(0)}$. For all cases, we have fixed $\alpha_s^{(0)}=10^{-4}$. In the lower panel, we present the predictions in the $\alpha_s^{({\rm II})}$--$n_s^{({\rm II})}$ plane. The shaded region corresponds to the latest bounds on the spectral index. We again observe that $\alpha_s$ grows and then decreases as $\xi_2$ increases.}
\label{fig:alphas-II}
\end{figure}

Figure~\ref{fig:alphas-I} shows the behaviour of the running of the spectral index as a function of the nonminimal coupling parameter (upper panel) and the spectral index (lower panel) for Class~I. Two values of $s_*$, 0.001 (left) and 0.1 (right), are considered, together with various choices of $\{n_s^{(0)},r^{(0)}\}$. For all cases, we have fixed $\alpha_s^{(0)}$ to be $10^{-4}$. We observe that the running of the spectral index first grows as the nonminimal coupling parameter increases. Eventually, the running of the spectral index decreases. This behaviour is similar to that of the nonlinearity parameter. 
For Class~II, the behaviour of the running of the spectral index remains the same as presented in Fig.~\ref{fig:alphas-II}. On top of the various choices for $s_*$, $n_s^{(0)}$, and $r^{(0)}$, we have considered two values, $10^{-2}$ and $10^{-5}$, for $\epsilon_e^{(0)}$.

The latest Planck experiment constrains the running of the spectral index as, {\it e.g.}, $-0.0158 \leq \alpha_s \leq -0.0012$ (Planck TT,TE,EE+lowEB+lensing) at 68\% C.L. On the other hand, once the running of the running of the spectral index is taken into account, the constraint becomes
\begin{align}
0.001 \leq \alpha_s \leq 0.025
\,,\quad 
\text{(Planck TT+lowE+lensing)}
\end{align}
or $-0.008 \leq  \alpha_s \leq 0.012$ (Planck TT,TE,EE$+$lowE$+$lensing), both at 68\% C.L. \cite{Planck:2018jri}. For all cases we have considered, $\alpha_s \lesssim 0.015$. We also note that the running of the spectral index tends to be smaller when $s_*$ takes a larger value, giving $\alpha_s \lesssim 0.007$, which is well within the allowed bounds obtained with the running of the running, while they are larger than the bounds where the running of the running is not taken into account.

Actually, the current constraint on the running is rather weak. It can be well improved in future observations such as the 21 cm line of neutral hydrogen \cite{Kohri:2013mxa,Pourtsidou:2016ctq,Sekiguchi:2017cdy} and galaxy surveys with CMB  \cite{Basse:2014qqa,Munoz:2016owz,Li:2018epc}, which may be useful to differentiate some cases that give the same predictions on $n_s$ and $r$.

%\paragraph{Note added.} This is also a good position for notes added
%after the paper has been written.

\bibliographystyle{JHEP}
%\bibliography{NMAssiInfGen}
\input{NMAssiInfGen.bbl}
%\begin{thebibliography}{99}
%\end{thebibliography}
\end{document}

%% file: NMAssiInfGen.bbl
\providecommand{\href}[2]{#2}\begingroup\raggedright\endgroup